%% file: main.tex
\pdfoutput=1
\documentclass[aps,prl,reprint,floatfix,superscriptaddress,amsmath,amssymb]{revtex4-2}
\input{header}

\begin{document}

\title{The kinematic Kelvin--Helmholtz instability-driven dynamo: \\ implications for merging compact objects}

\author{James R. Beattie$^{\dagger,}$}
\email{beattie@ias.edu \\ $^\dagger$NASA Hubble Fellow}
\affiliation{School of Natural Sciences, Institute for Advanced Study, 1 Einstein Drive, Princeton, NJ 08540, USA}
\affiliation{Department of Astrophysical Sciences, Princeton University, Princeton, NJ 08544, USA}
\affiliation{Canadian Institute for Theoretical Astrophysics, University of Toronto, Toronto, ON M5S 3H8, Canada}

\author{Amitava Bhattacharjee}
\affiliation{Department of Astrophysical Sciences, Princeton University, Princeton, NJ 08540, USA}

\author{Elias R. Most}
\affiliation{TAPIR, Mailcode 350-17, California Institute of Technology, Pasadena, CA 91125, USA}
\affiliation{Walter Burke Institute for Theoretical Physics, California Institute of Technology, Pasadena, CA 91125, USA}

\author{Bart Ripperda}
\affiliation{Canadian Institute for Theoretical Astrophysics, University of Toronto, Toronto, ON M5S 3H8, Canada}
\affiliation{Department of Physics, University of Toronto, 60 St. George St, Toronto, ON M5S 1A7, Canada}
\affiliation{David A. Dunlap Department of Astronomy \& Astrophysics, University of Toronto, 50 St. George St,Toronto, ON M5S 3H4, Canada}
\affiliation{Perimeter Institute for Theoretical Physics, 31 Caroline St. North, Waterloo, ON N2L 2Y5, Canada}

\author{Alexander Philippov}
\affiliation{Department of Physics, University of Maryland, College Park, MD 20742, USA}
\affiliation{Physics Department, Stanford University, Stanford, CA 94305, USA}

\author{Ralf S. Klessen}
\affiliation{Zentrum f\"ur Astronomie, Institut f\"ur Theoretische Astrophysik, Universit\"at Heidelberg, Heidelberg, 69120, Baden-W\"urttemberg, Germany}
\affiliation{Interdisziplin\"ares Zentrum f\"ur Wissenschaftliches Rechnen, Universit\"at Heidelberg, Heidelberg, 69120, Baden-W\"urttemberg, Germany} 

\author{Salvatore Cielo}
\affiliation{Leibniz Supercomputing Center of the Bavarian Academy of Sciences and Humanities, Garching, 85748, Bavaria, Germany}

\begin{abstract}
    Binary neutron star (BNS) mergers excite a Kelvin--Helmholtz instability (KHI) in the shear layer between them. It is parasitized upon by the magnetic field to form a powerful magnetic dynamo. Using a large ensemble of high-resolution visco-resistive KHI-driven dynamo simulations we construct analytical, asymptotic in Reynolds number ($\rm{Re}$) models for the magnetic field growth rates and mean vertical field dynamo onset times. Our models suggest that in the $\rm{Re} \gg 1$ limit, the mean vertical field is seeded immediately, kinematic growth rates become asymptotic in $\rm{Re}$, and the dynamo moves out of the kinematic stage on $\mathcal{O}(0.1\,\rm{ms})$ timescales, readily attaining magnetar-strength fields during this transition. This suggests that most global BNS merger simulations are not time-resolving the correct KHI-driven dynamo regime.
\end{abstract}

\maketitle

\vspace{-5em}
\section{Introduction}
\vspace{-1.5em}
    The merger of two compact objects leads to the formation of an intense shear layer that is susceptible to the Kelvin--Helmholtz instability (KHI)~\cite{Price2006_KHI_ultrastrong,Chabanov2023,Zhu2015,Kiuchi2015_BNS_KHI,Schneider2019,Pakmor2024_merging_white_dwarfs,Radice2024_BNS_review,Ohlmann2026,Kiuchi2026_magnetar,Gutierrez2026}. In binary neutron-star (BNS) mergers the shear layer has a thickness of order $\delta_0 \sim 50\,\rm{m}$ thick and shear velocity of $\Delta U \sim 0.3c$, where $c$ is the speed of light, leading to a shear timescale of $t_{\Delta} = \delta_0/\Delta U \sim 10^{-3}\,\rm{ms}$ \cite{Gutierrez2026}. Simulations show that coupling to the KHI amplifies magnetic fields from $B\sim10^{12}\,\mathrm{G}$ to $B\sim10^{15}$--$10^{16}\,\mathrm{G}$ the magnetic within a few $\rm{ms}$ \cite{Price2006_KHI_ultrastrong,Chabanov2023,Zhu2015,Kiuchi2015_BNS_KHI,Schneider2019,Pakmor2024_merging_white_dwarfs,Ohlmann2026,Kiuchi2026_magnetar,Gutierrez2026}. These strong, magnetar-strength fields \cite{Kiuchi2026_magnetar} can power relativistic jets and potentially connect the shear-layer dynamics (after further large-scale dynamo action \cite{Kiuchi2024,Reboul-Salze:2024jst,Ng:2026jev}) to the electromagnetic counterpart of BNS merger gravitational-wave events \cite{Blandford1977_bz_jets,LIGO2017_BNS_gr_gw,Arcavi2017_BNS_kilonova,Mosta:2020hlh,Most:2023sft,Combi2023_PRL_jets,Kiuchi2024,Hayashi2025_jet_launching,Kiuchi:2026thl}.
    
    BNS merger simulations provide evidence for the type of dynamo that is present -- a kinematic KHI-driven turbulence dynamo~\cite{Price2006_KHI_ultrastrong,Kiuchi2015_BNS_KHI,Radice2024_BNS_review,Kiuchi2026,Gutierrez2026,Kiuchi2026_magnetar}, followed by a large-scale Taylor--Spruit \cite{Margalit:2022rde,Reboul-Salze:2024jst,Ng:2026jev} or $\alpha\Omega$ dynamo on long post-merger timescales \cite{Kiuchi2024,Most:2023sme}. In a standard kinematic ($E_{\rm kin} \gg E_{\rm mag} = B^2/2$, where $E_{\rm kin}$ is the characteristic kinetic energy of the system) turbulence dynamo, $E_{\rm mag}$ grows exponentially without any Lorentz force backreaction to modify the viscous stretching  \cite{Schekochihin2002_nonlinear,Galishnikova2022_tearing,Beattie2023_growth_or_decay,Kriel2026_universal}. If the plasma is in a state of Kolmogorov-type turbulence $E_{\rm kin}(k)\propto k^{-5/3}$ (which is the case in BNS mergers \cite{Aguilera-Miret:2021fre,Gutierrez2026}), where $E_{\rm kin}(k)$ is the kinetic energy spectrum and $k$ is the wavenumber, the prediction for the exponential growth rate at high magnetic Prandtl number $\rm{Pm}$, relevant to BNS mergers \cite{Radice2024_BNS_review,Kiuchi2025_review}, is $\gamma_Et_0\propto\mathrm{Re}^{1/2}$, where $\mathrm{Re}=u_0\ell_0/\nu$ is the plasma Reynolds number with outer-scale characteristic velocity $u_0$, length $\ell_0$ and kinematic viscosity $\nu$~\cite{Schekochihin2004_dynamo,Rincon2019_dynamo_theories,Kriel2026_universal}. Another signature of a dynamo driven by fluctuations is contained within the magnetic energy spectrum, $E_{\rm mag}(k)$.

    In Fourier space, the Kazantsev--Kulsrud--Anderson theory predicts $E_{\rm mag}(k)\propto e^{2\gamma_{E}t/t_0}k^{3/2}K_0(k/k_\eta)$ \cite{Kazantsev1968,Kulsrud1992_stochastic_fields} by prescribing a random Gaussian velocity field, where $K_0(x)$ is the modified Bessel function of the second kind and $k_\eta\sim k_{\rm peak}$ is the Ohmic resistive cutoff and peak $E_{\rm mag}(k)$ scale~\cite{Kazantsev1968,Kulsrud1992_stochastic_fields,Schekochihin2004_dynamo,Rincon2019_dynamo_theories,Schober2012_SSD}. Both $k_{\rm peak}$ and $k_\eta$ are determined by $k_\eta/k_\nu\propto\mathrm{Pm}^{1/2}$, where $k_\nu$ is the viscous wavenumber and the $\mathrm{Pm}^{1/2}$ scaling comes from stretching on $k_\nu$ balancing dissipation on $k_\eta$. Detailed local simulations have confirmed these relations \cite{Schekochihin2004_dynamo,Kriel2022_kinematic_dynamo_scales,Kriel2025_fundamental_scales_II,Kriel2026_universal}. This makes the turbulence dynamo uniquely challenging for global simulations because they do not converge without explicit resistivity ($k_\eta\sim k_{\rm peak}$) and viscosity ($\gamma_Et_0\propto\mathrm{Re}^{1/2}$), the full and consistent inclusion of which has just become possible in GR \cite{Most:2026uga} (see also Ref. \cite{Dionysopoulou:2015tda,Miravet-Tenes:2026gno} for resistive-only approaches). Whether the KHI dynamo fits into this framework is an open and important question for understanding the nature of BNS merger dynamos \cite{Radice2024_BNS_review,Kiuchi2025_review,Gutierrez2026}. Answering it requires controlled first-principles simulations.

    In this Letter we unravel the complete kinematic KHI dynamo chronology. We find that the KHI-driven dynamo is not a turbulence dynamo, in that the kinematic growth rates become independent of $\mathrm{Re}$. Further, within the kinematic regime the dynamo moves through three different stages: (1) winding under planar shear, (2) secondary, three-dimensional instability-driven break-up and (3) fully-fledged KHI-driven dynamo, beyond planar shear, including a vertical field dynamo with an onset time $\propto \mathrm{Re}^{-1/3}$. For BNS merger parameters, we demonstrate that the dynamo immediately seeds a vertical mean field, growth rates become asymptotic (for $k_\nu > \delta_0^{-1}$, a condition not readily met in global simulations) and the dynamo passes through the kinematic stage and becomes nonlinear on timescales of $\mathcal{O}(0.1\,\rm{ms})$, much shorter than the dynamical merger time of the neutron stars.   

\vspace{-2em}
\section{Numerical Models}
\vspace{-1.5em}
    We evolve the three-dimensional, compressible, Newtonian MHD equations with explicit shear viscosity and Ohmic resistivity using the HLLD Riemann solver \cite{Miyoshi2005_HLLD} with first-order-flux correction \cite{Lemaster2009_fofc} and fifth-order TENO5 flux reconstruction \cite{Fu2016_high_order_reconstructions} in \textsc{athenak}~\cite{Stone2026_athenak}. In a periodic domain of size $L$ we initialize two freely decaying, perturbed shear layers with half-width $\delta_0=0.025L$, velocity jump $\Delta U$, density contrast $8/3$, and uniform pressure \cite{Lecoanet2016_KHI_benchmark}. We add a random, three-dimensional magnetic seed with $E_{\rm mag,0}\approx10^{-6}E_{\rm kin,0}$, where $E_{\rm kin,0}$ is the initial kinetic energy of the shear layer. We define $\ReD=\Delta U\delta_0/\nu$, magnetic Reynolds number $\RmD=\Pm\ReD$, where $\Pm=\nu/\eta$, and shear Mach number $\M_\Delta=2\Delta U/(c_{s,\rm in}+c_{s,\rm out})$, where $c_{s,\rm in}$ and $c_{s,\rm out}$ are the sound speeds on either side of the shear layer. Our simulations span $1,\!152^3\leq N_{\rm grid}^3 \leq 2,\!304^3$ grid cells, $64\leq\ReD\leq16,\!384$ at $\mathcal M_\Delta=0.5,1,2$ and $\Pm=1$, such that viscous and resistive scales are resolved at all times. At $\ReD=1,\!024$ and $\mathcal M_\Delta=0.5$, we additionally vary $\Pm=2,5,10$. Our fiducial runs are integrated for $10^3t_{\Delta}$ and use $N_{\rm grid}^3 = 1,\!152^3$. The \hyperref[sec:end_matter]{End Matter} gives the governing equations, initial conditions and numerical methods. The \hyperref[sec:sm_setup]{Supplemental Material}~\cite{SupplementalMaterial} details the growth-rate measurements and spectral diagnostics.

\vspace{-1.5em}
\section{Results}
\begin{figure}[t]
         \includegraphics[width=\columnwidth]{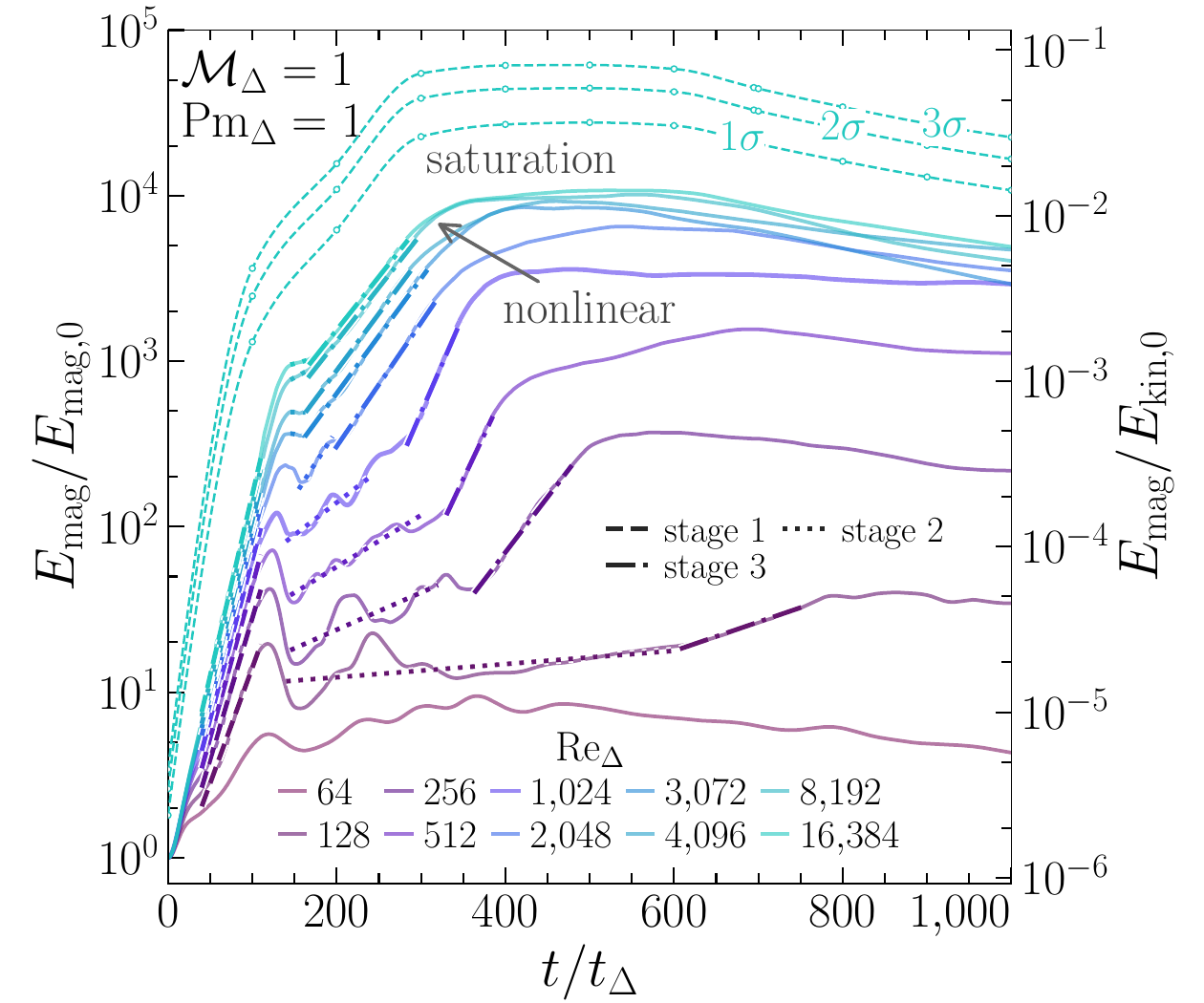}
         \vspace{-2em}
         \caption{\textbf{Volume-integrated magnetic-energy evolution for the KHI-driven dynamo with $\bm{\mathcal M_\Delta=1}$ and $\bm{\Pm=1}$.} Magnetic energy, $\emag$, is normalized by the initial magnetic energy (left axis) and initial kinetic energy (right axis). Additional lines mark stage 1 (winding; dashed), stage 2 (amplification pause; dotted), and stage 3 (exponential growth; dot-dashed). The aqua $1\sigma$, $2\sigma$ and $3\sigma$ curves show the evolution of the energy fluctuations in the $\ReD = 16,\!384$ simulation. The nonlinear and saturation labels mark the stages that come after the kinematic dynamo.}
         \label{fig:energy}
         \vspace{-2em}
    \end{figure}

\vspace{-1.5em}
    \noindent \textit{Volume-integrated magnetic-energy evolution}---\autoref{fig:energy} shows $\emag=\int B^2/2\,dV$, normalized by its initial value. At high $\ReD$, amplification proceeds through an extremely rapid initial stage (stage 1; dashed line), a pause (stage 2; dotted line), and a second period of rapid growth (stage 3; dot-dashed line) before transitioning into the nonlinear growth and saturation ($\partial_t \emag \approx 0)$ at $\emag \approx 10^{-2}\ekin$. The same amplification stages appear in many global BNS simulations~\cite{Palenzuela2022,Chabanov2023,Kiuchi2024,Kiuchi2026_magnetar}. We examine each stage below to identify its amplification mechanism and the conditions governing transitions.

    \noindent\textit{Stage 1: initial amplification by mean-shear stretching}--- Within a few hundred $\td$, the first stage of the $\ReD \gg 1$ KHI dynamo results in $\emag/E_{\rm mag,0} \gtrsim 10^3$. For a BNS merger, this is equivalent to reaching $B\sim10^{16}\,\mathrm G$ in $t \sim 0.1\,\rm{ms}$ (following \cite{Radice2024_BNS_review}). We outline in the following how this is from the initial winding of the resistive magnetic field by the viscous mean-shear layer.
    
    For ideal ($\RmD \rightarrow \infty$) induction in the stationary initial shear $U_0(x)$, the energy obeys $E_{\rm mag}(t)/E_{\rm mag,0}=1+a(t/\td)+b(t/\td)^2,$ where $a=2\td\langle(\partial_xU_0)B_{x0}B_{y0}\rangle/\langle B_0^2\rangle$ and $b=\td^2\langle(\partial_xU_0)^2B_{x0}^2\rangle/\langle B_0^2\rangle$, with angle brackets denoting volume averages, and subscript 0, initial values (see \hyperref[sec:sm_setup]{Supplemental Material}~\cite{SupplementalMaterial}). However, viscosity broadens the shear layer and magnetic flux is not perfectly frozen into the plasma, so the ideal induction equation is insufficient.
    
\begin{figure}[!tbp]         \includegraphics[width=\columnwidth]{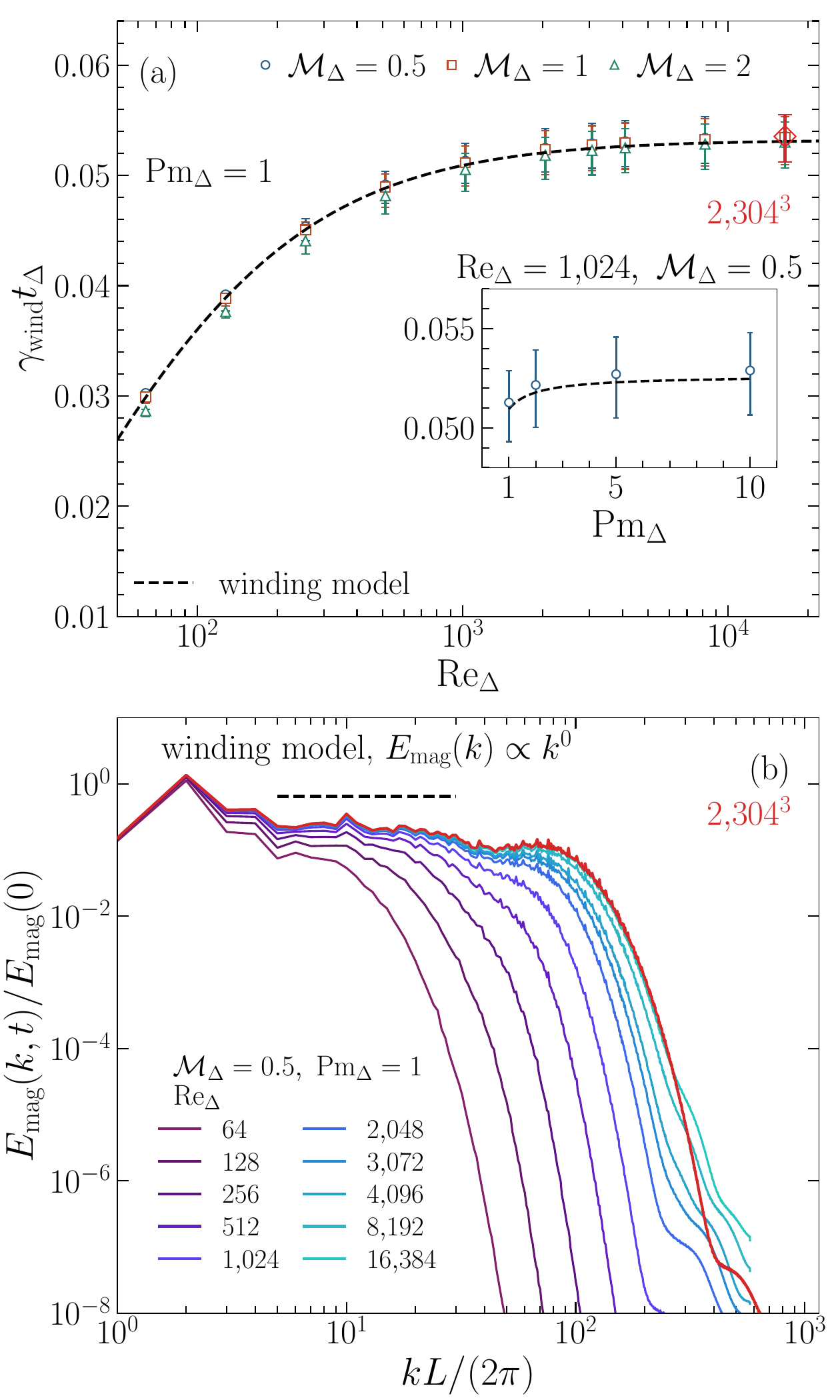}
         \vspace{-2em}
         \caption{\textbf{Initial magnetic amplification by shear winding.} \textbf{(a)} Magnetic-energy growth rates $\gamma_{\rm wind}$ measured during stage 1 in \autoref{fig:energy}. Symbols distinguish $\M_{\Delta}$, and uncertainty bars show sensitivity to the measurement interval. Dashed black curves show the analytical model, \autoref{eq:closed_winding}, evaluated over the same interval. The inset varies $\Pm$ at $\ReD=1,\!024$ and $\mathcal M_\Delta=0.5$. \textbf{(b)} Magnetic spectra towards the end of stage 1, normalized to the initial magnetic energy, for $\mathcal M_\Delta=0.5$ and $\Pm=1$. The dashed line shows the predicted $k^0$ winding dependence (\autoref{eq:shear_spectrum}). In both panels, red markers and curves denote the $2,\!304^3$ run at $\ReD=16,\!384$, $\mathcal M_\Delta=0.5$, and $\Pm=1$, for comparison with the $1,\!152^3$ simulation.}
         \label{fig:growth}
         \vspace{-2em}
    \end{figure}

    Including the viscous broadening of the layer and Ohmic resistive effects gives
    \begin{multline}\label{eq:closed_winding}
         \frac{E_{\rm mag}(t)}{E_{\rm mag,0}}
         \simeq e^{-\alpha t/(\RmD\td)}
         \Bigg[1+a\frac{t}{\td}\\
         +\frac{b(t/\td)^2}
         {\sqrt{1+2(t/\td)(\kappa_\nu/\ReD+\kappa_\eta/\RmD)}}\\
         \times\exp\!\left(-\frac{\beta(t/\td)^3}
         {\RmD[1+2(t/\td)(\kappa_\nu/\ReD+\kappa_\eta/\RmD)]}\right)\Bigg],
    \end{multline}
    with coefficients ($\alpha$, $\beta$, $\kappa_\nu$, $\kappa_\eta$) calculated analytically from the initial $\bm{B}$ field and shear profiles (\hyperref[sec:sm_setup]{Supplemental Material}~\cite{SupplementalMaterial}).     The model recovers purely ideal winding as $\ReD,\RmD\to\infty$. We compare slopes from \autoref{eq:closed_winding} and the simulations, $\gamma_{\rm wind}$, over the same $1\leq t/t_{\Delta} \leq 10$ interval in \autoref{fig:growth}~(a). The model provides excellent agreement with the measured stage~1 $\gamma_{\rm stretch}$ across all sampled $\mathcal M_\Delta$, $\ReD$, and $\Pm$. We therefore identify stage 1 as non-ideal winding by the mean-shear layer, which is insufficient for a sustained dynamo \cite{Zeldovich1957,Cowling1933}.

    Winding transfers magnetic energy to higher $k$. In a locally constant shear, induction gives $k\simeq|\partial_xU|t|k_{y0}|$ and wound-mode energy $\propto t^2$. These determine the spectral transport derived in the \hyperref[sec:sm_energy]{Supplemental Material}. An isotropic, solenoidal seed concentrated initially near $k_0$ admits to
    \begin{equation}\label{eq:shear_spectrum}
         E_{\rm mag}(k,t)\simeq\frac{E_{\rm mag,0}}{4k_0}|\partial_xU|t
         \propto k^0,
    \end{equation}
    for $k_0\ll k\ll\min[|\partial_xU|tk_0,(3/2\eta t)^{1/2}]$, where $|\partial_xU|tk_0$ is the expanding front of wound modes and $k_\eta = (3/2\eta t)^{1/2}$. In the ideal limit, the spectral height and bandwidth each grow as $t$, recovering $\emag \propto t^2$. Hence, as the magnetic field is stretched by the mean-shear, it populates all of the available $k$ linearly in time and uniformly in $\emag(k)$ (see \autoref{fig:growth}~b), which sets the initial magnetic field conditions for the next fully-fledged dynamo stage.

\begin{figure}[!tp]
     \includegraphics[width=\columnwidth]{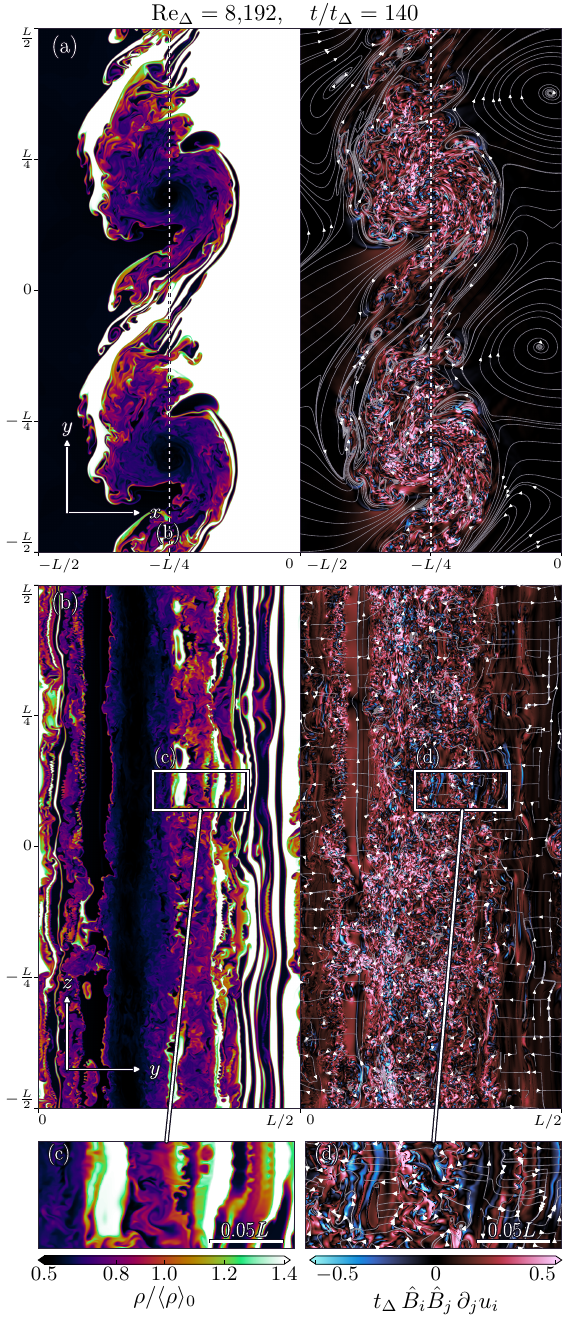}
        \vspace{-2.5em}
     \caption{\textbf{Three-dimensional instability of KHI billows.} All panels show mean-normalized mass density (left) and field-aligned stretching, i.e., dynamo flux, $\td\hat B_i\hat B_j\partial_j u_i$ (right) at $t/\td=140$, $\ReD=8,\!192$, $\Pm=1$. \textbf{(a)} a) \(xy\) slices at $z=0$, $-L/2<x<0$; \textbf{(b)} $yz$ slices at $x=-L/4$ (dashed lines in a), $0<y<L/2$. \textbf{(c,d)} The same transverse patch enlarged to reveal Rayleigh--Taylor instability fingers within the billows. Streamlines trace $(B_x,B_y)$ in (a) and $(B_y,B_z)$ in (b,d). Fly-throughs of the unstable billows are available online in 8K resolution \href{https://www.youtube.com/watch?v=t-6yRoGyFhk}{here}.}
     \label{fig:morphology}
\end{figure}

    \textit{Stage 2: secondary RTI and three-dimensional breakup}--- As the KHI becomes nonlinear, billows wind the density interface into columns of dense and light fluid along $z$. Further, each billow supplies an effective centrifugal acceleration $u_\theta^2/R$, where $u_\theta$ is its circulation speed and $R$ its radius. Small displacements, potentially seeded by the three-dimensional $B$ field, grow via the Rayleigh--Taylor instability (RTI), disrupting the billows \cite{Matsumoto2004,Dixit2010,Skoutnev2021_RTI_dynamo}. \autoref{fig:morphology} shows high-$k$ RTI fingers in the wound density layers, alongside magnetic-field stretching, $\hat{B}_i\hat{B}_j \partial_j u_i$, where $\hat{B}_i\hat{B}_j$ is the unit stress tensor. This shows the RTI excites $k_z$ velocity modes, breaking the planar symmetry of the KHI \footnote{Movies showing different slice orientations and three-dimensional fly-throughs are available \href{https://astro-beattie.com/movies/}{here}.}. The resulting three-dimensional motions facilitate both kinematic large- and small-scale dynamo action, examined below. 

    Under RTI, shorter wavelengths grow faster ($\gamma_{\rm RTI} \propto \sqrt{k}$), but viscosity damps them more strongly ($\propto - k^2$). With local Atwood number $A$ \footnote{Initially $A= (\rho_h - \rho_l)/ (\rho_h + \rho_l) = 5/11$, where $\rho_l$ and $\rho_h$ are the densities of the light and dense fluids.} and destabilizing acceleration $g_{\rm eff}\simeq u_\theta^2/R$ normal to the interface, the linear theory exposes this competition as
    \begin{equation}\label{eq:rti}
         \gamma_{\rm RTI}\td\simeq
         \left[\frac{A g_{\rm eff}\delta_0}{(\Delta U)^2}(k\delta_0)\right]^{1/2}
         -\frac{(k\delta_0)^2}{\ReD}.
    \end{equation}
    Maximizing $\gamma_{\rm RTI}$ to find the fastest-growing mode gives
    \begin{equation}
         \gamma_{\rm RTI}^\infty\td=\frac{3}{4^{4/3}}
         \left[\frac{A g_{\rm eff}\delta_0}{(\Delta U)^2}\right]^{2/3}
         \ReD^{1/3}.\label{eq:rti_max}
    \end{equation}
    Hence, at fixed $g_{\rm eff}$ and $A$, increasing $\ReD$ permits smaller, faster-growing disturbances. We will next show that the $\ReD^{1/3}$ growth-rate scaling sets the subsequent dynamo onset time.
    
    \begin{figure}[!t]
         \includegraphics[width=\columnwidth]{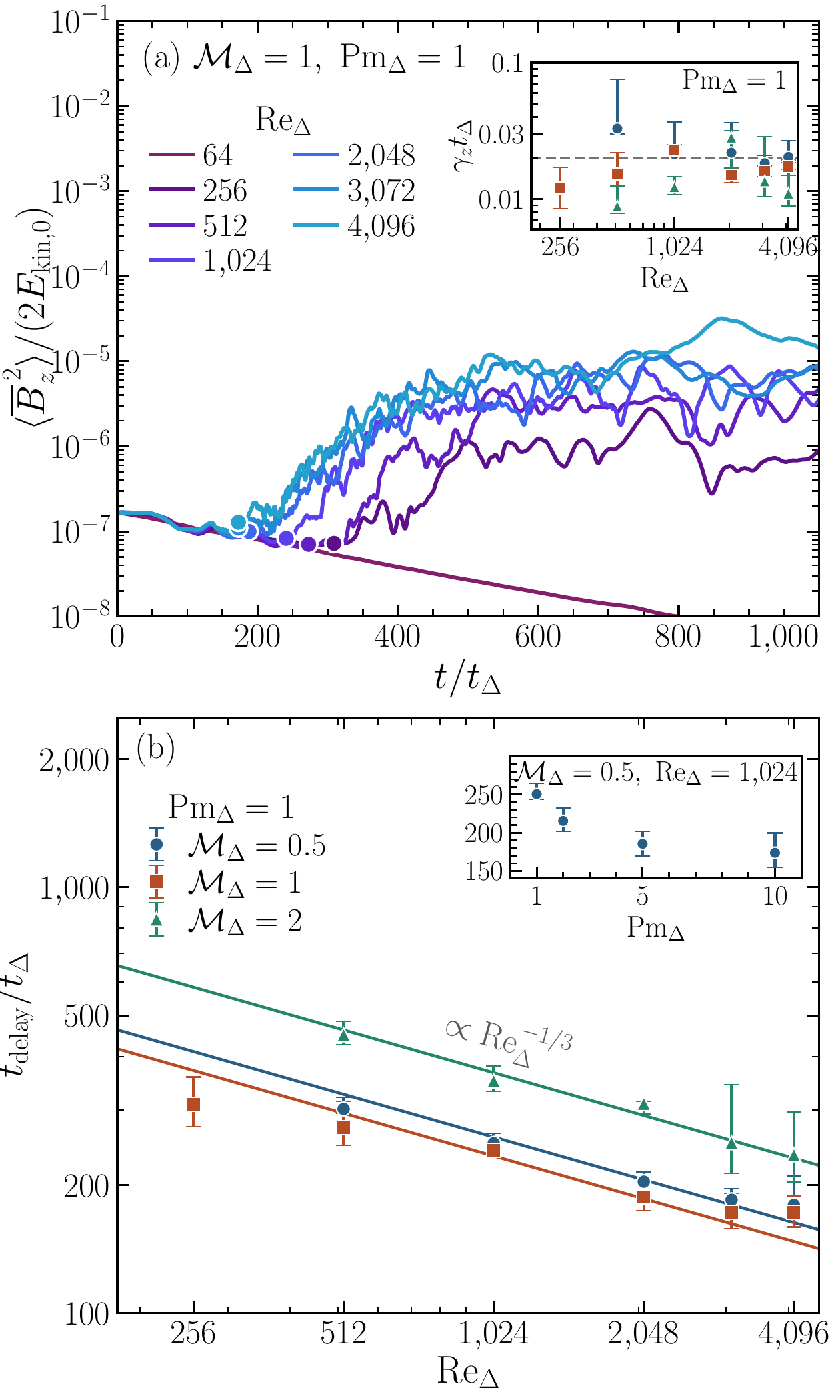}
         \vspace{-2.5em}
         \caption{\textbf{Delayed amplification of the large-scale vertical field.} \textbf{(a)} Energy of the \(yz\)-averaged field $(k_x,k_y=0,k_z=0)$, \(\langle\overline B_z^{\,2}\rangle/(2E_{\rm kin,0})\), for \(\Md=1\) and \(\Pm=1\). Dots mark the growth onset \(t_{\rm delay}\). The inset shows logarithmic growth rates, with a reference line at $\gamma_{\rm z}\td=0.02$. \textbf{(b)} Onset delays $t_{\rm delay}/\td$ at $\Pm=1$. Lines show the RTI prediction $t_{\rm delay} \propto \ReD^{-1/3}$ fitted for $\ReD\leq4,\!096$. The inset varies $\Pm$ at $\Md=0.5$ and $\ReD=1,\!024$. Uncertainty bars show fitting sensitivity (see \hyperref[sec:sm_setup]{Supplemental Material}~\cite{SupplementalMaterial}).}
         \vspace{-1.8em}
         \label{fig:pol}
         \label{fig:onset}
    \end{figure}

    \textit{Stage 3: delayed vertical-field amplification}--- In our geometry, the mean vertical field, $B_z$, relevant for some jet-launching scenarios \cite{Blandford1977_bz_jets} (though subsequent Parker processes can convert toroidal field \cite{Most:2023sft,Musolino:2024sju}), is orthogonal to the $xy$ plane containing the initial flow and its shear gradient. Averaging over \(yz\) planes isolates $\overline B_z(x,t)$, the $(k_x,0,0)$ component. We then measure the layer-averaged $\langle\overline B_z^{\,2}\rangle_x$. \autoref{fig:pol}~(a) shows $\langle\overline B_z^{\,2}\rangle$ growing after the initial winding phase. Because the original shear cannot amplify $B_z$ by winding alone, this later growth provides evidence for a vertical-field dynamo in the three-dimensional flow.

    The dynamo growth rate, $\gamma_z=d\ln\langle\overline B_z^{\,2}\rangle/dt\sim0.02/t_\Delta$, depends weakly on $\ReD$ across all three $\Md$ ensembles (inset~\autoref{fig:pol}~a). Meanwhile, the velocity spectra $E_{\rm kin}(k)$ extend to higher $k$ as $\ReD$ increases and follow a standard $k^{-5/3}$ scaling (Supplemental \autoref{fig:sm_velocity_spectra}), consistent with a shifting $k_{\nu}$ and shorter $t_{\nu}$ for higher $\ReD$. This suggests that $k_{\nu}$ motions do not control the amplification as they do in the standard kinematic turbulence dynamo \citep{Kriel2022_kinematic_dynamo_scales,Kriel2025_fundamental_scales_II}. 
    
    We show in \autoref{fig:onset}~(b) that the onset occurs earlier at successively larger $\ReD$. We interpret the delay as the time required for \(k_z\) modes to break planar symmetry and enable sustained dynamo growth \cite{Cowling1933,Zeldovich1957}. For RTI-driven onset, billows first form over \(t_{\rm roll}\), after which the fastest-growing \(k_z\) modes grow from \(\xi_0\) to the disruptive amplitude \(\xi_{\rm nl}\). The resulting delay is
    \begin{equation}\label{eq:delay}
         \frac{t_{\rm delay}}{\td}\simeq\frac{\ln(\xi_{\rm nl}/\xi_0)}{\gamma_{\rm RTI}^\infty\td} \propto\ReD^{-1/3},
    \end{equation}
    because $\gamma_{\rm RTI}^\infty\td\propto\ReD^{1/3}$. \autoref{fig:onset}~(b) provides a test for this scenario, where we plot $t_{\rm delay}/\td$ as a function of $\ReD$. We show that this provides excellent agreement with the data, with different normalizations for $\Md$. At $\ReD=1,\!024$, onset time shrinks further as $\Pm$ increases from $1$ to $10$ (inset in \autoref{fig:onset}~b), indicating that the magnetization ($E_{\rm mag} \propto \Pm^{2/3}$ by the end of the winding, shown in the \hyperref[sec:end_matter]{End Matter}) also shortens $t_{\rm delay}/\td$. The \hyperref[sec:sm_setup]{Supplemental Material}~\cite{SupplementalMaterial} details the fits and vertical-field induction equations.

\begin{figure}[!tbp]
         \includegraphics[width=\columnwidth]{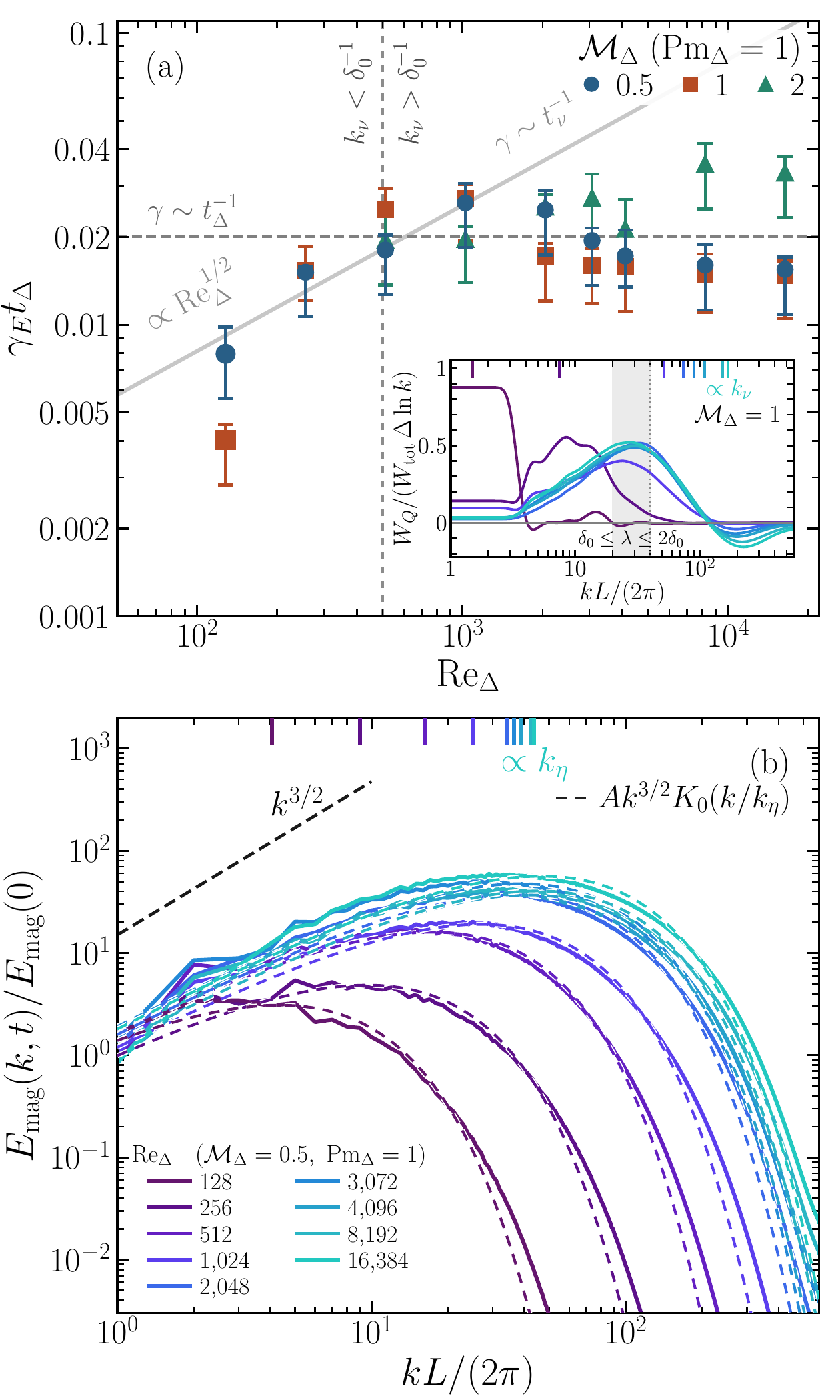}
         \vspace{-2.5em}
         \caption{\textbf{Magnetic amplification during the second growth stage.} \textbf{(a)} Exponential magnetic-energy growth rates at $\Pm=1$. The inset shows signed Lorentz work per logarithmic Fourier shell for $\Md=1$ and $\Pm=1$, with colors as in (b). Shading marks $\delta_0\leq\lambda\leq2\delta_0$, and top ticks mark the viscous scale. Uncertainty bars show sensitivity to the fitting interval (\hyperref[sec:sm_setup]{Supplemental Material}~\cite{SupplementalMaterial}). Lines mark $\gamma_E\td=0.02$ (gray, dashed) and the turbulence dynamo $\gamma_Et_\Delta$ prediction (gray, solid). \textbf{(b)} Magnetic spectra (solid) at 75\% of each measured growth interval for $\mathcal M_\Delta=0.5$ and $\Pm=1$, normalized by the initial total magnetic energy and fit (dashed) to the resistive Kazantsev-Kulsrud-Anderson model, showing good agreement across all $k$.}
         \label{fig:dynamo}
         \vspace{-2em}
    \end{figure}

    \textit{Stage 3 continued: small-scale field amplification}---Three-dimensional motions excited during secondary breakup enable small-scale magnetic amplification alongside the mean-field dynamo. After the pause, which shortens with increasing RTI growth rates, $E_{\rm mag}$ grows exponentially, with the spectra approaching the standard Kazantsev-Kulsrud-Anderson prediction $E_{\rm mag}(k)\propto k^{3/2}$, shown in \autoref{fig:dynamo}~(b). Standard theory then gives $k_{\rm peak}\sim k_{\eta}$ \cite{Schekochihin2004_dynamo,Kriel2022_kinematic_dynamo_scales}. 
    
    In \autoref{fig:dynamo}~(a) we plot the growth rates, $\gamma_E$, 
    by fitting the exponential growth segments for all simulations that exhibit this behavior. At $\ReD < 512$, we find $\gamma_E\td \sim \mathrm{Re}^{1/2}$, as predicted for $\M < 1$ turbulence dynamos~\cite{Schober2012_highPm,Kriel2022_kinematic_dynamo_scales,Kriel2025_fundamental_scales_II,Kriel2026_universal}. For $\ReD \geq 512$ we find $\gamma_{\rm E}\td\sim\gamma_{\rm z}\td\sim0.02$, i.e., independent of $\ReD$, with factor-of-a-few scatter in $\gamma_E$ set by $\M_\Delta$. 
    
    We further compute the Lorentz work as a function of $k$, $W_Q = \langle{\bm{u}_Q}\cdot\bm{J}\times\bm{B}/\rho\rangle$, for logarithmically binned wavenumber shell $Q$ \cite{Grete2017_transfer_funcs,Beattie2025_slk}. We plot $W_Q$ in inset \autoref{fig:dynamo}~(a) and determine that for \(\ReD\gg1\) the peak \(U\rightarrow B\) flux shifts from \(k_\nu\) to \(\approx\delta_0^{-1}\). Further, by measuring the $k_\nu$ via the strain tensor spectrum \cite{Merlini2026_shock_interaction}, we find $\ReD \lesssim 256$ simulations are in a $k_\nu \lesssim (\delta_0)^{-1}$ regime, and conversely for $\ReD \gtrsim 512$. This implies an $\ReD$ independent kinematic growth rate that scales as $\hat{B}_i\hat{B}_j\partial_j u_i \sim t_\Delta^{-1} \sim \Delta U/\delta_0 $, with a dynamo efficiency typical of turbulence dynamos, $\mathcal{O}(10^{-2})$ \cite{Beresnyak2012_universal_ssd,Kriel2026_universal}.

\vspace{-2em}
\section{Discussion}
\vspace{-1.5em}
    Now we discuss the implications this has for the dynamo in compact object mergers. To directly compare to our theory, let us first consider $\ReD$ for a BNS system. In dense parts of the merger, the viscosity is set by neutrinos and is of order $\nu_\nu\sim10^7$--$10^8\,\mathrm{cm^2\,s^{-1}}$~\cite{Guilet:2016sqd,Gavassino2025_shear_viscosity_radiation,ReboulSalze2025_neutrino_viscosity,NgMost2026}. With $\delta_0=50\,\mathrm{m}$ set by the scale height at the stellar surface, and $\Delta U=0.3c$~\cite{Gutierrez2026} gives $\ReD\simeq4.5\times10^5$--$4.5\times10^6$ and $\td\simeq0.56\,\mu\mathrm{s}$. We therefore use the high-$\ReD$ limit of our model.

    At high $\ReD$, magnetic energy grows by four orders of magnitude before nonlinear backreaction produces saturation at \(E_{\rm mag}/E_{\rm kin,0}\sim10^{-2}\). Using $B\sim10^{17}\,\mathrm G \,(E_{\rm mag}/E_{\rm kin,0}/0.1)^{1/2}$ from \cite{Radice2024_BNS_review}, this corresponds to growing to a strength of $B\sim3\times10^{16}\,\mathrm{G}$ (as in  \cite{Kiuchi2026}) in $t \lesssim 0.5\,\rm{ms}$. Further, \autoref{fig:energy} shows that the $3\sigma$ energy fluctuations, occupying \(<1\%\) of the volume, approach $B\sim10^{17}\,\mathrm{G}$. The KHI-driven dynamo alone in BNS mergers can therefore create the strongest known magnetic fields in the Universe.

    Driven mean-shear layers can generate a mean magnetic field aligned with the flow~\cite{Tripathi2026_jets}, but amplification perpendicular to the shear-flow plane has remained unclear. This vertical (seed) field is a critical ingredient for jet-launching in BNS ~\cite{Kiuchi2024,Chan2026_jets_from_scratch}, likely in addition to subsequent large-scale dynamo action for long-lived merger remnants \cite{Combi2023_PRL_jets,Kiuchi2024,Most:2023sme,Ng:2026jev,Kiuchi:2026thl}. We demonstrate that the KHI generates such a field as three-dimensional modes become unstable to RTI, with onset following our $t_{\rm delay}$ relation (\autoref{fig:pol}) and subsequent growth at $\gamma_z\sim0.02\td^{-1}$, matching the mean-shear prediction from \autoref{fig:growth}. Using the \cite{Radice2024_BNS_review} scaling and \autoref{fig:pol}, the saturated strength is $\overline{B}_z \sim 10^{14}-10^{15}\,\rm{G}$. Furthermore, because $t_{\rm delay} \propto \ReD^{-1/3}$ due to the visco-resistive RTI, disruption of the stage~1 winding ends almost immediately as $\ReD \rightarrow \infty$ \footnote{The corresponding wavelength is
    \begin{align}
        \lambda_{\mathrm{RTI}}^{\infty}
        &= 2\pi\left(\frac{16\nu^2 R}{A u_\theta^2}\right)^{1/3}.
    \end{align}
    Using $\Delta U=0.3c$, $u_\theta=\Delta U/2$, $\delta_0=50\,\mathrm{m}$, density ratio $\rho_h/\rho_l=8/3$, Atwood number $A=5/11\simeq0.4545$, and kinematic viscosity $\nu_\nu=10^7\,\mathrm{cm^2\,s^{-1}}$ \cite{Skoutnev2021_RTI,Gavassino2025_shear_viscosity_radiation,ReboulSalze2025_neutrino_viscosity,NgMost2026}, for the BNS system we obtain \(\lambda_{\rm RTI}^{\infty}\approx6\,\mathrm{cm}\), below the spatial resolution of current global and even local zoom-in simulations \cite{Gutierrez2026,Kiuchi2026}.}. Hence, in this limit, the KHI-driven dynamo proceeds directly to stage 3 where it grows both planar-averaged vertical (i.e., $B_z(k_x,k_y=0,k_z=0)$) and fluctuating $(k\neq0)$ magnetic fields.

    We show in both \autoref{fig:onset} and \autoref{fig:dynamo} that the dynamo growth rates become asymptotically independent of $\ReD$, with the peak in Lorentz work fixed to $\approx \delta_0^{-1}$. In this limit $k_\nu \gtrsim \delta_0^{-1}$, which may be used as a criterion for $\gamma_E$ convergence in global merger simulations. 
    
    Replacing the measured \(\gamma_E\simeq11\,\mathrm{ms}^{-1}\)~\cite{Gutierrez2026} with our asymptotic \(\gamma_E\simeq36\,\mathrm{ms}^{-1}\) increases the magnetic energy by \(\exp[(36-11)\Delta t]\) over the same interval \(\Delta t\), i.e., an additional factor of 100 within \(\Delta t\simeq0.18\,\mathrm{ms}\). This is sufficient to raise an initial $E_{\rm mag}/E_{\rm kin}\sim 10^{-3}$ field well into the nonlinear dynamo regime $E_{\rm mag}/E_{\rm kin} \sim 10^{-1}$ in only $ t\simeq0.18\,\mathrm{ms}$. This suggests that global merger simulations are time-resolving the wrong KHI-driven dynamo stage, and either require properly calibrated subgrid recipes to skip stages 1 and 2 or resolve $\Delta x = \mathcal{O}(\rm cm)$ such that $\delta_0$ and the fastest-growing RTI modes are resolved.

    The same KHI dynamo mechanism may operate in unstable shear layers behind core-collapse supernova shocks~\cite{Guilet2010_SASI} and in galaxy-cluster cold fronts~\cite{Roediger2013_Virgo}. In these hot, diffuse systems, where $\ReD \propto n_eT^{-5/2} \sim 10^3$ \cite{Ferriere2020_plasma_turbulence}, $n_e$ is the electron number density and $\ReD$ is defined assuming collisional viscosity, the first stages of the kinematic KHI dynamo may become relevant.

\vspace{-2em}
\section*{Acknowledgments}
\vspace{-1.5em}
    We acknowledge productive discussions with Luciano Combi, Bindesh Tripathi, Uddipan Banik, Chris Thompson, Adam Burrows, Jim Stone, Jacob Fields, Alexander Dittman, Romain Teyssier and Christoph Federrath. J.~R.~B., R.~S.~K. and S.~C. acknowledge high-performance computing resources provided by the Leibniz Rechenzentrum and the Gauss Centre for Supercomputing through grants pr73fi, pn76gi and pn76ga, and GCS large-scale project 10391. J.~R.~B. acknowledges allocation rrg-essick from the Digital Research Alliance of Canada and funding from the Natural Sciences and Engineering Research Council of Canada (NSERC) (funding reference number 568580). J.~R.~B. is supported by NASA through NASA Hubble Fellowship grant HST-HF2-51597.001-A, awarded by the Space Telescope Science Institute and administered by AURA for NASA under contract NAS5-26555. J.~R.~B. and A.~B. acknowledge support from NSF Award 2206756. E.~R.~M. acknowledges support from the U.S. National Science Foundation {(NSF)} under Grant No. PHY-2541792. E.~R.~M. is also supported by a Research Fellowship from the Sloan Foundation and a William H. Hurt Scholarship at Caltech. B.~R. acknowledges the support of the Natural Sciences and Engineering Research Council of Canada (NSERC), funding reference numbers RGPIN-2023-04844 and ALLRP 592744. B.~R. and A.~P. are supported by a grant from the Simons Foundation (MP-SCMPS-00001470). B.~R. is supported by the Canadian Space Agency (23JWGO2A01). R.~S.~K. acknowledges financial support from the ERC via Synergy Grant ``ECOGAL'' (project ID 855130) and from DFG in project ``STARCLUSTERS'' (funding ID KL 1358/22-1). This research was supported in part by grant NSF PHY-2309135 to the Kavli Institute for Theoretical Physics (KITP).

\input{supplemental_citations.tex}

\input{main.bbl}
\clearpage

\section*{End Matter}
\label{sec:end_matter}
\raggedbottom
\setcounter{equation}{0}
\renewcommand{\theequation}{A\arabic{equation}}
\def\theHequation{end.\arabic{equation}}
\vspace{-1em}

\subsection*{\normalfont\bfseries Equations and numerical methods.}
\vspace{-1em}
    We solve the compressible, Newtonian magnetohydrodynamics equations in the triply periodic domain $[-L/2,L/2]^3$,
    \begin{gather}
     \partial_t\rho+\bnab\cdot(\rho\bm u)=0,\\
     \begin{split}
     \partial_t(\rho\bm u)+\bnab\cdot[\rho\bm u\bm u
     +(P+B^2/2)\mathbb I-\bm B\bm B]=\bnab\cdot\bm\Pi,
     \end{split}\\
     \begin{split}
     \partial_t\mathcal E+\bnab\cdot[(\mathcal E+P+B^2/2)\bm u
     -(\bm u\cdot\bm B)\bm B]\\
     =\bnab\cdot[\bm\Pi\cdot\bm u+\eta\bm B\times\bm J],
     \end{split}\\
     \partial_t\bm B=\bnab\times(\bm u\times\bm B)+\eta\nabla^2\bm B,
     \qquad\bnab\cdot\bm B=0,\label{eq:em_setup_induction}
    \end{gather}
    where $\bm J=\bnab\times\bm B$, $\mathcal E=\rho\epsilon+\rho u^2/2+B^2/2$, and $P=(\gamma_{\rm ad}-1)\rho\epsilon$ with $\gamma_{\rm ad}=5/3$. The magnetic permeability to set to unity. We include only shear viscosity, with stress tensor $\Pi_{ij}=2\rho\nu S_{ij}$, where $S_{ij}=(\partial_j u_i+\partial_i u_j)/2-\bnab\cdot\bm u\delta_{ij}/3$. The constant coefficients are $\nu=\Delta U\delta_0/\ReD$ and $\eta=\nu/\Pm$. We use \textsc{athenak}~\cite{Stone2026_athenak} with fifth-order targeted essentially non-oscillatory (TENO5) reconstruction \cite{Fu2016_high_order_reconstructions}, HLLD fluxes, third-order Runge--Kutta integration, constrained transport, and first-order flux correction. TENO5 reduces numerical diffusion by factors of approximately $1.5$ relative to WENO-Z and $3.5$ relative to PLM (used in \cite{Beattie2023_growth_or_decay,Beattie2025_bulk_viscosity}). Hence, in terms of resolved modes, our fiducial $\sim 1,\!000^3$ domains are equivalent to $\sim 3,\!500^3$ domains for PLM and $\sim 1,\!500^3$ for WENO-Z.
    
    All simulations ran on LRZ’s \textsc{SuperMUC-NG} Phase 1 and SciNet’s \textsc{Trillium}, using approximately $10^8$ core-hours.

\begin{figure}
    \centering
    \includegraphics[width=\linewidth]{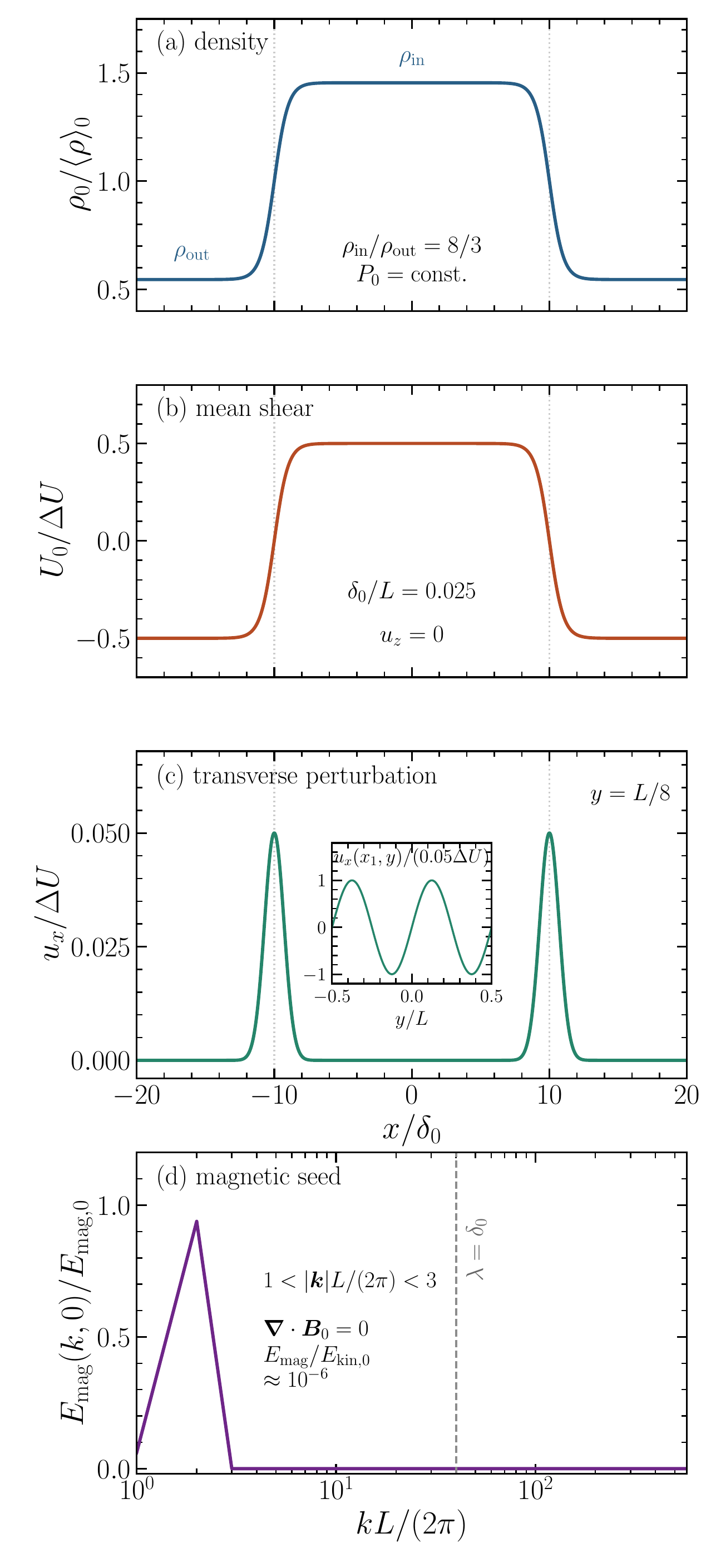}
    \caption{Initial conditions shared by all simulations. \textbf{(a)} The initial plasma density profile, e.g., the stratification for each shear layer. \textbf{(b)} The mean shear profile. \textbf{(c)} The transverse perturbation (inset) as well as damping function to isolate the perturbation the shear layer. \textbf{(d)} The magnetic energy power spectrum, $\emag(k)$ of the initial seed magnetic field, shown in isotropic wavenumber $k$. }
    \label{fig:ics}
\end{figure}

\vspace{-1em}
\subsection*{\normalfont\bfseries Shear, density, and thermodynamic state.}
\vspace{-1em}
    The layers are centered at $x_{1,2}=\pm L/4$, following the smooth construction of Refs.~\cite{Lecoanet2016_KHI_benchmark,Berlok2019_KHI_linear_theory},
    \begin{align}
     H(x)&=\tfrac12\left[\tanh\frac{x+L/4}{\delta_0}
                      -\tanh\frac{x-L/4}{\delta_0}\right],\\
     U_0(x)&=\Delta U[H(x)-\tfrac12],\quad
     \rho_0(x)=\rho_{\rm out}[1+\tfrac53H(x)],
    \end{align}
    shown visually in \autoref{fig:ics}~(a) and (b). We use $\delta_0/L=0.025$, uniform initial pressure, and a density contrast $\rho_{\rm in}/\rho_{\rm out}=8/3$. In units with $\delta_0=\Delta U=\langle\rho\rangle_0=1$, we choose $P_0$ to give the specified $\mathcal M_\Delta$. The in-plane velocity perturbation, visualized in \autoref{fig:ics}~(c), is
    \begin{equation}
     u_x(x,y)=0.05\Delta U\sum_{j=1}^{2}
     e^{-(x-x_j)^2/\delta_0^2}\sin(4\pi y/L).
    \end{equation}
    The velocity is initially planar, with $u_z=0$. We prescribe no independent out-of-plane velocity perturbation. The weak three-dimensional Lorentz force can seed secondary perturbations that excite the RTI. In Supplemental \autoref{sec:sm_vertical} we discuss how resistivity may affect this seeding and the measured onset delays.

\vspace{-1em}
\subsection*{\normalfont\bfseries Magnetic seed.}
\vspace{-1em}
    We initialize a divergence-free field, $\bm B=\bnab\times\bm A$, using a random vector potential $\bm A$ with a parabolic spectral envelope over $1<|\bm k|L/(2\pi)<3$. We plot the $\emag(k)$ of the field in \autoref{fig:ics}~(d). This places the magnetic energy at wavelengths $\gg\delta_0$, representing the merging stars’ initial dipole fields. We normalize the field to $E_{\rm mag,0}/E_{\rm kin,0}\simeq1.32\times10^{-6}$, placing it initially in the kinematic regime ($\ekin\gg\emag$), as expected initially in BNS mergers \cite{Radice2024_BNS_review}.

\vspace{-1em}
\subsection*{\normalfont\bfseries Magnetic-energy dependence on $\Pm$.}
\label{sec:em_pm_pause}
\begin{figure}[!tbp]
     \centering
     \includegraphics[width=\columnwidth]{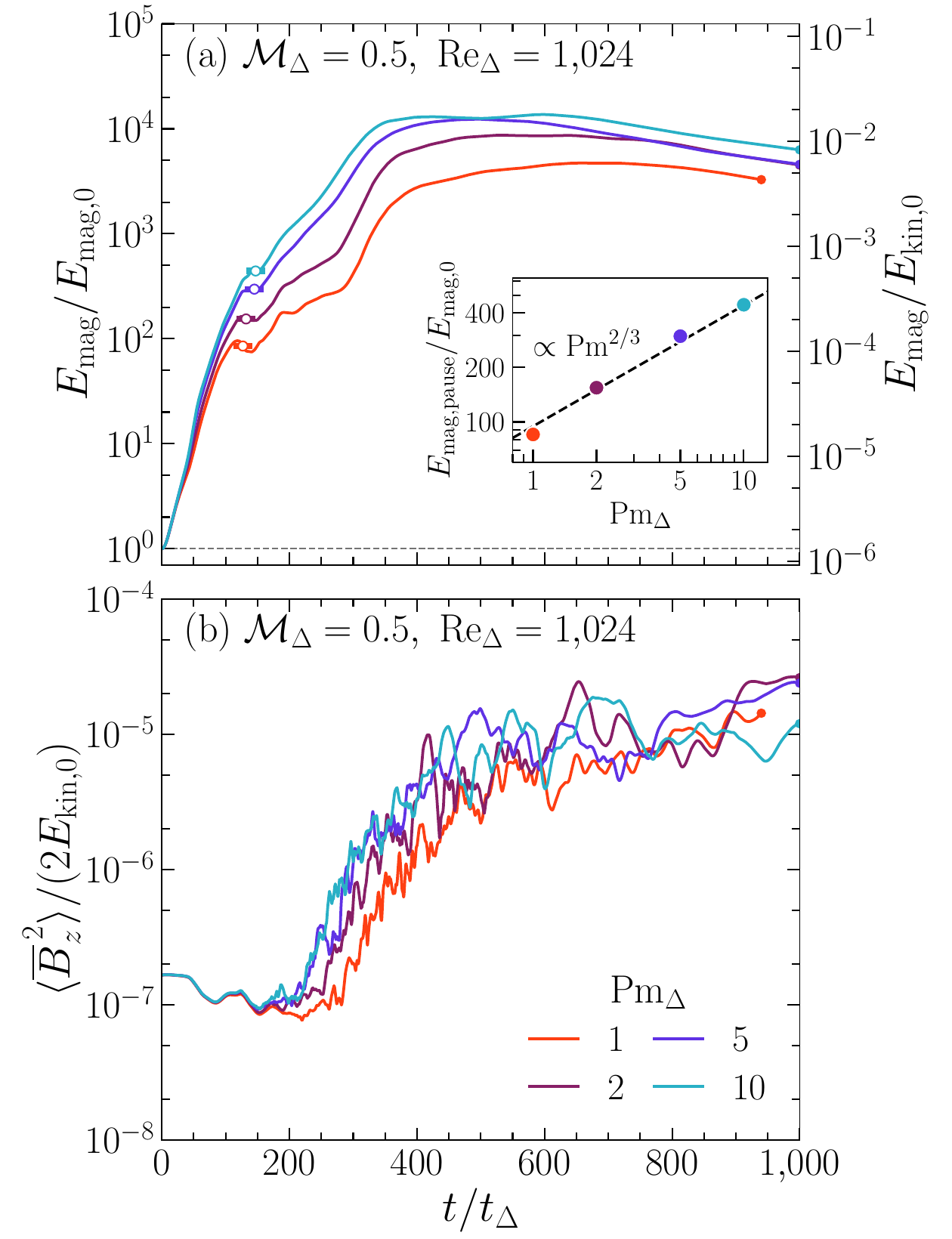}
     \caption{\textbf{Magnetic amplification at varying $\bm{\Pm}$.} All runs have $\mathcal M_\Delta=0.5$ and $\ReD=1,\!024$. Colors identify $\Pm$ in both panels. \textbf{(a)} Total magnetic energy normalized by $E_{\rm mag,0}$ (left axis) and $E_{\rm kin,0}$ (right axis). Open markers show mean pause energies; horizontal segments show the averaging intervals. The inset shows the energies as a function of $\Pm$, averaged over $20\td$ around the pause in growth. The dashed curve shows $E_{\rm wound, pause}\propto \Pm^{2/3}$, predicted by \autoref{eq:em_pm_peak}. \textbf{(b)} Energy of the $yz$-averaged vertical field, normalized by $E_{\rm kin,0}$.}
     \label{fig:pm_histories}
\end{figure}

\vspace{-1em}
    At fixed $\ReD$, increasing $\Pm = \nu/\eta = \RmD/\ReD$ reduces resistive diffusion while leaving the viscosity unchanged. The insets of \autoref{fig:growth} and \autoref{fig:onset} show the $\Pm>1$ growth rates and onset delays. Here we examine the $\emag$ evolution of the same simulations. \autoref{fig:pm_histories}~(a) shows that more $\emag$ is generated directly after the winding at higher $\Pm$, and in general higher $\emag/E_{\rm mag,0}$ within saturation, at least until the largest two $\Pm$. We articulate below how the higher winding energies, may account for the differences in late-time saturated energy.

    What follows is a simple argument that models the energy saturation for stage 1 (the winding; \autoref{fig:energy}) for different $\Pm$. Winding increases the magnetic energy as $E_{\rm wound}/E_{\rm mag,0}\sim(t/\td)^2$ while reducing the transverse magnetic scale as $\ell_\perp/\ell_{\perp,0}\sim\td/t$, where $\ell_{\perp,0}$ is the initial transverse magnetic scale. Here $\ell_\perp$ measures variation across the wound field, analogous to the reversal scale in the standard kinematic turbulence dynamo~\cite{Schekochihin2004_dynamo}. When resistive diffusion balances winding, $\ell_\eta\sim\ell_{\perp,0}t_\Delta/t_\eta$, halting transfer to higher $k$. Using $t_\eta\sim\ell_\eta^2/\eta$ gives $t_\eta \sim\ell_{\perp,0}^2t_\Delta^2/(t_\eta^2\eta)$ and hence 
    \begin{align}
        \frac{t_\eta}{t_\Delta}
        \sim\left(\frac{\ell_{\perp,0}^2}{\eta t_\Delta}\right)^{1/3},
    \end{align}
    and therefore
    \begin{equation}\label{eq:em_pm_peak}
     \frac{E_{\rm wound,peak}}{E_{\rm mag,0}}
     \sim\left(\frac{t_\eta}{\td}\right)^2
     \sim\left(\frac{\ell_{\perp,0}^2}{\eta\td}\right)^{2/3}
     \propto\Pm^{2/3},
    \end{equation}
    where $\RmD = \Delta U\delta_0 /\eta = \delta_0^2/(\eta t_\Delta) = \ReD\Pm$ gives the $\Pm$ proportionality. Higher $\Pm$ (at fixed $\ReD$) therefore allows the field to be wound into higher-$k$ modes before resistive truncation, as expected. The dashed curve in the inset of \autoref{fig:pm_histories}~(a) shows that $E_{\rm wound,peak}\propto\Pm^{2/3}$ agrees perfectly with the data.

    The vertical-field energies in \autoref{fig:pm_histories}~(b) depend weakly on $\Pm$, although their onset occurs earlier at higher $\Pm$, as discussed in the main text and shown in \autoref{fig:onset}. Because the KHI billows, where the winding occurs, become unstable on timescales $\propto \ReD^{-1/3}$ as $\ReD \rightarrow \infty$, the winding stage becomes so short-lived that we expect it to fundamentally disappear. We therefore suspect that this $\Pm^{2/3}$ dependence weakens at $\ReD\gg1$. Testing this requires a systematic study that is beyond the scope of this study.

\label{sec:end_matter_end}

\clearpage
\flushbottom
\onecolumngrid
\setcounter{page}{1}
\renewcommand{\thepage}{S\arabic{page}}
\begin{center}
 {\large\bfseries Supplemental Material for\\[0.4em]
 The kinematic Kelvin--Helmholtz instability-driven dynamo: \\ implications for merging compact objects\par}
 \medskip
 James R. Beattie, Amitava Bhattacharjee, Elias R. Most,\\
 Bart Ripperda, Alexander Philippov, Ralf S. Klessen, and Salvatore Cielo
 \medskip

 {\small Citation numbers refer to the reference list of the main text.}
\end{center}
\label{sec:supplemental_start}
\twocolumngrid
\setcounter{secnumdepth}{1}
\setcounter{section}{0}
\setcounter{subsection}{0}
\setcounter{figure}{0}
\setcounter{equation}{0}
\setcounter{table}{0}
\renewcommand{\thesection}{\Roman{section}}
\renewcommand{\thefigure}{S\arabic{figure}}
\renewcommand{\theequation}{S\arabic{equation}}
\renewcommand{\thetable}{S\Roman{table}}
\def\theHsection{supp.\arabic{section}}
\def\theHequation{supp.\arabic{equation}}
\def\theHfigure{supp.\arabic{figure}}
\def\theHtable{supp.\arabic{table}}
\input{supplemental_body.tex}

\end{document}

%% file: header.tex
\usepackage{graphicx,bm,xcolor}
\makeatletter
\renewcommand{\fnum@figure}{\textbf{\figurename~\thefigure}}
\makeatother
\usepackage[colorlinks=true,allcolors=blue]{hyperref}
\providecommand{\apjs}{Astrophys. J. Suppl. Ser.}
\providecommand{\aap}{Astron. Astrophys.}
\providecommand{\apjl}{Astrophys. J. Letters.}
\providecommand{\mnras}{Mon. Not. R. Astron. Soc.}
\newcommand{\ReD}{\mathrm{Re}_\Delta}
\newcommand{\RmD}{\mathrm{Rm}_\Delta}
\newcommand{\Pm}{\mathrm{Pm}_\Delta}
\newcommand{\td}{t_\Delta}
\newcommand{\Md}{\mathcal{M}_\Delta}
\newcommand{\M}{\mathcal{M}}
\newcommand{\emag}{E_{{\rm mag}}}
\newcommand{\ekin}{E_{{\rm kin}}}
\newcommand{\bnab}{\bm{\nabla}}

%% file: supplemental_citations.tex
\nocite{Dixit2010,FontaneJoly2008,Matsumoto2004,MatsumotoSeki2007}

%% file: supplemental_body.tex
\vspace{-1em}
\section{The phases of the KHI-driven dynamo in $k$ space at high Reynolds number}\label{sec:sm_energy}
\begin{figure}
    \centering
    \includegraphics[width=\linewidth]{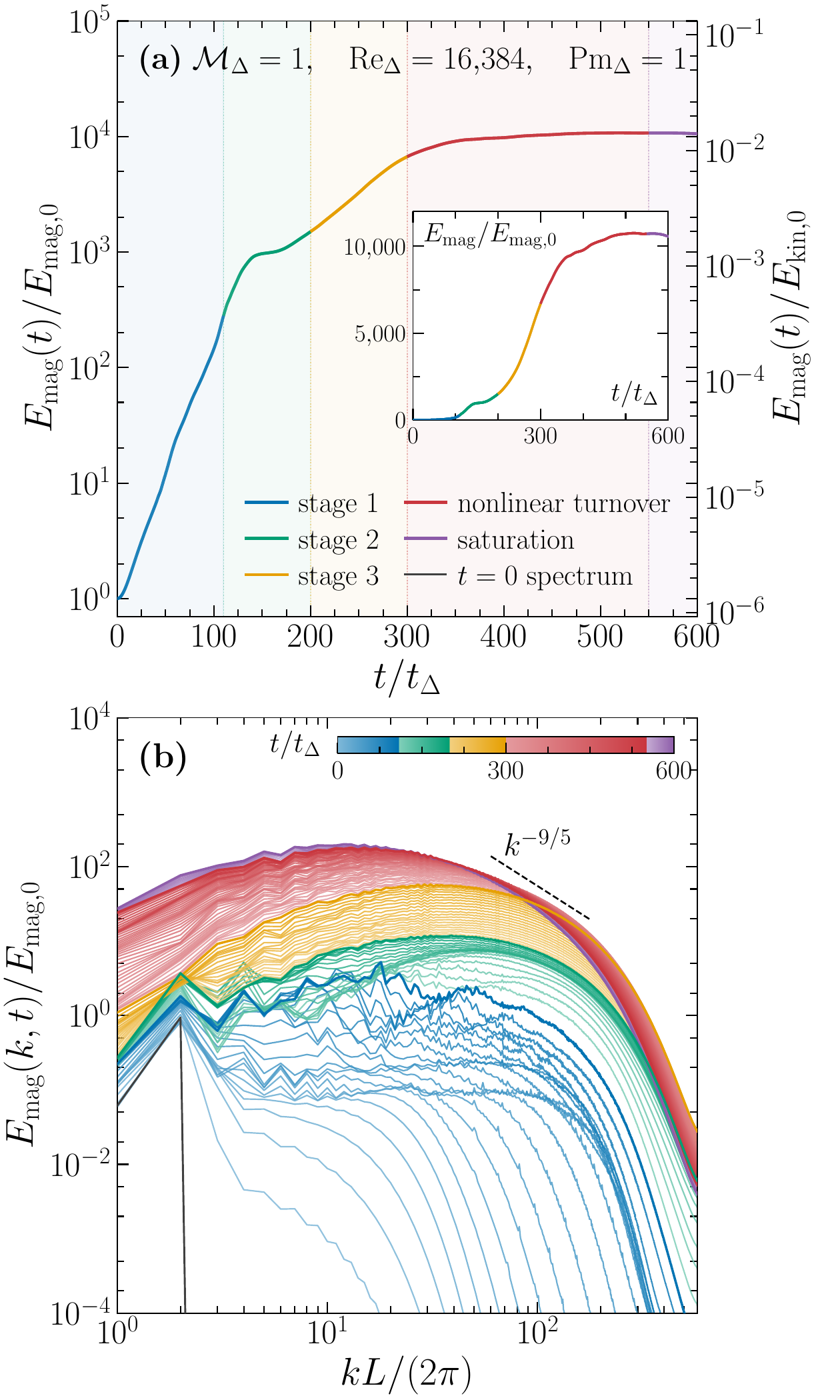}
\vspace{-1.5em}
    \caption{Evolution of the volume-integrated magnetic energy ($\emag$, panel a) and magnetic spectrum ($\emag(k)$, panel b) for the $\M_{\Delta}= 1$, $\ReD = 16,\!384$, $\Pm = 1$ simulation. Annotations in panel (a) and the time color map in panel (b) identify the dynamo stages. During the nonlinear stage, magnetic energy shifts toward longer wavelengths \cite{Schekochihin2002_nonlinear,Galishnikova2022_tearing,Kriel2026_universal,Kiuchi2026}. The $k^{-9/5}$ spectrum, recently measured in high-resolution MHD turbulence simulations, is plotted for reference \cite{Fielding2023_plasmoid,Beattie2025_nature_astronomy}.}
\vspace{-1.5em}
    \label{fig:spectra_by_stage}
\end{figure}

\vspace{-1em}
    The main text follows magnetic spectra and volume-integrated energies across our parameter range. Here we focus on $\ReD=16,!384$, $\M_\Delta=1$, and $\Pm=1$. Here we focus on a single simulation, $\ReD = 16,384$, $\M_{\Delta} = 1$, and $\Pm = 1$. We plot the $\emag$ evolution and $\emag(k)$ evolution in the top and bottom panels of \autoref{fig:spectra_by_stage}, respectively. We also show the $\emag$ evolution on a linear-linear axes in the inset of panel (a), which can be revealing for episodes of algebraic growth (e.g., \cite{Kriel2025_fundamental_scales_II,Kriel2026_universal}). 
    
    The first stage (blue) exhibits the winding spectrum, $\emag(k) \propto k^0$ from \autoref{eq:closed_winding}. Stage 2 marks the transition into a three-dimensional dynamo, via the RTI as discussed in the main text (and shown in \autoref{fig:morphology} and \autoref{fig:clear_RTI}). The RTI excites magnetic modes across $k$ space, from the long wavelengths to the short wavelengths, establishing the $\emag(k) \propto k^{3/2}$ Kazantsev–Kulsrud–Anderson spectrum \cite{Kazantsev1968,Kulsrud1992_stochastic_fields} in stage 3. Stage 3 (yellow) progresses until reaching the nonlinear dynamo stage, where the scale-by-scale backreaction of the Lorentz on the flow suppresses the kinematic dynamo and makes the induction equation nonlinear in $\bm{B}$ \cite{Schekochihin2002_nonlinear,Galishnikova2022_tearing,Kriel2026_universal,Beattie2026_sda_prl,Kiuchi2026}. We will examine the nonlinear regime (red) in subsequent work. Here, magnetic energy shifts toward longer wavelengths, departing from $\emag(k)\propto k^{3/2}$ and reaching slopes as steep as $\emag(k)\propto k^{-9/5}$ (annotated in \autoref{fig:spectra_by_stage}) within the inertial range \cite{Fielding2023_plasmoid,Beattie2025_nature_astronomy}.

\vspace{-2em}
\section{The three-dimensional RTI at moderate Reynolds number}\label{sec:sm_rti}
\begin{figure}
    \centering
    \includegraphics[width=\linewidth]{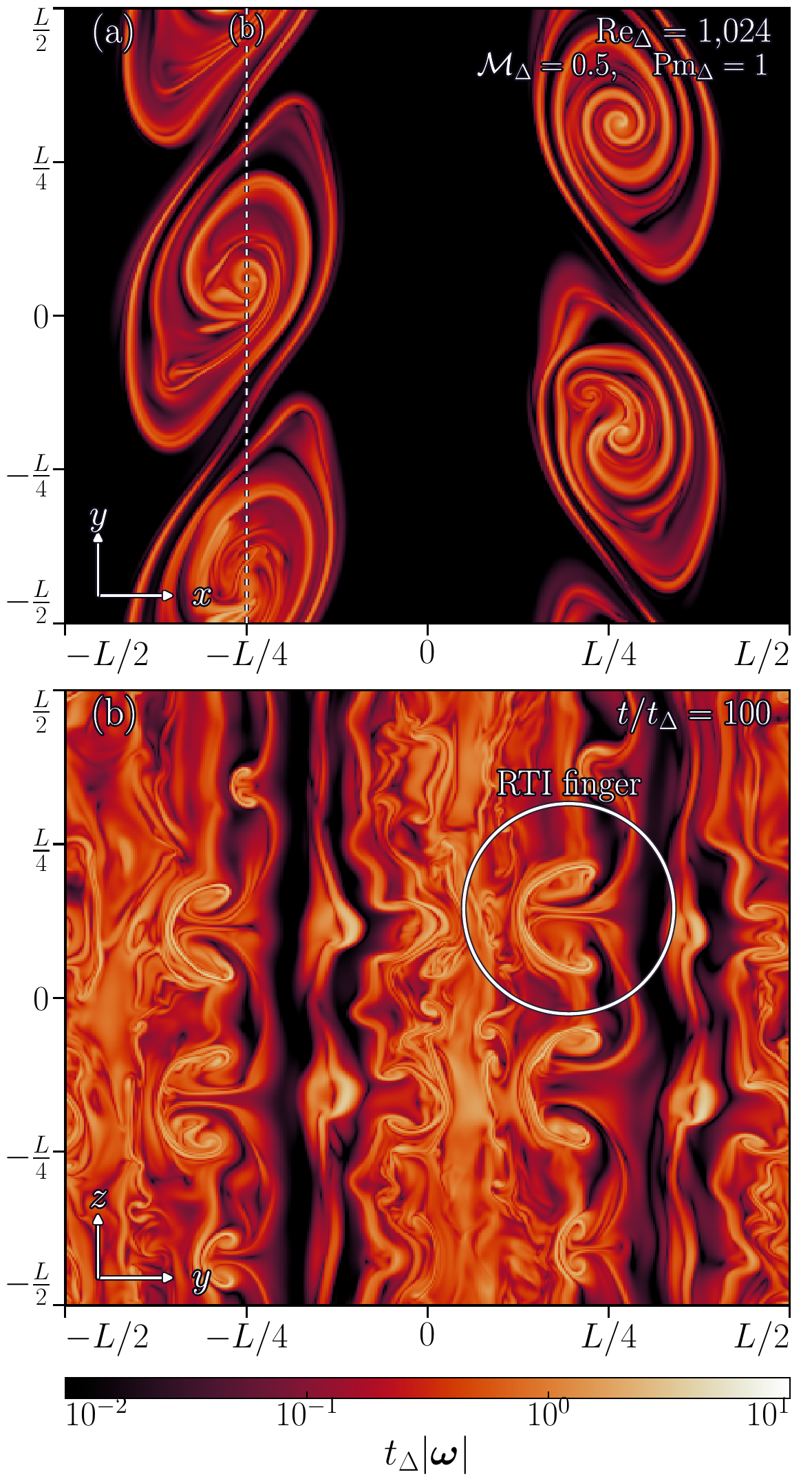}
    \vspace{-1.5em}
    \caption{Vorticity-magnitude slices, $|\bm\omega|=|\bnab\times\bm u|$, for the $\ReD=1,\!024$ simulation with enhanced magnetization to suppress secondary shear instabilities \cite{Stone2007_RTI}. Coordinates are shown in the bottom-left corner of each panel. \textbf{(a)} $|\bm{\omega}|$ in the $xy$ plane, showing mature KHI billows during stage 2 (\autoref{fig:spectra_by_stage}). \textbf{(b)} The $x=-L/4$ slice of $|\bm{\omega}|$ in the $zy$ plane (dashed line in panel a), showing prominent, long-wavelength RTI modes growing along $z$ through the billows. The white circle highlights one RTI finger.}
    \vspace{-2em}
    \label{fig:clear_RTI}
\end{figure}

\vspace{-1em}
    \autoref{fig:morphology} shows extremely short-wavelength RTI modes, which grow fastest in the inviscid limit, $\gamma_{\rm RTI}\propto\sqrt{k}$, disrupting the billows. Increasing viscosity lowers $\ReD$ and shifts the fastest-growing modes toward longer wavelengths. Further, by increasing the magnetization from $\emag \approx 10^{-6}E_{\rm{kin},0}$ to $E_{\rm mag,0} \approx 10^{-3}E_{\rm{kin},0}$ we can suppress secondary shear instabilities that disrupt the RTI fingers \cite{Stone2007_RTI}. Hence we run a $\emag \approx 10^{-3}E_{\rm{kin},0}$, $\ReD = 1,\!024$, $\Pm=1$, $\M_{\Delta} = 0.5$ simulation and show $|\bm{\omega}| = |\bnab\times\bm{u}|$ in the $xy$ and $yz$ planes as three-dimensional modes become unstable in \autoref{fig:clear_RTI}. 

    The top panel shows large KHI billows with less secondary structure than at $\ReD=8,\!192$ in \autoref{fig:morphology}. The bottom panel shows the $x=-L/4$ slice, marked by the white dashed line above. Because the simulation is more viscous and more magnetized than the main text simulations, we see very well-preserved RTI fingers on long wavelengths. The fingers grow inward from the billows’ outer edges, protruding into the small radii direction. The same process occurs at higher $k$ in the main-text simulations, giving $t_{\rm delay}\propto\ReD^{-1/3}$ from the viscous RTI growth-rate scaling, \autoref{eq:delay}.

    \vspace{-1.5em}
    \section{Winding energy and spectrum}\label{sec:sm_winding}
    \vspace{-1em}
        We derive the energy evolution and magnetic spectrum during stage~1, retaining only the planar shear $\bm u=U(x,t)\bm e_y$. We neglect the imposed perturbation, billow roll-up, three-dimensional motions, and magnetic feedback, and show that this model reproduces the growth rates and spectrum.
        
        We assume an incompressible flow, $\bnab\cdot\bm{u} = 0$, so the initial density $\rho_0(x)$ remains fixed. The initial profiles and constant viscosity $\nu$ and resistivity $\eta$ are specified in the End Matter. The complete derivation, which includes every step and the coefficient evaluations, is available in \href{https://astro-beattie.com/lecture-notes/Beattie2026_winding_derivation_notes.pdf}{\textit{Magnetic-energy evolution and spectrum under shear winding}}.

    \paragraph*{\normalfont\bfseries Ideal winding.}
        For a stationary shear without resistivity, induction gives $B_x=B_{x0}$, $B_z=B_{z0}$, and $B_y=B_{y0}+t(\partial_xU_0)B_{x0}$, with initial fields evaluated at $(x,y-U_0(x)t,z)$. Squaring and averaging over the periodic volume gives $E_{\rm mag}/E_{\rm mag,0}=1+a(t/\td)+b(t/\td)^2$, where
        \begin{align}
         a&=2\td\frac{\langle(\partial_xU_0)B_{x0}B_{y0}\rangle}{\langle B_0^2\rangle},\label{eq:a}\\
         b&=\td^2\frac{\langle(\partial_xU_0)^2B_{x0}^2\rangle}{\langle B_0^2\rangle}.\label{eq:b}
        \end{align}
        The linear term $(a)$ is the cross term between the initial and generated streamwise fields and can have either sign. The quadratic term $(b)$ is the generated field’s energy.

    \paragraph*{\normalfont\bfseries Resistive losses.}
        Winding also creates magnetic gradients, which are subject to resistive loss. In a locally constant shear, a magnetic mode with initial wavevector $\bm p$ has $k_x=p_x-t(\partial_xU_0)p_y$, $k_y=p_y$, and $k_z=p_z$. Resistivity multiplies its energy by $\exp[-2\eta\int_0^t k^2dt']$, where
        \begin{equation}\label{eq:shear_wave_damping}
         \int_0^t k^2dt'=p^2t-(\partial_xU_0)p_xp_yt^2
         +\tfrac13(\partial_xU_0)^2p_y^2t^3.
        \end{equation}
        We represent diffusion of the initial field and of winding-created gradients by $e^{-\alpha\eta t/\delta_0^2}$ and $e^{-\beta\eta t^3/(\delta_0^2\td^2)}$. Matching the initial resistive loss, $-\eta\langle J_0^2\rangle$, and the cubic loss weighted by the generated energy gives
        \begin{align}
         \alpha&=2\delta_0^2\frac{\langle J_0^2\rangle}{\langle B_0^2\rangle},\qquad \bm J_0=\bm\nabla\times\bm B_0,\label{eq:alpha}\\
         \beta&=\frac{2\delta_0^2\td^2}{3}
         \frac{\langle(\partial_xU_0)^4|\partial_yB_{x0}|^2\rangle}
         {\langle(\partial_xU_0)^2B_{x0}^2\rangle}.\label{eq:beta}
        \end{align}
        These exponentials approximate the distribution of local losses and omit the orientation-dependent $t^2$ term (see \href{https://astro-beattie.com/lecture-notes/Beattie2026_winding_derivation_notes.pdf}{\textit{the full derivation}}). The additional cubic loss, which comes directly from \autoref{eq:shear_wave_damping}, is applied to the quadratic winding term.

    \paragraph*{\normalfont\bfseries Viscous layer broadening.}
        Resistivity spreads the generated field, while viscosity weakens its source through $\partial_tU=\nu\rho_0^{-1}\partial_x(\rho_0\partial_xU)$. Taking $B_{x0}$ approximately constant across an isolated layer, the induction equation gives
        \begin{align}
         B_{y,\rm wound}\simeq B_{x0}t\bigg\{&\partial_xU_0+\frac{t}{2}\bigg[\eta\partial_x^3U_0\nonumber\\
         &+\nu\partial_x[\rho_0^{-1}\partial_x(\rho_0\partial_xU_0)]\bigg]\bigg\}.
        \end{align}
        The factor $1/2$ follows from integrating the changing shear and diffusion over time. Squaring and integrating across the layer gives a relative energy correction $1-t(\kappa_\nu\nu+\kappa_\eta\eta)/\delta_0^2+\cdots$. Integration by parts, with vanishing boundary terms at infinity, gives analytical expressions for the diffusion coefficients,
        \begin{align}
             \kappa_\eta&=\delta_0^2
             \frac{\displaystyle\int_{-\infty}^{\infty}(\partial_x^2U_0)^2dx}
             {\displaystyle\int_{-\infty}^{\infty}(\partial_xU_0)^2dx},\label{eq:kappa_eta}\\
             \kappa_\nu&=\kappa_\eta-\frac{\delta_0^2}{2}
             \frac{\displaystyle\int_{-\infty}^{\infty}(\partial_x^2\ln\rho_0)(\partial_xU_0)^2dx}
             {\displaystyle\int_{-\infty}^{\infty}(\partial_xU_0)^2dx}.\label{eq:kappa_nu}
        \end{align}
        We evaluate these coefficients using the tanh profiles with density contrast $8/3$, neglecting layer overlap, and use them to calculate the growth rates in \autoref{fig:growth}.

    \paragraph*{\normalfont\bfseries Finite-time energy model.}
        To extend the initial correction, we approximate the wound field by a fixed-shape profile of width $w(t)$, preserving $\int B_{y,\rm wound}dx=B_{x0}t\Delta U$. Its energy relative to ideal winding is then $\delta_0/w(t)$. A diffusive width that matches the initial correction is
        \begin{equation}
            w^2(t)=\delta_0^2+2t(\kappa_\nu\nu+\kappa_\eta\eta).
        \end{equation}
        The broader layer also weakens the representative squared shear by $\delta_0^2/w^2(t)$, motivating the cubic-loss factor $e^{-\beta\eta t^3/[\td^2w^2(t)]}$. Multiplying the ideal polynomial by the initial-field loss, and its quadratic term by these broadening and cubic-loss factors, gives \autoref{eq:closed_winding} after substituting $\nu=\delta_0^2/(\ReD\td)$ and $\eta=\delta_0^2/(\RmD\td)$.

        All coefficients listed throughout this section are evaluated from the initial conditions of $\bm{B}$ or the shear profile. Through direct evaluation of \autoref{eq:a}, \autoref{eq:b}, \autoref{eq:alpha}, \autoref{eq:beta}, \autoref{eq:kappa_nu} and \autoref{eq:kappa_eta} we find
        \begin{align}
             a&=-0.00106582,\\
             b&=0.00345363,\\
             \alpha&=0.236542,\\
             \beta&=0.00629781,\\
             \kappa_\nu &=\frac{152,\!064}{3,\!125}\ln\!\frac83-\frac{29,\!298}{625}
                 \simeq0.850822,\\
                \kappa_\eta&=\frac45.
        \end{align}

    \paragraph*{\normalfont\bfseries Ideal winding spectrum.}
        The same mode evolution determines the energy per unit wavenumber, $E_{\rm mag}(t)=\int E_{\rm mag}(k,t)dk$. For $|\partial_xU|t\gg1$ and $k\gg k_0$, a magnetic mode has $k\simeq|\partial_xU|tq$, where $q=|p_y|$, while its generated energy grows as $t^2$. If $F(q)dq$ is the seed’s $x$-component energy in $q,q+dq$, then
        \begin{align}
             E_{\rm mag}(k,t)dk&\simeq(\partial_xU)^2t^2F(q)dq,\\
             dk&=|\partial_xU|t\,dq.
        \end{align}
        For a homogeneous, isotropic seed concentrated at $|\bm p|=k_0$, $q$ is uniform on $[0,k_0]$. A solenoidal field lies perpendicular to $\bm p$. With equal energy in the two directions in this plane, its $x$-component fraction is $(1-p_x^2/k_0^2)/2$. Averaging over azimuth about the flow direction at fixed $q$ gives $\langle p_x^2\rangle=(k_0^2-q^2)/2$, hence a mean fraction $[1+(q/k_0)^2]/4$. Thus
        \begin{align}
             F(q)&=\frac{E_{\rm mag,0}}{4k_0}\left[1+\left(\frac{q}{k_0}\right)^2\right],\\
             E_{\rm mag}(k,t)&\simeq\frac{E_{\rm mag,0}}{4k_0}|\partial_xU|t
             \left[1+\left(\frac{k}{|\partial_xU|tk_0}\right)^2\right].
        \end{align}
        The bracket approaches unity for $k\ll|\partial_xU|tk_0$, giving the $k^0$ spectrum in \autoref{eq:shear_spectrum}. Spectral height and bandwidth each grow as $t$, and the full expression integrates to the leading generated energy $E_{\rm mag,0}(\partial_xU)^2t^2/3$. Restoring resistivity adds $e^{-2\eta k^2t/3}$ from the cubic term in \autoref{eq:shear_wave_damping}, restricting the flat band to $k_0\ll k\ll\min[|\partial_xU|tk_0,(3/2\eta t)^{1/2}]$.

\vspace{-1.5em}
\section{Growth-rate measurements}\label{sec:sm_setup}
\vspace{-1em}
\paragraph*{\normalfont\bfseries Stage 1.}
    We measure $\gamma_{\rm wind}$ in \autoref{fig:growth}(a) by fitting $\ln E_{\rm mag}$ over $1\leq t/\td\leq10$, with a free intercept and equal weighting in time. We apply the same fit to the winding model. Uncertainty bars span fits over windows of width $1,2,3,4,5\td$, centered at each $\td$ within the $1\leq t/\td\leq10$ interval.

\paragraph*{\normalfont\bfseries Stage 3.}
    We measure stage-3 growth by fitting $\ln E_{\rm mag}=c+\gamma_Et$, weighting all times equally. We identify sustained three-dimensional motions when $[\langle u_z^2\rangle-\langle u_z\rangle^2]^{1/2}/\Delta U\geq0.003$ persists for $20\td$. We locate the pause from minima in the growth rate measured over windows of $10,20,30\td$. For each width, we select the first minimum after $100\td$ with a rate below $0.015/\td$ and a drop of at least $0.003/\td$ relative to the surrounding slopes. Growth fits begin at least $10\td$ after the latest identified pause and after three-dimensional motions develop.
    
    Accepted intervals span at least $20\td$ and one energy e-folding. We require positive growth and an rms departure from the fitted $\ln E_{\rm mag}$ of at most $0.05$. The rates in the two halves must differ by at most $30\%$ of the full-interval rate, and each quarter-interval rate must lie within $30\%$ of it. Among qualifying intervals, we select the one with the largest magnetic-energy amplification. Uncertainty bars span fits to the full interval, its halves and quarters, and qualifying intervals with endpoints shifted by up to $5\td$. Hence, our uncertainty bars measure fitting sensitivity across a range of parameters around the interval.

\begin{figure*}[tbp]
     \includegraphics[width=\textwidth]{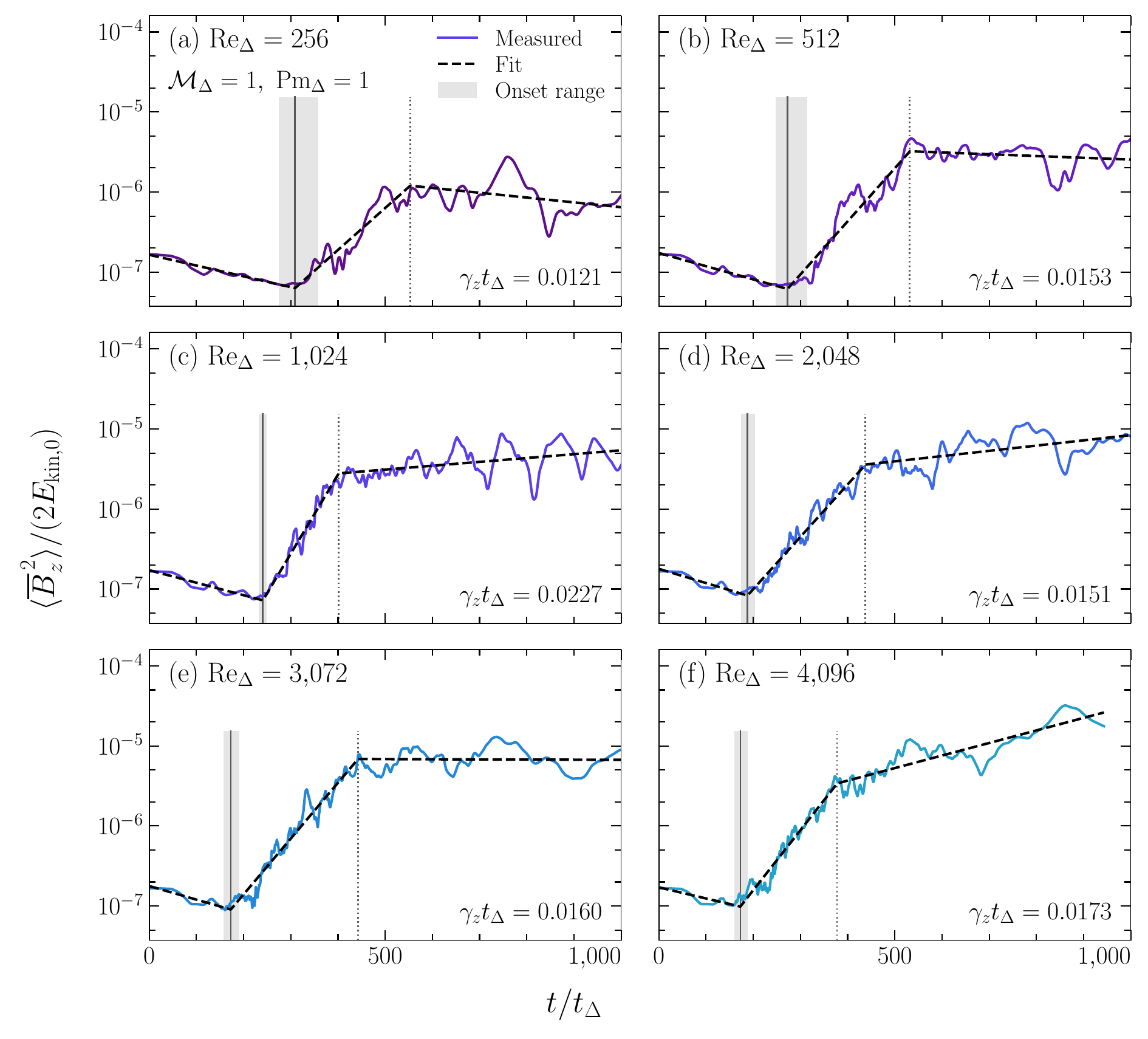}
     \caption{Vertical-field onset and growth fits at $\mathcal M_\Delta=1$, $\Pm=1$, and $256\leq\ReD\leq4,\!096$. Solid colored curves show $\langle\overline B_z^{,2}\rangle/(2E_{\rm kin,0})$; black dashed curves show the three-segment fits to its logarithm. Solid vertical lines mark $t_{\rm delay}$, gray bands show the uncertainty ranges defined in the text. Each panel identifies $\ReD$ (top left) and the middle-segment growth rate $\gamma_z\td$ (bottom right).}
     \label{fig:sm_onset_fits}
\end{figure*}

    \vspace{-1em}
    \section{Vertical-field onset and growth}\label{sec:sm_vertical}
    \vspace{-1em}
    \paragraph*{\normalfont\bfseries Induction and geometry.}
    Our analysis in the main text pertains to simulations for $\ReD \leq 4,\!096$, where we have data for the large-scale field at high cadence. We average $B_z$ over $yz$ planes to obtain $\overline B_z(x,t)$, then average $\overline B_z^{\,2}$ over $-L/2<x<0$, denoted by angle brackets. Averaging the induction equation over these planes gives
    \begin{equation}
         \partial_t\overline B_z=\partial_x\overline{\left(u_zB_x-u_xB_z\right)}+\eta\partial_x^2\overline B_z,
    \end{equation}
    hence, the unperturbed shear, $\bm u=U_0(x)\bm e_y$, supplies no source for $\overline B_z$. Its amplification therefore requires motions beyond the original shear, initially seeded by the RTI, as we show throughout the study.

    \paragraph*{\normalfont\bfseries Extracting onset times and uncertainty.}
    We fit three joined straight segments to the logarithm of the mean vertical-field energy. The first change in slope defines $t_{\rm delay}$, and the middle slope gives $\gamma_z$. Each segment lasts at least $20\td$. The middle slope must be positive, exceed both neighboring slopes, and span at least one energy e-folding. \autoref{fig:sm_onset_fits} shows the fits and onset uncertainties for $\M_\Delta=1$. We determine uncertainty bars using the same ``fit sensitivity" method as in stage 3.

    \paragraph*{\normalfont\bfseries Delay scaling.}
    Neglecting the roll-up time, we fit $t_{\rm delay}/\td=C\ReD^{-1/3}$ over $256\leq\ReD\leq4,\!096$, fixing the predicted exponent and fitting $C$ separately for each Mach ensemble with equal weighting in logarithmic delay. At fixed billow structure and initial disturbance, this hydrodynamic RTI estimate has no explicit $\Pm$ dependence but reproduces the viscosity scaling in \autoref{fig:onset}.

    \section{Additional power spectra}\label{sec:spectral}
    \vspace{-1em}

\begin{figure*}[p]
     \includegraphics[width=\textwidth]{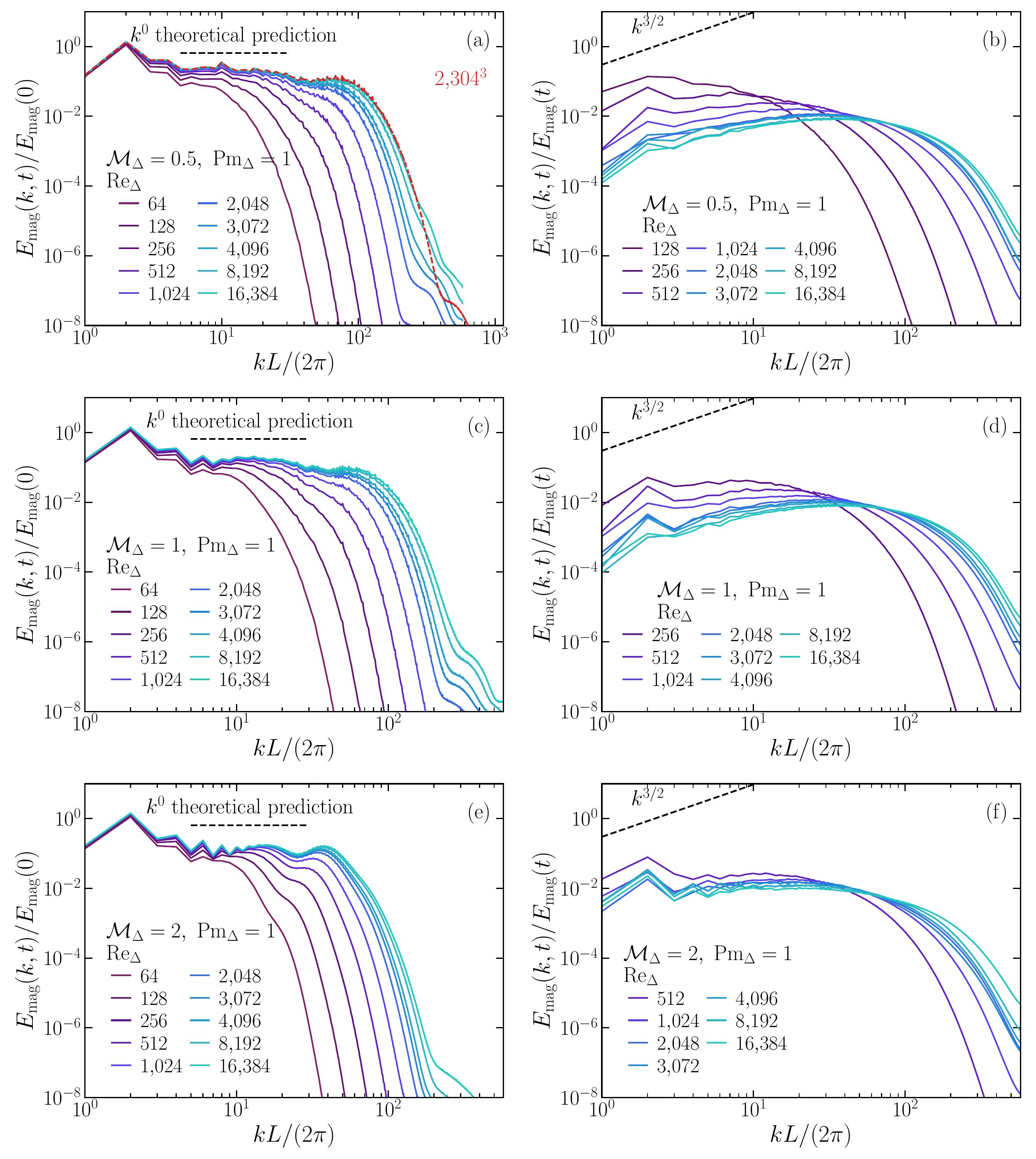}
     \caption{Magnetic spectra, $\emag(k)$, at $\Pm=1$ for $\mathcal M_\Delta=0.5$ (a,b), 1 (c,d), and 2 (e,f). Left: spectra at $t/\td\simeq50$, normalized by $E_{\rm mag,0}$. Right: spectra at the midpoints of the stage-3 growth intervals, normalized by $\emag(t)$ to compare spectral shapes. Colors identify $\ReD$. Dashed lines show the $k^0$ (winding spectrum; \autoref{eq:shear_spectrum}) and $k^{3/2}$ slopes. The red dashed curve in (a) shows the $2,\!304^3$ comparison at $\ReD=16,\!384$, at the same time and with the same normalization. Note not all of the simulations that have a winding phase become exponentially growing within the $10^{3}\td$ that we simulate (e.g., either too viscous or never dynamo super-critical).}
     \label{fig:sm_mach_spectra}
\end{figure*}

    \autoref{fig:sm_mach_spectra} shows the magnetic power spectra $\emag(k)$ for all three $\M_{\Delta}$ ensembles, colored by $\ReD$. Early snapshots (left column) lie near $50\td$, and later spectra are extracted approximately halfway through the stage 3 kinematic growth interval. We normalize early spectra by $E_{\rm mag,0}$, and later spectra, by $E_{\rm mag}(t)$. This allows us to compare spectral morphologies for different $\ReD$ and $\M_{\Delta}$. Reference lines show $\emag(k)\propto k^0$ on the left and $\emag(k)\propto k^{3/2}$ on the right. All of the spectra behave as expected. 

\begin{figure*}[p]
     \includegraphics[width=\textwidth]{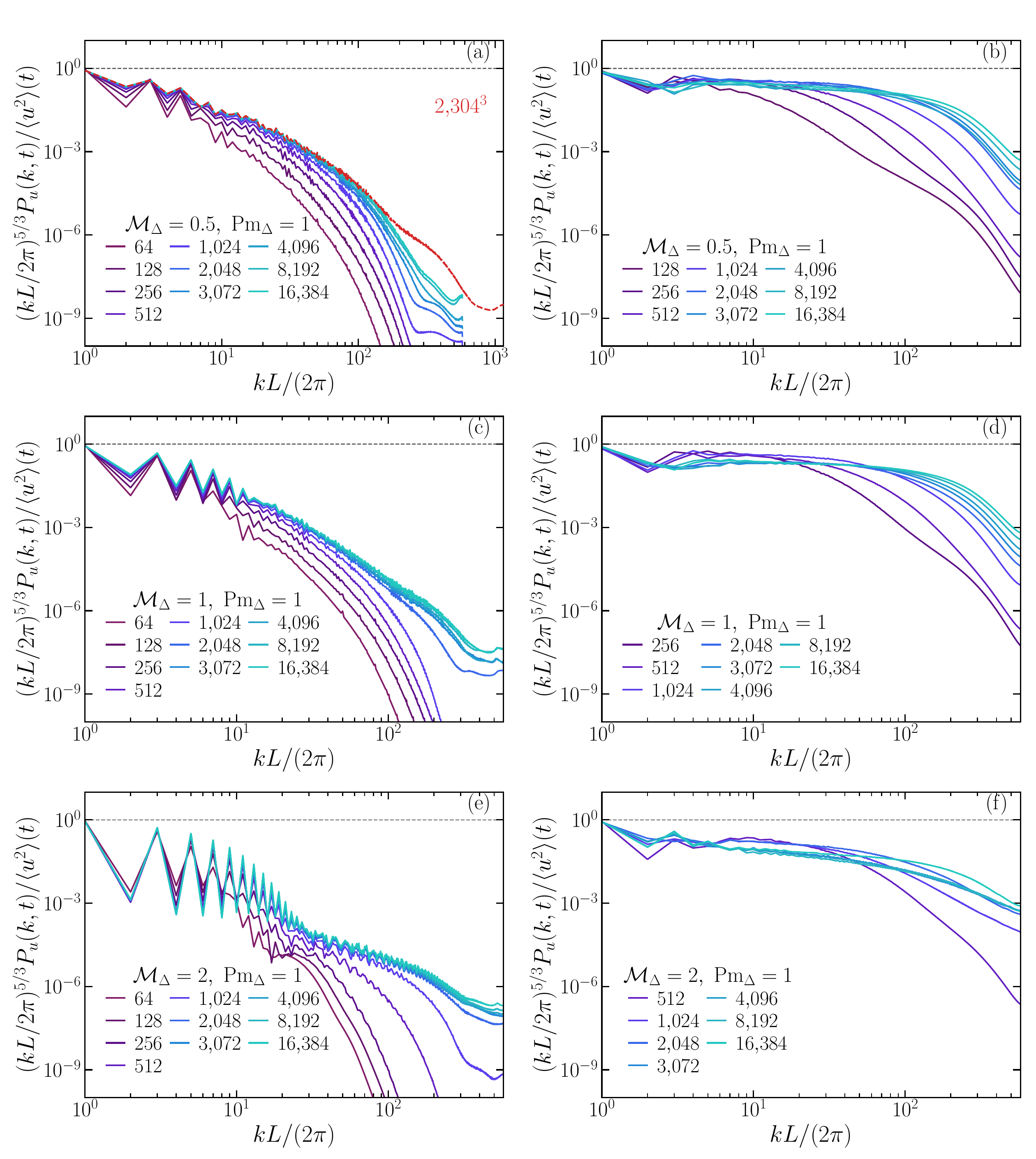}
     \caption{Velocity spectra, $P_u(k)$, for the runs and times in \autoref{fig:sm_mach_spectra}, colored by $\ReD$. Each spectrum is normalized by $\langle u^2\rangle(t)=\int P_u(k,t),dk$ and multiplied by $(kL/2\pi)^{5/3}$, making the Kolmogorov slope horizontal (which is found in many high-resolution global and zoom-in BNS merger simulations, e.g., \cite{Gutierrez2026,Kiuchi2026_magnetar} for the latest). The horizontal dashed line marks the compensated $P_u(k)\propto k^{-5/3}$ slope. The late-time spectra in the right column approximately follow this slope. The red dashed curve in (a) shows the same $2,\!304^3$ comparison as in \autoref{fig:sm_mach_spectra}~(a).}
     \label{fig:sm_velocity_spectra}
\end{figure*}

    \autoref{fig:sm_velocity_spectra} uses the same runs, times as colors as \autoref{fig:sm_mach_spectra}. We sum velocity power over spherical shells to obtain $P_u(k,t)$ and normalize by $\langle u^2\rangle(t)=\int P_u(k,t)\,dk$. In both columns we multiply by $(kL/2\pi)^{5/3}$ to make a $k^{-5/3}$ slope horizontal. As expected from high-resolution global and zoom-in models of BNS mergers \cite{Gutierrez2026,Kiuchi2026_magnetar}, we find a Kolmogorov-style spectrum in the late time evolution, and a steep, spiky spectrum at early times. The spikes most likely originate from the harmonics of the initial shear spectrum. Under $x\rightarrow x+L/2$, each Fourier mode acquires a factor $(-1)^{n_x}$. The antisymmetry $U_y(x+L/2)=-U_y(x)$ therefore selects odd modes, $n_x=1,3,5,\ldots$, which dominate the early-time spectrum. The same spiky initial spectra are shown in the zoom-in simulations in \cite{Gutierrez2026}.

\vspace{-1.5em}
\section{Volume renderings}
\vspace{-1.5em}
    We provide two volume renders of the KHI-driven dynamo: \href{https://youtu.be/t-6yRoGyFhk}{360$^{\circ}$ fly-through} and a \href{https://youtu.be/m3jhJwnkq5I}{magnetic field fly-through}. Each rendering used \textsc{PyVista} \cite{Sullivan2019pyvista} and approximately one hour on $1,\!000$ cores of \textsc{SuperMUC-NG} Phase 1. Both volumetric visualizations are taken during the kinematic dynamo stage (stage 3) in the $\ReD = 16,\!384$, $\Pm = 1$, $\M_{\Delta} = 0.5$ simulation on the $1,\!152^3$ grid.
    
\begin{figure*}
    \centering
    \includegraphics[width=\linewidth]{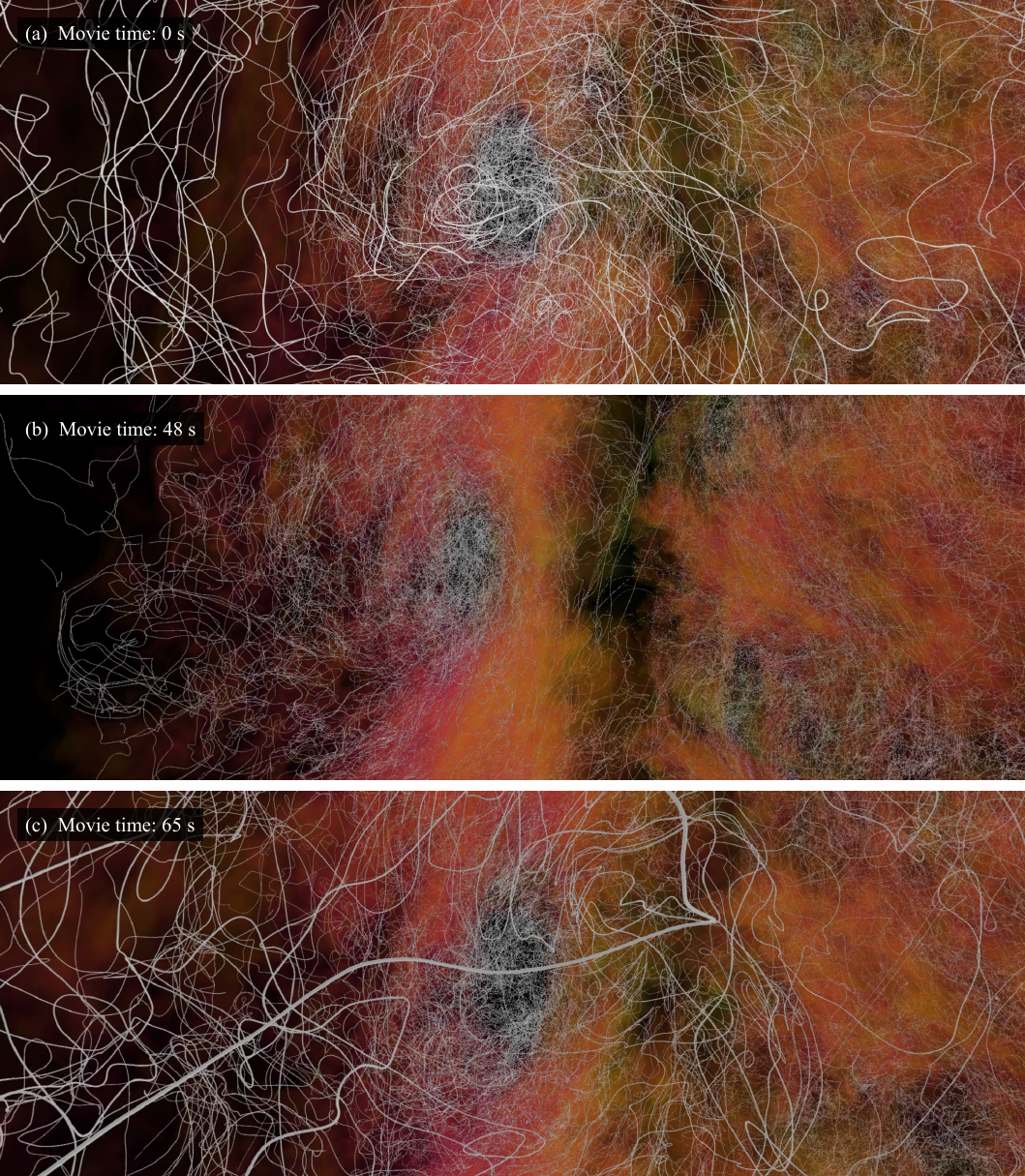}
    \caption{Three frames from \href{https://youtu.be/t-6yRoGyFhk}{the three-dimensional volume rendering of the KHI-driven dynamo}. The camera passes through one billow, turns, and follows a second. Yellow–red shows mass density. The rendering uses a fixed snapshot taken during the transition to the stage-3 kinematic dynamo. White lines trace the magnetic field. In this stage the spectrum is transitioning between $\emag(k) \propto k^{0}$ (winding spectrum; \autoref{eq:shear_spectrum}) and $\emag(k) \propto k^{3/2}$ (Kazantsev-Kulsrud-Anderson spectrum \cite{Kazantsev1968,Kulsrud1992_stochastic_fields}), and so the total magnetic field is energetically dominated by the small-scale (high-$k$) fluctuations.}
    \label{fig:3d_renders}
\end{figure*}

    The \href{https://youtu.be/t-6yRoGyFhk}{first} is a three-dimensional fly-through where we rendered a full spherical view for every frame, allowing viewers to rotate the view in 3D as the camera moves through a billow. In \autoref{fig:3d_renders} we show three frames as the camera rotates 360$^\circ$ around a great circle, and moves through the billow. The plasma density in the billow is shown in orange, and the magnetic field streamlines in white. The magnetic field is dominated by small-scale structure, as expected during stage 3.

    The \href{https://youtu.be/m3jhJwnkq5I}{second} is a three-dimensional fly-through of a billow, where we fade in the magnetic field and then fade out the plasma. A second scene follows a magnetic field line along a smoothed camera path.

%% file: main.bbl
\begin{thebibliography}{80}%
\makeatletter
\providecommand \@ifxundefined [1]{%
 \@ifx{#1\undefined}
}%
\providecommand \@ifnum [1]{%
 \ifnum #1\expandafter \@firstoftwo
 \else \expandafter \@secondoftwo
 \fi
}%
\providecommand \@ifx [1]{%
 \ifx #1\expandafter \@firstoftwo
 \else \expandafter \@secondoftwo
 \fi
}%
\providecommand \natexlab [1]{#1}%
\providecommand \enquote  [1]{``#1''}%
\providecommand \bibnamefont  [1]{#1}%
\providecommand \bibfnamefont [1]{#1}%
\providecommand \citenamefont [1]{#1}%
\providecommand \href@noop [0]{\@secondoftwo}%
\providecommand \href [0]{\begingroup \@sanitize@url \@href}%
\providecommand \@href[1]{\@@startlink{#1}\@@href}%
\providecommand \@@href[1]{\endgroup#1\@@endlink}%
\providecommand \@sanitize@url [0]{\catcode `\\12\catcode `\$12\catcode
  `\&12\catcode `\#12\catcode `\^12\catcode `\_12\catcode `\%12\relax}%
\providecommand \@@startlink[1]{}%
\providecommand \@@endlink[0]{}%
\providecommand \url  [0]{\begingroup\@sanitize@url \@url }%
\providecommand \@url [1]{\endgroup\@href {#1}{\urlprefix }}%
\providecommand \urlprefix  [0]{URL }%
\providecommand \Eprint [0]{\href }%
\providecommand \doibase [0]{https://doi.org/}%
\providecommand \selectlanguage [0]{\@gobble}%
\providecommand \bibinfo  [0]{\@secondoftwo}%
\providecommand \bibfield  [0]{\@secondoftwo}%
\providecommand \translation [1]{[#1]}%
\providecommand \BibitemOpen [0]{}%
\providecommand \bibitemStop [0]{}%
\providecommand \bibitemNoStop [0]{.\EOS\space}%
\providecommand \EOS [0]{\spacefactor3000\relax}%
\providecommand \BibitemShut  [1]{\csname bibitem#1\endcsname}%
\let\auto@bib@innerbib\@empty
\bibitem [{\citenamefont {{Price}}\ and\ \citenamefont
  {{Rosswog}}(2006)}]{Price2006_KHI_ultrastrong}%
  \BibitemOpen
  \bibfield  {author} {\bibinfo {author} {\bibfnamefont {D.~J.}\ \bibnamefont
  {{Price}}}\ and\ \bibinfo {author} {\bibfnamefont {S.}~\bibnamefont
  {{Rosswog}}},\ }\href {https://doi.org/10.1126/science.1125201} {\bibfield
  {journal} {\bibinfo  {journal} {Science}\ }\textbf {\bibinfo {volume}
  {312}},\ \bibinfo {pages} {719} (\bibinfo {year} {2006})},\ \Eprint
  {https://arxiv.org/abs/astro-ph/0603845} {arXiv:astro-ph/0603845 [astro-ph]}
  \BibitemShut {NoStop}%
\bibitem [{\citenamefont {Chabanov}\ \emph {et~al.}(2023)\citenamefont
  {Chabanov}, \citenamefont {Tootle}, \citenamefont {Most},\ and\ \citenamefont
  {Rezzolla}}]{Chabanov2023}%
  \BibitemOpen
  \bibfield  {author} {\bibinfo {author} {\bibfnamefont {M.}~\bibnamefont
  {Chabanov}}, \bibinfo {author} {\bibfnamefont {S.~D.}\ \bibnamefont
  {Tootle}}, \bibinfo {author} {\bibfnamefont {E.~R.}\ \bibnamefont {Most}},\
  and\ \bibinfo {author} {\bibfnamefont {L.}~\bibnamefont {Rezzolla}},\ }\href
  {https://doi.org/10.3847/2041-8213/acbbc5} {\bibfield  {journal} {\bibinfo
  {journal} {Astrophys. J. Lett.}\ }\textbf {\bibinfo {volume} {945}},\
  \bibinfo {pages} {L14} (\bibinfo {year} {2023})},\ \Eprint
  {https://arxiv.org/abs/2211.13661} {arXiv:2211.13661 [astro-ph.HE]}
  \BibitemShut {NoStop}%
\bibitem [{\citenamefont {Zhu}\ \emph {et~al.}(2015)\citenamefont {Zhu},
  \citenamefont {Pakmor}, \citenamefont {van Kerkwijk},\ and\ \citenamefont
  {Chang}}]{Zhu2015}%
  \BibitemOpen
  \bibfield  {author} {\bibinfo {author} {\bibfnamefont {C.}~\bibnamefont
  {Zhu}}, \bibinfo {author} {\bibfnamefont {R.}~\bibnamefont {Pakmor}},
  \bibinfo {author} {\bibfnamefont {M.~H.}\ \bibnamefont {van Kerkwijk}},\ and\
  \bibinfo {author} {\bibfnamefont {P.}~\bibnamefont {Chang}},\ }\href
  {https://doi.org/10.1088/2041-8205/806/1/L1} {\bibfield  {journal} {\bibinfo
  {journal} {Astrophys. J. Lett.}\ }\textbf {\bibinfo {volume} {806}},\
  \bibinfo {pages} {L1} (\bibinfo {year} {2015})}\BibitemShut {NoStop}%
\bibitem [{\citenamefont {{Kiuchi}}\ \emph {et~al.}(2015)\citenamefont
  {{Kiuchi}}, \citenamefont {{Cerd{\'a}-Dur{\'a}n}}, \citenamefont {{Kyutoku}},
  \citenamefont {{Sekiguchi}},\ and\ \citenamefont
  {{Shibata}}}]{Kiuchi2015_BNS_KHI}%
  \BibitemOpen
  \bibfield  {author} {\bibinfo {author} {\bibfnamefont {K.}~\bibnamefont
  {{Kiuchi}}}, \bibinfo {author} {\bibfnamefont {P.}~\bibnamefont
  {{Cerd{\'a}-Dur{\'a}n}}}, \bibinfo {author} {\bibfnamefont {K.}~\bibnamefont
  {{Kyutoku}}}, \bibinfo {author} {\bibfnamefont {Y.}~\bibnamefont
  {{Sekiguchi}}},\ and\ \bibinfo {author} {\bibfnamefont {M.}~\bibnamefont
  {{Shibata}}},\ }\href {https://doi.org/10.1103/PhysRevD.92.124034} {\bibfield
   {journal} {\bibinfo  {journal} {Physical Review D}\ }\textbf {\bibinfo
  {volume} {92}},\ \bibinfo {pages} {124034} (\bibinfo {year} {2015})},\
  \Eprint {https://arxiv.org/abs/1509.09205} {arXiv:1509.09205 [astro-ph.HE]}
  \BibitemShut {NoStop}%
\bibitem [{\citenamefont {Schneider}\ \emph {et~al.}(2019)\citenamefont
  {Schneider}, \citenamefont {Ohlmann}, \citenamefont {Podsiadlowski},
  \citenamefont {R{\"o}pke}, \citenamefont {Balbus}, \citenamefont {Pakmor},\
  and\ \citenamefont {Springel}}]{Schneider2019}%
  \BibitemOpen
  \bibfield  {author} {\bibinfo {author} {\bibfnamefont {F.~R.~N.}\
  \bibnamefont {Schneider}}, \bibinfo {author} {\bibfnamefont {S.~T.}\
  \bibnamefont {Ohlmann}}, \bibinfo {author} {\bibfnamefont {P.}~\bibnamefont
  {Podsiadlowski}}, \bibinfo {author} {\bibfnamefont {F.~K.}\ \bibnamefont
  {R{\"o}pke}}, \bibinfo {author} {\bibfnamefont {S.~A.}\ \bibnamefont
  {Balbus}}, \emph {et~al.},\ }\href
  {https://doi.org/10.1038/s41586-019-1621-5} {\bibfield  {journal} {\bibinfo
  {journal} {Nature}\ }\textbf {\bibinfo {volume} {574}},\ \bibinfo {pages}
  {211} (\bibinfo {year} {2019})}\BibitemShut {NoStop}%
\bibitem [{\citenamefont {{Pakmor}}\ \emph {et~al.}(2024)\citenamefont
  {{Pakmor}}, \citenamefont {{Pelisoli}}, \citenamefont {{Justham}},
  \citenamefont {{Rajamuthukumar}}, \citenamefont {{R{\"o}pke}}, \citenamefont
  {{Schneider}}, \citenamefont {{de Mink}}, \citenamefont {{Ohlmann}},
  \citenamefont {{Podsiadlowski}}, \citenamefont {{Mor{\'a}n-Fraile}},
  \citenamefont {{Vetter}},\ and\ \citenamefont
  {{Andrassy}}}]{Pakmor2024_merging_white_dwarfs}%
  \BibitemOpen
  \bibfield  {author} {\bibinfo {author} {\bibfnamefont {R.}~\bibnamefont
  {{Pakmor}}}, \bibinfo {author} {\bibfnamefont {I.}~\bibnamefont
  {{Pelisoli}}}, \bibinfo {author} {\bibfnamefont {S.}~\bibnamefont
  {{Justham}}}, \bibinfo {author} {\bibfnamefont {A.~S.}\ \bibnamefont
  {{Rajamuthukumar}}}, \bibinfo {author} {\bibfnamefont {F.~K.}\ \bibnamefont
  {{R{\"o}pke}}}, \emph {et~al.},\ }\href
  {https://doi.org/10.1051/0004-6361/202451352} {\bibfield  {journal} {\bibinfo
   {journal} {Astronomy and Astrophysics}\ }\textbf {\bibinfo {volume} {691}},\
  \bibinfo {pages} {A179} (\bibinfo {year} {2024})},\ \Eprint
  {https://arxiv.org/abs/2407.02566} {arXiv:2407.02566 [astro-ph.SR]}
  \BibitemShut {NoStop}%
\bibitem [{\citenamefont {{Radice}}\ and\ \citenamefont
  {{Hawke}}(2024)}]{Radice2024_BNS_review}%
  \BibitemOpen
  \bibfield  {author} {\bibinfo {author} {\bibfnamefont {D.}~\bibnamefont
  {{Radice}}}\ and\ \bibinfo {author} {\bibfnamefont {I.}~\bibnamefont
  {{Hawke}}},\ }\href {https://doi.org/10.1007/s41115-023-00019-9} {\bibfield
  {journal} {\bibinfo  {journal} {Living Reviews in Computational
  Astrophysics}\ }\textbf {\bibinfo {volume} {10}},\ \bibinfo {eid} {1}
  (\bibinfo {year} {2024})},\ \Eprint {https://arxiv.org/abs/2402.03224}
  {arXiv:2402.03224 [astro-ph.HE]} \BibitemShut {NoStop}%
\bibitem [{\citenamefont {Ohlmann}\ \emph {et~al.}(2026)\citenamefont
  {Ohlmann}, \citenamefont {Schneider}, \citenamefont {R{\"o}pke},
  \citenamefont {Pakmor}, \citenamefont {Podsiadlowski},\ and\ \citenamefont
  {Springel}}]{Ohlmann2026}%
  \BibitemOpen
  \bibfield  {author} {\bibinfo {author} {\bibfnamefont {S.~T.}\ \bibnamefont
  {Ohlmann}}, \bibinfo {author} {\bibfnamefont {F.~R.~N.}\ \bibnamefont
  {Schneider}}, \bibinfo {author} {\bibfnamefont {F.~K.}\ \bibnamefont
  {R{\"o}pke}}, \bibinfo {author} {\bibfnamefont {R.}~\bibnamefont {Pakmor}},
  \bibinfo {author} {\bibfnamefont {P.}~\bibnamefont {Podsiadlowski}}, \emph
  {et~al.},\ }\href {https://doi.org/10.1051/0004-6361/202659060} {\bibfield
  {journal} {\bibinfo  {journal} {Astronomy \& Astrophysics}\ }\textbf
  {\bibinfo {volume} {713}},\ \bibinfo {pages} {A6} (\bibinfo {year} {2026})},\
  \Eprint {https://arxiv.org/abs/2512.13424} {arXiv:2512.13424} \BibitemShut
  {NoStop}%
\bibitem [{\citenamefont {{Kiuchi}}\ \emph {et~al.}(2026)\citenamefont
  {{Kiuchi}}, \citenamefont {{Reboul-Salze}}, \citenamefont {{Sekiguchi}},\
  and\ \citenamefont {{Shibata}}}]{Kiuchi2026_magnetar}%
  \BibitemOpen
  \bibfield  {author} {\bibinfo {author} {\bibfnamefont {K.}~\bibnamefont
  {{Kiuchi}}}, \bibinfo {author} {\bibfnamefont {A.}~\bibnamefont
  {{Reboul-Salze}}}, \bibinfo {author} {\bibfnamefont {Y.}~\bibnamefont
  {{Sekiguchi}}},\ and\ \bibinfo {author} {\bibfnamefont {M.}~\bibnamefont
  {{Shibata}}},\ }\href {https://doi.org/10.48550/arXiv.2606.11299} {\bibfield
  {journal} {\bibinfo  {journal} {arXiv e-prints}\ ,\ \bibinfo {eid}
  {arXiv:2606.11299}} (\bibinfo {year} {2026})},\ \Eprint
  {https://arxiv.org/abs/2606.11299} {arXiv:2606.11299 [astro-ph.HE]}
  \BibitemShut {NoStop}%
\bibitem [{\citenamefont {Guti{\'e}rrez}\ \emph {et~al.}(2026)\citenamefont
  {Guti{\'e}rrez}, \citenamefont {Radice}, \citenamefont {Fields},\ and\
  \citenamefont {Stone}}]{Gutierrez2026}%
  \BibitemOpen
  \bibfield  {author} {\bibinfo {author} {\bibfnamefont {E.~M.}\ \bibnamefont
  {Guti{\'e}rrez}}, \bibinfo {author} {\bibfnamefont {D.}~\bibnamefont
  {Radice}}, \bibinfo {author} {\bibfnamefont {J.}~\bibnamefont {Fields}},\
  and\ \bibinfo {author} {\bibfnamefont {J.~M.}\ \bibnamefont {Stone}},\
  }\bibfield  {journal} {\bibinfo  {journal} {arXiv e-prints}\ }\href
  {https://doi.org/10.48550/arXiv.2601.20953} {10.48550/arXiv.2601.20953}
  (\bibinfo {year} {2026}),\ \Eprint {https://arxiv.org/abs/2601.20953}
  {arXiv:2601.20953 [astro-ph.HE]} \BibitemShut {NoStop}%
\bibitem [{\citenamefont {Kiuchi}\ \emph {et~al.}(2024)\citenamefont {Kiuchi},
  \citenamefont {Reboul-Salze}, \citenamefont {Shibata},\ and\ \citenamefont
  {Sekiguchi}}]{Kiuchi2024}%
  \BibitemOpen
  \bibfield  {author} {\bibinfo {author} {\bibfnamefont {K.}~\bibnamefont
  {Kiuchi}}, \bibinfo {author} {\bibfnamefont {A.}~\bibnamefont
  {Reboul-Salze}}, \bibinfo {author} {\bibfnamefont {M.}~\bibnamefont
  {Shibata}},\ and\ \bibinfo {author} {\bibfnamefont {Y.}~\bibnamefont
  {Sekiguchi}},\ }\href {https://doi.org/10.1038/s41550-024-02194-y} {\bibfield
   {journal} {\bibinfo  {journal} {Nature Astronomy}\ }\textbf {\bibinfo
  {volume} {8}},\ \bibinfo {pages} {298} (\bibinfo {year} {2024})}\BibitemShut
  {NoStop}%
\bibitem [{\citenamefont {Reboul-Salze}\ \emph {et~al.}(2025)\citenamefont
  {Reboul-Salze}, \citenamefont {Barr{\`e}re}, \citenamefont {Kiuchi},
  \citenamefont {Guilet}, \citenamefont {Raynaud}, \citenamefont
  {Fujibayashi},\ and\ \citenamefont {Shibata}}]{Reboul-Salze:2024jst}%
  \BibitemOpen
  \bibfield  {author} {\bibinfo {author} {\bibfnamefont {A.}~\bibnamefont
  {Reboul-Salze}}, \bibinfo {author} {\bibfnamefont {P.}~\bibnamefont
  {Barr{\`e}re}}, \bibinfo {author} {\bibfnamefont {K.}~\bibnamefont {Kiuchi}},
  \bibinfo {author} {\bibfnamefont {J.}~\bibnamefont {Guilet}}, \bibinfo
  {author} {\bibfnamefont {R.}~\bibnamefont {Raynaud}}, \emph {et~al.},\ }\href
  {https://doi.org/10.1051/0004-6361/202453126} {\bibfield  {journal} {\bibinfo
   {journal} {Astron. Astrophys.}\ }\textbf {\bibinfo {volume} {699}},\
  \bibinfo {pages} {A4} (\bibinfo {year} {2025})},\ \Eprint
  {https://arxiv.org/abs/2411.19328} {arXiv:2411.19328 [astro-ph.HE]}
  \BibitemShut {NoStop}%
\bibitem [{\citenamefont {Ng}\ and\ \citenamefont
  {Most}(2026{\natexlab{a}})}]{Ng:2026jev}%
  \BibitemOpen
  \bibfield  {author} {\bibinfo {author} {\bibfnamefont {H.~H.-Y.}\
  \bibnamefont {Ng}}\ and\ \bibinfo {author} {\bibfnamefont {E.~R.}\
  \bibnamefont {Most}},\ }\href@noop {} {\  (\bibinfo {year}
  {2026}{\natexlab{a}})},\ \Eprint {https://arxiv.org/abs/2607.28556}
  {arXiv:2607.28556 [astro-ph.HE]} \BibitemShut {NoStop}%
\bibitem [{\citenamefont {{Blandford}}\ and\ \citenamefont
  {{Znajek}}(1977)}]{Blandford1977_bz_jets}%
  \BibitemOpen
  \bibfield  {author} {\bibinfo {author} {\bibfnamefont {R.~D.}\ \bibnamefont
  {{Blandford}}}\ and\ \bibinfo {author} {\bibfnamefont {R.~L.}\ \bibnamefont
  {{Znajek}}},\ }\href {https://doi.org/10.1093/mnras/179.3.433} {\bibfield
  {journal} {\bibinfo  {journal} {\mnras}\ }\textbf {\bibinfo {volume} {179}},\
  \bibinfo {pages} {433} (\bibinfo {year} {1977})}\BibitemShut {NoStop}%
\bibitem [{\citenamefont {{Abbott}}\ \emph {et~al.}(2017)\citenamefont
  {{Abbott}}, \citenamefont {{Abbott}}, \citenamefont {{Abbott}}, \citenamefont
  {{Acernese}}, \citenamefont {{Ackley}}, \citenamefont {{Adams}},
  \citenamefont {{Adams}}, \citenamefont {{Addesso}}, \citenamefont
  {{Adhikari}}, \citenamefont {{Adya}}, \citenamefont {{Affeldt}},
  \citenamefont {{Afrough}}, \citenamefont {{Agarwal}}, \citenamefont
  {{Agathos}}, \citenamefont {{Agatsuma}}, \citenamefont {{Aggarwal}},
  \citenamefont {{Aguiar}}, \citenamefont {{Aiello}}, \citenamefont {{Ain}},
  \citenamefont {{Ajith}}, \citenamefont {{Allen}}, \citenamefont {{Allen}},
  \citenamefont {{Allocca}}, \citenamefont {{Aloy}}, \citenamefont {{Altin}},
  \citenamefont {{Amato}}, \citenamefont {{Ananyeva}}, \citenamefont
  {{Anderson}}, \citenamefont {{Anderson}}, \citenamefont {{Angelova}},
  \citenamefont {{Antier}}, \citenamefont {{Appert}}, \citenamefont {{Arai}},
  \citenamefont {{Araya}}, \citenamefont {{Areeda}}, \citenamefont {{Arnaud}},
  \citenamefont {{Arun}}, \citenamefont {{Ascenzi}}, \citenamefont {{Ashton}},
  \citenamefont {{Ast}}, \citenamefont {{Aston}}, \citenamefont {{Astone}},
  \citenamefont {{Atallah}}, \citenamefont {{Aufmuth}}, \citenamefont
  {{Aulbert}}, \citenamefont {{AultONeal}}, \citenamefont {{Austin}},
  \citenamefont {{Avila-Alvarez}}, \citenamefont {{Babak}}, \citenamefont
  {{Bacon}}, \citenamefont {{Bader}}, \citenamefont {{Bae}}, \citenamefont
  {{Baker}}, \citenamefont {{Baldaccini}}, \citenamefont {{Ballardin}},
  \citenamefont {{Ballmer}}, \citenamefont {{Banagiri}}, \citenamefont
  {{Barayoga}}, \citenamefont {{Barclay}}, \citenamefont {{Barish}},
  \citenamefont {{Barker}}, \citenamefont {{Barkett}}, \citenamefont
  {{Barone}}, \citenamefont {{Barr}}, \citenamefont {{Barsotti}}, \citenamefont
  {{Barsuglia}}, \citenamefont {{Barta}}, \citenamefont {{Bartlett}},
  \citenamefont {{Bartos}}, \citenamefont {{Bassiri}}, \citenamefont {{Basti}},
  \citenamefont {{Batch}}, \citenamefont {{Bawaj}}, \citenamefont {{Bayley}},
  \citenamefont {{Bazzan}}, \citenamefont {{B{\'e}csy}}, \citenamefont
  {{Beer}}, \citenamefont {{Bejger}}, \citenamefont {{Belahcene}},
  \citenamefont {{Bell}}, \citenamefont {{Berger}}, \citenamefont {{Bergmann}},
  \citenamefont {{Bero}}, \citenamefont {{Berry}}, \citenamefont
  {{Bersanetti}}, \citenamefont {{Bertolini}}, \citenamefont {{Betzwieser}},
  \citenamefont {{Bhagwat}}, \citenamefont {{Bhandare}}, \citenamefont
  {{Bilenko}}, \citenamefont {{Billingsley}}, \citenamefont {{Billman}},
  \citenamefont {{Birch}}, \citenamefont {{Birney}}, \citenamefont
  {{Birnholtz}}, \citenamefont {{Biscans}}, \citenamefont {{Biscoveanu}},
  \citenamefont {{Bisht}}, \citenamefont {{Bitossi}}, \citenamefont {{Biwer}},
  \citenamefont {{Bizouard}}, \citenamefont {{Blackburn}}, \citenamefont
  {{Blackman}}, \citenamefont {{Blair}}, \citenamefont {{Blair}}, \citenamefont
  {{Blair}}, \citenamefont {{Bloemen}}, \citenamefont {{Bock}}, \citenamefont
  {{Bode}}, \citenamefont {{Boer}}, \citenamefont {{Bogaert}}, \citenamefont
  {{Bohe}}, \citenamefont {{Bondu}}, \citenamefont {{Bonilla}}, \citenamefont
  {{Bonnand}}, \citenamefont {{Boom}}, \citenamefont {{Bork}}, \citenamefont
  {{Boschi}}, \citenamefont {{Bose}}, \citenamefont {{Bossie}}, \citenamefont
  {{Bouffanais}}, \citenamefont {{Bozzi}}, \citenamefont {{Bradaschia}},
  \citenamefont {{Brady}}, \citenamefont {{Branchesi}}, \citenamefont {{Brau}},
  \citenamefont {{Briant}}, \citenamefont {{Brillet}}, \citenamefont
  {{Brinkmann}}, \citenamefont {{Brisson}}, \citenamefont {{Brockill}},
  \citenamefont {{Broida}}, \citenamefont {{Brooks}}, \citenamefont {{Brown}},
  \citenamefont {{Brown}}, \citenamefont {{Brunett}}, \citenamefont
  {{Buchanan}}, \citenamefont {{Buikema}}, \citenamefont {{Bulik}},
  \citenamefont {{Bulten}}, \citenamefont {{Buonanno}}, \citenamefont
  {{Buskulic}}, \citenamefont {{Buy}}, \citenamefont {{Byer}}, \citenamefont
  {{Cabero}}, \citenamefont {{Cadonati}}, \citenamefont {{Cagnoli}},
  \citenamefont {{Cahillane}}, \citenamefont {{Calder{\'o}n Bustillo}},
  \citenamefont {{Callister}}, \citenamefont {{Calloni}}, \citenamefont
  {{Camp}}, \citenamefont {{Canepa}}, \citenamefont {{Canizares}},
  \citenamefont {{Cannon}}, \citenamefont {{Cao}}, \citenamefont {{Cao}},
  \citenamefont {{Capano}}, \citenamefont {{Capocasa}}, \citenamefont
  {{Carbognani}}, \citenamefont {{Caride}}, \citenamefont {{Carney}},
  \citenamefont {{Casanueva Diaz}}, \citenamefont {{Casentini}}, \citenamefont
  {{Caudill}}, \citenamefont {{Cavagli{\`a}}}, \citenamefont {{Cavalier}},
  \citenamefont {{Cavalieri}}, \citenamefont {{Cella}}, \citenamefont
  {{Cepeda}}, \citenamefont {{Cerd{\'a}-Dur{\'a}n}}, \citenamefont
  {{Cerretani}}, \citenamefont {{Cesarini}}, \citenamefont {{Chamberlin}},
  \citenamefont {{Chan}}, \citenamefont {{Chao}}, \citenamefont {{Charlton}},
  \citenamefont {{Chase}}, \citenamefont {{Chassande-Mottin}}, \citenamefont
  {{Chatterjee}}, \citenamefont {{Chatziioannou}}, \citenamefont
  {{Cheeseboro}}, \citenamefont {{Chen}}, \citenamefont {{Chen}}, \citenamefont
  {{Chen}}, \citenamefont {{Cheng}}, \citenamefont {{Chia}}, \citenamefont
  {{Chincarini}}, \citenamefont {{Chiummo}}, \citenamefont {{Chmiel}},
  \citenamefont {{Cho}}, \citenamefont {{Cho}}, \citenamefont {{Chow}},
  \citenamefont {{Christensen}}, \citenamefont {{Chu}}, \citenamefont {{Chua}},
  \citenamefont {{Chua}}, \citenamefont {{Chung}}, \citenamefont {{Chung}},\
  and\ \citenamefont {{Ciani}}}]{LIGO2017_BNS_gr_gw}%
  \BibitemOpen
  \bibfield  {author} {\bibinfo {author} {\bibfnamefont {B.~P.}\ \bibnamefont
  {{Abbott}}}, \bibinfo {author} {\bibfnamefont {R.}~\bibnamefont {{Abbott}}},
  \bibinfo {author} {\bibfnamefont {T.~D.}\ \bibnamefont {{Abbott}}}, \bibinfo
  {author} {\bibfnamefont {F.}~\bibnamefont {{Acernese}}}, \bibinfo {author}
  {\bibfnamefont {K.}~\bibnamefont {{Ackley}}}, \emph {et~al.},\ }\href
  {https://doi.org/10.3847/2041-8213/aa920c} {\bibfield  {journal} {\bibinfo
  {journal} {\apjl}\ }\textbf {\bibinfo {volume} {848}},\ \bibinfo {eid} {L13}
  (\bibinfo {year} {2017})},\ \Eprint {https://arxiv.org/abs/1710.05834}
  {arXiv:1710.05834 [astro-ph.HE]} \BibitemShut {NoStop}%
\bibitem [{\citenamefont {{Arcavi}}\ \emph {et~al.}(2017)\citenamefont
  {{Arcavi}}, \citenamefont {{Hosseinzadeh}}, \citenamefont {{Howell}},
  \citenamefont {{McCully}}, \citenamefont {{Poznanski}}, \citenamefont
  {{Kasen}}, \citenamefont {{Barnes}}, \citenamefont {{Zaltzman}},
  \citenamefont {{Vasylyev}}, \citenamefont {{Maoz}},\ and\ \citenamefont
  {{Valenti}}}]{Arcavi2017_BNS_kilonova}%
  \BibitemOpen
  \bibfield  {author} {\bibinfo {author} {\bibfnamefont {I.}~\bibnamefont
  {{Arcavi}}}, \bibinfo {author} {\bibfnamefont {G.}~\bibnamefont
  {{Hosseinzadeh}}}, \bibinfo {author} {\bibfnamefont {D.~A.}\ \bibnamefont
  {{Howell}}}, \bibinfo {author} {\bibfnamefont {C.}~\bibnamefont {{McCully}}},
  \bibinfo {author} {\bibfnamefont {D.}~\bibnamefont {{Poznanski}}}, \emph
  {et~al.},\ }\href {https://doi.org/10.1038/nature24291} {\bibfield  {journal}
  {\bibinfo  {journal} {\nat}\ }\textbf {\bibinfo {volume} {551}},\ \bibinfo
  {pages} {64} (\bibinfo {year} {2017})},\ \Eprint
  {https://arxiv.org/abs/1710.05843} {arXiv:1710.05843 [astro-ph.HE]}
  \BibitemShut {NoStop}%
\bibitem [{\citenamefont {M{\"o}sta}\ \emph {et~al.}(2020)\citenamefont
  {M{\"o}sta}, \citenamefont {Radice}, \citenamefont {Haas}, \citenamefont
  {Schnetter},\ and\ \citenamefont {Bernuzzi}}]{Mosta:2020hlh}%
  \BibitemOpen
  \bibfield  {author} {\bibinfo {author} {\bibfnamefont {P.}~\bibnamefont
  {M{\"o}sta}}, \bibinfo {author} {\bibfnamefont {D.}~\bibnamefont {Radice}},
  \bibinfo {author} {\bibfnamefont {R.}~\bibnamefont {Haas}}, \bibinfo {author}
  {\bibfnamefont {E.}~\bibnamefont {Schnetter}},\ and\ \bibinfo {author}
  {\bibfnamefont {S.}~\bibnamefont {Bernuzzi}},\ }\href
  {https://doi.org/10.3847/2041-8213/abb6ef} {\bibfield  {journal} {\bibinfo
  {journal} {Astrophys. J. Lett.}\ }\textbf {\bibinfo {volume} {901}},\
  \bibinfo {pages} {L37} (\bibinfo {year} {2020})},\ \Eprint
  {https://arxiv.org/abs/2003.06043} {arXiv:2003.06043 [astro-ph.HE]}
  \BibitemShut {NoStop}%
\bibitem [{\citenamefont {Most}\ and\ \citenamefont
  {Quataert}(2023)}]{Most:2023sft}%
  \BibitemOpen
  \bibfield  {author} {\bibinfo {author} {\bibfnamefont {E.~R.}\ \bibnamefont
  {Most}}\ and\ \bibinfo {author} {\bibfnamefont {E.}~\bibnamefont
  {Quataert}},\ }\href {https://doi.org/10.3847/2041-8213/acca84} {\bibfield
  {journal} {\bibinfo  {journal} {Astrophys. J. Lett.}\ }\textbf {\bibinfo
  {volume} {947}},\ \bibinfo {pages} {L15} (\bibinfo {year} {2023})},\ \Eprint
  {https://arxiv.org/abs/2303.08062} {arXiv:2303.08062 [astro-ph.HE]}
  \BibitemShut {NoStop}%
\bibitem [{\citenamefont {{Combi}}\ and\ \citenamefont
  {{Siegel}}(2023)}]{Combi2023_PRL_jets}%
  \BibitemOpen
  \bibfield  {author} {\bibinfo {author} {\bibfnamefont {L.}~\bibnamefont
  {{Combi}}}\ and\ \bibinfo {author} {\bibfnamefont {D.~M.}\ \bibnamefont
  {{Siegel}}},\ }\href {https://doi.org/10.1103/PhysRevLett.131.231402}
  {\bibfield  {journal} {\bibinfo  {journal} {\prl}\ }\textbf {\bibinfo
  {volume} {131}},\ \bibinfo {eid} {231402} (\bibinfo {year} {2023})},\ \Eprint
  {https://arxiv.org/abs/2303.12284} {arXiv:2303.12284 [astro-ph.HE]}
  \BibitemShut {NoStop}%
\bibitem [{\citenamefont {{Hayashi}}\ \emph {et~al.}(2025)\citenamefont
  {{Hayashi}}, \citenamefont {{Kiuchi}}, \citenamefont {{Kyutoku}},
  \citenamefont {{Sekiguchi}},\ and\ \citenamefont
  {{Shibata}}}]{Hayashi2025_jet_launching}%
  \BibitemOpen
  \bibfield  {author} {\bibinfo {author} {\bibfnamefont {K.}~\bibnamefont
  {{Hayashi}}}, \bibinfo {author} {\bibfnamefont {K.}~\bibnamefont {{Kiuchi}}},
  \bibinfo {author} {\bibfnamefont {K.}~\bibnamefont {{Kyutoku}}}, \bibinfo
  {author} {\bibfnamefont {Y.}~\bibnamefont {{Sekiguchi}}},\ and\ \bibinfo
  {author} {\bibfnamefont {M.}~\bibnamefont {{Shibata}}},\ }\href
  {https://doi.org/10.1103/PhysRevLett.134.211407} {\bibfield  {journal}
  {\bibinfo  {journal} {\prl}\ }\textbf {\bibinfo {volume} {134}},\ \bibinfo
  {eid} {211407} (\bibinfo {year} {2025})},\ \Eprint
  {https://arxiv.org/abs/2410.10958} {arXiv:2410.10958 [astro-ph.HE]}
  \BibitemShut {NoStop}%
\bibitem [{\citenamefont {Kiuchi}\ \emph
  {et~al.}(2026{\natexlab{a}})\citenamefont {Kiuchi}, \citenamefont
  {Fujibayashi}, \citenamefont {Hayashi}, \citenamefont {Kawaguchi},
  \citenamefont {Pognan}, \citenamefont {Reboul-Salze}, \citenamefont
  {Sekiguchi}, \citenamefont {Shibata},\ and\ \citenamefont
  {Wanajo}}]{Kiuchi:2026thl}%
  \BibitemOpen
  \bibfield  {author} {\bibinfo {author} {\bibfnamefont {K.}~\bibnamefont
  {Kiuchi}}, \bibinfo {author} {\bibfnamefont {S.}~\bibnamefont {Fujibayashi}},
  \bibinfo {author} {\bibfnamefont {K.}~\bibnamefont {Hayashi}}, \bibinfo
  {author} {\bibfnamefont {K.}~\bibnamefont {Kawaguchi}}, \bibinfo {author}
  {\bibfnamefont {Q.}~\bibnamefont {Pognan}}, \emph {et~al.},\ }\href@noop {}
  {\  (\bibinfo {year} {2026}{\natexlab{a}})},\ \Eprint
  {https://arxiv.org/abs/2608.27555} {arXiv:2608.27555 [astro-ph.HE]}
  \BibitemShut {NoStop}%
\bibitem [{\citenamefont {Kiuchi}\ \emph
  {et~al.}(2026{\natexlab{b}})\citenamefont {Kiuchi}, \citenamefont
  {Reboul-Salze}, \citenamefont {Sekiguchi},\ and\ \citenamefont
  {Shibata}}]{Kiuchi2026}%
  \BibitemOpen
  \bibfield  {author} {\bibinfo {author} {\bibfnamefont {K.}~\bibnamefont
  {Kiuchi}}, \bibinfo {author} {\bibfnamefont {A.}~\bibnamefont
  {Reboul-Salze}}, \bibinfo {author} {\bibfnamefont {Y.}~\bibnamefont
  {Sekiguchi}},\ and\ \bibinfo {author} {\bibfnamefont {M.}~\bibnamefont
  {Shibata}},\ }\bibfield  {journal} {\bibinfo  {journal} {arXiv e-prints}\
  }\href {https://doi.org/10.48550/arXiv.2606.11299}
  {10.48550/arXiv.2606.11299} (\bibinfo {year} {2026}{\natexlab{b}}),\ \Eprint
  {https://arxiv.org/abs/2606.11299} {arXiv:2606.11299} \BibitemShut {NoStop}%
\bibitem [{\citenamefont {Margalit}\ \emph {et~al.}(2022)\citenamefont
  {Margalit}, \citenamefont {Jermyn}, \citenamefont {Metzger}, \citenamefont
  {Roberts},\ and\ \citenamefont {Quataert}}]{Margalit:2022rde}%
  \BibitemOpen
  \bibfield  {author} {\bibinfo {author} {\bibfnamefont {B.}~\bibnamefont
  {Margalit}}, \bibinfo {author} {\bibfnamefont {A.~S.}\ \bibnamefont
  {Jermyn}}, \bibinfo {author} {\bibfnamefont {B.~D.}\ \bibnamefont {Metzger}},
  \bibinfo {author} {\bibfnamefont {L.~F.}\ \bibnamefont {Roberts}},\ and\
  \bibinfo {author} {\bibfnamefont {E.}~\bibnamefont {Quataert}},\ }\href
  {https://doi.org/10.3847/1538-4357/ac8b01} {\bibfield  {journal} {\bibinfo
  {journal} {Astrophys. J.}\ }\textbf {\bibinfo {volume} {939}},\ \bibinfo
  {pages} {51} (\bibinfo {year} {2022})},\ \Eprint
  {https://arxiv.org/abs/2206.10645} {arXiv:2206.10645 [astro-ph.HE]}
  \BibitemShut {NoStop}%
\bibitem [{\citenamefont {Most}(2023)}]{Most:2023sme}%
  \BibitemOpen
  \bibfield  {author} {\bibinfo {author} {\bibfnamefont {E.~R.}\ \bibnamefont
  {Most}},\ }\href {https://doi.org/10.1103/PhysRevD.108.123012} {\bibfield
  {journal} {\bibinfo  {journal} {Phys. Rev. D}\ }\textbf {\bibinfo {volume}
  {108}},\ \bibinfo {pages} {123012} (\bibinfo {year} {2023})},\ \Eprint
  {https://arxiv.org/abs/2311.03333} {arXiv:2311.03333 [astro-ph.HE]}
  \BibitemShut {NoStop}%
\bibitem [{\citenamefont {{Schekochihin}}\ \emph {et~al.}(2002)\citenamefont
  {{Schekochihin}}, \citenamefont {{Cowley}}, \citenamefont {{Hammett}},
  \citenamefont {{Maron}},\ and\ \citenamefont
  {{McWilliams}}}]{Schekochihin2002_nonlinear}%
  \BibitemOpen
  \bibfield  {author} {\bibinfo {author} {\bibfnamefont {A.~A.}\ \bibnamefont
  {{Schekochihin}}}, \bibinfo {author} {\bibfnamefont {S.~C.}\ \bibnamefont
  {{Cowley}}}, \bibinfo {author} {\bibfnamefont {G.~W.}\ \bibnamefont
  {{Hammett}}}, \bibinfo {author} {\bibfnamefont {J.~L.}\ \bibnamefont
  {{Maron}}},\ and\ \bibinfo {author} {\bibfnamefont {J.~C.}\ \bibnamefont
  {{McWilliams}}},\ }\href {https://doi.org/10.1088/1367-2630/4/1/384}
  {\bibfield  {journal} {\bibinfo  {journal} {New Journal of Physics}\ }\textbf
  {\bibinfo {volume} {4}},\ \bibinfo {pages} {84} (\bibinfo {year} {2002})},\
  \Eprint {https://arxiv.org/abs/astro-ph/0207503} {arXiv:astro-ph/0207503
  [astro-ph]} \BibitemShut {NoStop}%
\bibitem [{\citenamefont {{Galishnikova}}\ \emph {et~al.}(2022)\citenamefont
  {{Galishnikova}}, \citenamefont {{Kunz}},\ and\ \citenamefont
  {{Schekochihin}}}]{Galishnikova2022_tearing}%
  \BibitemOpen
  \bibfield  {author} {\bibinfo {author} {\bibfnamefont {A.~K.}\ \bibnamefont
  {{Galishnikova}}}, \bibinfo {author} {\bibfnamefont {M.~W.}\ \bibnamefont
  {{Kunz}}},\ and\ \bibinfo {author} {\bibfnamefont {A.~A.}\ \bibnamefont
  {{Schekochihin}}},\ }\href {https://doi.org/10.1103/PhysRevX.12.041027}
  {\bibfield  {journal} {\bibinfo  {journal} {Physical Review X}\ }\textbf
  {\bibinfo {volume} {12}},\ \bibinfo {eid} {041027} (\bibinfo {year}
  {2022})},\ \Eprint {https://arxiv.org/abs/2201.07757} {arXiv:2201.07757
  [astro-ph.HE]} \BibitemShut {NoStop}%
\bibitem [{\citenamefont {{Beattie}}\ \emph {et~al.}(2023)\citenamefont
  {{Beattie}}, \citenamefont {{Federrath}}, \citenamefont {{Kriel}},
  \citenamefont {{Mocz}},\ and\ \citenamefont
  {{Seta}}}]{Beattie2023_growth_or_decay}%
  \BibitemOpen
  \bibfield  {author} {\bibinfo {author} {\bibfnamefont {J.~R.}\ \bibnamefont
  {{Beattie}}}, \bibinfo {author} {\bibfnamefont {C.}~\bibnamefont
  {{Federrath}}}, \bibinfo {author} {\bibfnamefont {N.}~\bibnamefont
  {{Kriel}}}, \bibinfo {author} {\bibfnamefont {P.}~\bibnamefont {{Mocz}}},\
  and\ \bibinfo {author} {\bibfnamefont {A.}~\bibnamefont {{Seta}}},\ }\href
  {https://doi.org/10.1093/mnras/stad1863} {\bibfield  {journal} {\bibinfo
  {journal} {\mnras}\ }\textbf {\bibinfo {volume} {524}},\ \bibinfo {pages}
  {3201} (\bibinfo {year} {2023})},\ \Eprint {https://arxiv.org/abs/2209.10749}
  {arXiv:2209.10749 [astro-ph.GA]} \BibitemShut {NoStop}%
\bibitem [{\citenamefont {Kriel}\ \emph {et~al.}(2026)\citenamefont {Kriel},
  \citenamefont {Krumholz}, \citenamefont {Armstrong}, \citenamefont
  {Beattie},\ and\ \citenamefont {Schober}}]{Kriel2026_universal}%
  \BibitemOpen
  \bibfield  {author} {\bibinfo {author} {\bibfnamefont {N.}~\bibnamefont
  {Kriel}}, \bibinfo {author} {\bibfnamefont {M.~R.}\ \bibnamefont {Krumholz}},
  \bibinfo {author} {\bibfnamefont {P.~J.}\ \bibnamefont {Armstrong}}, \bibinfo
  {author} {\bibfnamefont {J.~R.}\ \bibnamefont {Beattie}},\ and\ \bibinfo
  {author} {\bibfnamefont {J.}~\bibnamefont {Schober}},\ }\href
  {https://doi.org/10.1103/8qjf-8gg4} {\bibfield  {journal} {\bibinfo
  {journal} {Physical Review E}\ }\textbf {\bibinfo {volume} {113}},\ \bibinfo
  {pages} {045208} (\bibinfo {year} {2026})},\ \Eprint
  {https://arxiv.org/abs/2509.09949} {arXiv:2509.09949} \BibitemShut {NoStop}%
\bibitem [{\citenamefont {Aguilera-Miret}\ \emph {et~al.}(2022)\citenamefont
  {Aguilera-Miret}, \citenamefont {Vigan{\`o}},\ and\ \citenamefont
  {Palenzuela}}]{Aguilera-Miret:2021fre}%
  \BibitemOpen
  \bibfield  {author} {\bibinfo {author} {\bibfnamefont {R.}~\bibnamefont
  {Aguilera-Miret}}, \bibinfo {author} {\bibfnamefont {D.}~\bibnamefont
  {Vigan{\`o}}},\ and\ \bibinfo {author} {\bibfnamefont {C.}~\bibnamefont
  {Palenzuela}},\ }\href {https://doi.org/10.3847/2041-8213/ac50a7} {\bibfield
  {journal} {\bibinfo  {journal} {Astrophys. J. Lett.}\ }\textbf {\bibinfo
  {volume} {926}},\ \bibinfo {pages} {L31} (\bibinfo {year} {2022})},\ \Eprint
  {https://arxiv.org/abs/2112.08406} {arXiv:2112.08406 [gr-qc]} \BibitemShut
  {NoStop}%
\bibitem [{\citenamefont {Kiuchi}(2025)}]{Kiuchi2025_review}%
  \BibitemOpen
  \bibfield  {author} {\bibinfo {author} {\bibfnamefont {K.}~\bibnamefont
  {Kiuchi}},\ }in\ \href {https://doi.org/10.1007/978-981-97-8522-3_16} {\emph
  {\bibinfo {booktitle} {New Frontiers in {GRMHD} Simulations}}},\ \bibinfo
  {series and number} {Springer Series in Astrophysics and Cosmology},\
  \bibinfo {editor} {edited by\ \bibinfo {editor} {\bibfnamefont
  {C.}~\bibnamefont {Bambi}}, \bibinfo {editor} {\bibfnamefont
  {Y.}~\bibnamefont {Mizuno}}, \bibinfo {editor} {\bibfnamefont
  {S.}~\bibnamefont {Shashank}},\ and\ \bibinfo {editor} {\bibfnamefont
  {F.}~\bibnamefont {Yuan}}}\ (\bibinfo  {publisher} {Springer},\ \bibinfo
  {address} {Singapore},\ \bibinfo {year} {2025})\ pp.\ \bibinfo {pages}
  {529--572},\ \Eprint {https://arxiv.org/abs/2405.10081} {arXiv:2405.10081
  [astro-ph.HE]} \BibitemShut {NoStop}%
\bibitem [{\citenamefont {Schekochihin}\ \emph {et~al.}(2004)\citenamefont
  {Schekochihin}, \citenamefont {Cowley}, \citenamefont {Taylor}, \citenamefont
  {Maron},\ and\ \citenamefont {McWilliams}}]{Schekochihin2004_dynamo}%
  \BibitemOpen
  \bibfield  {author} {\bibinfo {author} {\bibfnamefont {A.~A.}\ \bibnamefont
  {Schekochihin}}, \bibinfo {author} {\bibfnamefont {S.~C.}\ \bibnamefont
  {Cowley}}, \bibinfo {author} {\bibfnamefont {S.~F.}\ \bibnamefont {Taylor}},
  \bibinfo {author} {\bibfnamefont {J.~L.}\ \bibnamefont {Maron}},\ and\
  \bibinfo {author} {\bibfnamefont {J.~C.}\ \bibnamefont {McWilliams}},\ }\href
  {https://doi.org/10.1086/422547} {\bibfield  {journal} {\bibinfo  {journal}
  {The Astrophysical Journal}\ }\textbf {\bibinfo {volume} {612}},\ \bibinfo
  {pages} {276} (\bibinfo {year} {2004})},\ \Eprint
  {https://arxiv.org/abs/astro-ph/0312046} {arXiv:astro-ph/0312046}
  \BibitemShut {NoStop}%
\bibitem [{\citenamefont {Rincon}(2019)}]{Rincon2019_dynamo_theories}%
  \BibitemOpen
  \bibfield  {author} {\bibinfo {author} {\bibfnamefont {F.}~\bibnamefont
  {Rincon}},\ }\href {https://doi.org/10.1017/S0022377819000539} {\bibfield
  {journal} {\bibinfo  {journal} {Journal of Plasma Physics}\ }\textbf
  {\bibinfo {volume} {85}},\ \bibinfo {pages} {205850401} (\bibinfo {year}
  {2019})},\ \Eprint {https://arxiv.org/abs/1903.07829} {arXiv:1903.07829}
  \BibitemShut {NoStop}%
\bibitem [{\citenamefont {Kazantsev}(1968)}]{Kazantsev1968}%
  \BibitemOpen
  \bibfield  {author} {\bibinfo {author} {\bibfnamefont {A.~P.}\ \bibnamefont
  {Kazantsev}},\ }\href {https://jetp.ras.ru/cgi-bin/dn/e_026_05_1031.pdf}
  {\bibfield  {journal} {\bibinfo  {journal} {Sov. Phys. JETP}\ }\textbf
  {\bibinfo {volume} {26}},\ \bibinfo {pages} {1031} (\bibinfo {year}
  {1968})}\BibitemShut {NoStop}%
\bibitem [{\citenamefont {{Kulsrud}}\ and\ \citenamefont
  {{Anderson}}(1992)}]{Kulsrud1992_stochastic_fields}%
  \BibitemOpen
  \bibfield  {author} {\bibinfo {author} {\bibfnamefont {R.~M.}\ \bibnamefont
  {{Kulsrud}}}\ and\ \bibinfo {author} {\bibfnamefont {S.~W.}\ \bibnamefont
  {{Anderson}}},\ }\href {https://doi.org/10.1086/171743} {\bibfield  {journal}
  {\bibinfo  {journal} {\apj}\ }\textbf {\bibinfo {volume} {396}},\ \bibinfo
  {pages} {606} (\bibinfo {year} {1992})}\BibitemShut {NoStop}%
\bibitem [{\citenamefont {{Schober}}\ \emph {et~al.}(2012)\citenamefont
  {{Schober}}, \citenamefont {{Schleicher}}, \citenamefont {{Federrath}},
  \citenamefont {{Klessen}},\ and\ \citenamefont
  {{Banerjee}}}]{Schober2012_SSD}%
  \BibitemOpen
  \bibfield  {author} {\bibinfo {author} {\bibfnamefont {J.}~\bibnamefont
  {{Schober}}}, \bibinfo {author} {\bibfnamefont {D.}~\bibnamefont
  {{Schleicher}}}, \bibinfo {author} {\bibfnamefont {C.}~\bibnamefont
  {{Federrath}}}, \bibinfo {author} {\bibfnamefont {R.}~\bibnamefont
  {{Klessen}}},\ and\ \bibinfo {author} {\bibfnamefont {R.}~\bibnamefont
  {{Banerjee}}},\ }\href {https://doi.org/10.1103/PhysRevE.85.026303}
  {\bibfield  {journal} {\bibinfo  {journal} {\pre}\ }\textbf {\bibinfo
  {volume} {85}},\ \bibinfo {eid} {026303} (\bibinfo {year} {2012})},\ \Eprint
  {https://arxiv.org/abs/1109.4571} {arXiv:1109.4571 [astro-ph.CO]}
  \BibitemShut {NoStop}%
\bibitem [{\citenamefont {{Kriel}}\ \emph {et~al.}(2022)\citenamefont
  {{Kriel}}, \citenamefont {{Beattie}}, \citenamefont {{Seta}},\ and\
  \citenamefont {{Federrath}}}]{Kriel2022_kinematic_dynamo_scales}%
  \BibitemOpen
  \bibfield  {author} {\bibinfo {author} {\bibfnamefont {N.}~\bibnamefont
  {{Kriel}}}, \bibinfo {author} {\bibfnamefont {J.~R.}\ \bibnamefont
  {{Beattie}}}, \bibinfo {author} {\bibfnamefont {A.}~\bibnamefont {{Seta}}},\
  and\ \bibinfo {author} {\bibfnamefont {C.}~\bibnamefont {{Federrath}}},\
  }\href {https://doi.org/10.1093/mnras/stac969} {\bibfield  {journal}
  {\bibinfo  {journal} {Monthly Notices of the Royal Astronomical Society}\
  }\textbf {\bibinfo {volume} {513}},\ \bibinfo {pages} {2457} (\bibinfo {year}
  {2022})},\ \Eprint {https://arxiv.org/abs/2204.00828} {arXiv:2204.00828
  [astro-ph.SR]} \BibitemShut {NoStop}%
\bibitem [{\citenamefont {{Kriel}}\ \emph {et~al.}(2025)\citenamefont
  {{Kriel}}, \citenamefont {{Beattie}}, \citenamefont {{Federrath}},
  \citenamefont {{Krumholz}},\ and\ \citenamefont
  {{Hew}}}]{Kriel2025_fundamental_scales_II}%
  \BibitemOpen
  \bibfield  {author} {\bibinfo {author} {\bibfnamefont {N.}~\bibnamefont
  {{Kriel}}}, \bibinfo {author} {\bibfnamefont {J.~R.}\ \bibnamefont
  {{Beattie}}}, \bibinfo {author} {\bibfnamefont {C.}~\bibnamefont
  {{Federrath}}}, \bibinfo {author} {\bibfnamefont {M.~R.}\ \bibnamefont
  {{Krumholz}}},\ and\ \bibinfo {author} {\bibfnamefont {J.~K.~J.}\
  \bibnamefont {{Hew}}},\ }\href {https://doi.org/10.1093/mnras/staf188}
  {\bibfield  {journal} {\bibinfo  {journal} {\mnras}\ }\textbf {\bibinfo
  {volume} {537}},\ \bibinfo {pages} {2602} (\bibinfo {year} {2025})},\ \Eprint
  {https://arxiv.org/abs/2310.17036} {arXiv:2310.17036 [astro-ph.GA]}
  \BibitemShut {NoStop}%
\bibitem [{\citenamefont {Most}\ and\ \citenamefont
  {Dunham}(2026)}]{Most:2026uga}%
  \BibitemOpen
  \bibfield  {author} {\bibinfo {author} {\bibfnamefont {E.~R.}\ \bibnamefont
  {Most}}\ and\ \bibinfo {author} {\bibfnamefont {S.~J.}\ \bibnamefont
  {Dunham}},\ }\href@noop {} {\  (\bibinfo {year} {2026})},\ \Eprint
  {https://arxiv.org/abs/2609.21923} {arXiv:2609.21923 [astro-ph.HE]}
  \BibitemShut {NoStop}%
\bibitem [{\citenamefont {Dionysopoulou}\ \emph {et~al.}(2015)\citenamefont
  {Dionysopoulou}, \citenamefont {Alic},\ and\ \citenamefont
  {Rezzolla}}]{Dionysopoulou:2015tda}%
  \BibitemOpen
  \bibfield  {author} {\bibinfo {author} {\bibfnamefont {K.}~\bibnamefont
  {Dionysopoulou}}, \bibinfo {author} {\bibfnamefont {D.}~\bibnamefont
  {Alic}},\ and\ \bibinfo {author} {\bibfnamefont {L.}~\bibnamefont
  {Rezzolla}},\ }\href {https://doi.org/10.1103/PhysRevD.92.084064} {\bibfield
  {journal} {\bibinfo  {journal} {Phys. Rev. D}\ }\textbf {\bibinfo {volume}
  {92}},\ \bibinfo {pages} {084064} (\bibinfo {year} {2015})},\ \Eprint
  {https://arxiv.org/abs/1502.02021} {arXiv:1502.02021 [gr-qc]} \BibitemShut
  {NoStop}%
\bibitem [{\citenamefont {Miravet-Ten{\'e}s}\ \emph {et~al.}(2026)\citenamefont
  {Miravet-Ten{\'e}s}, \citenamefont {Hawke}, \citenamefont {Matur},\ and\
  \citenamefont {Wright}}]{Miravet-Tenes:2026gno}%
  \BibitemOpen
  \bibfield  {author} {\bibinfo {author} {\bibfnamefont {M.}~\bibnamefont
  {Miravet-Ten{\'e}s}}, \bibinfo {author} {\bibfnamefont {I.}~\bibnamefont
  {Hawke}}, \bibinfo {author} {\bibfnamefont {R.}~\bibnamefont {Matur}},\ and\
  \bibinfo {author} {\bibfnamefont {A.~J.}\ \bibnamefont {Wright}},\
  }\href@noop {} {\  (\bibinfo {year} {2026})},\ \Eprint
  {https://arxiv.org/abs/2609.12998} {arXiv:2609.12998 [astro-ph.HE]}
  \BibitemShut {NoStop}%
\bibitem [{\citenamefont {{Miyoshi}}\ and\ \citenamefont
  {{Kusano}}(2005)}]{Miyoshi2005_HLLD}%
  \BibitemOpen
  \bibfield  {author} {\bibinfo {author} {\bibfnamefont {T.}~\bibnamefont
  {{Miyoshi}}}\ and\ \bibinfo {author} {\bibfnamefont {K.}~\bibnamefont
  {{Kusano}}},\ }\href {https://doi.org/10.1016/j.jcp.2005.02.017} {\bibfield
  {journal} {\bibinfo  {journal} {Journal of Computational Physics}\ }\textbf
  {\bibinfo {volume} {208}},\ \bibinfo {pages} {315} (\bibinfo {year}
  {2005})}\BibitemShut {NoStop}%
\bibitem [{\citenamefont {{Lemaster}}\ and\ \citenamefont
  {{Stone}}(2009)}]{Lemaster2009_fofc}%
  \BibitemOpen
  \bibfield  {author} {\bibinfo {author} {\bibfnamefont {M.~N.}\ \bibnamefont
  {{Lemaster}}}\ and\ \bibinfo {author} {\bibfnamefont {J.~M.}\ \bibnamefont
  {{Stone}}},\ }\href {https://doi.org/10.1088/0004-637X/691/2/1092} {\bibfield
   {journal} {\bibinfo  {journal} {\apj}\ }\textbf {\bibinfo {volume} {691}},\
  \bibinfo {pages} {1092} (\bibinfo {year} {2009})},\ \Eprint
  {https://arxiv.org/abs/0809.4005} {arXiv:0809.4005 [astro-ph]} \BibitemShut
  {NoStop}%
\bibitem [{\citenamefont {Fu}\ \emph {et~al.}(2016)\citenamefont {Fu},
  \citenamefont {Hu},\ and\ \citenamefont
  {Adams}}]{Fu2016_high_order_reconstructions}%
  \BibitemOpen
  \bibfield  {author} {\bibinfo {author} {\bibfnamefont {L.}~\bibnamefont
  {Fu}}, \bibinfo {author} {\bibfnamefont {X.~Y.}\ \bibnamefont {Hu}},\ and\
  \bibinfo {author} {\bibfnamefont {N.~A.}\ \bibnamefont {Adams}},\ }\href
  {https://doi.org/https://doi.org/10.1016/j.jcp.2015.10.037} {\bibfield
  {journal} {\bibinfo  {journal} {Journal of Computational Physics}\ }\textbf
  {\bibinfo {volume} {305}},\ \bibinfo {pages} {333} (\bibinfo {year}
  {2016})}\BibitemShut {NoStop}%
\bibitem [{\citenamefont {{Stone}}\ \emph {et~al.}(2026)\citenamefont
  {{Stone}}, \citenamefont {{Mullen}}, \citenamefont {{Fielding}},
  \citenamefont {{Grete}}, \citenamefont {{Guo}}, \citenamefont {{Kempski}},
  \citenamefont {{Most}}, \citenamefont {{White}},\ and\ \citenamefont
  {{Wong}}}]{Stone2026_athenak}%
  \BibitemOpen
  \bibfield  {author} {\bibinfo {author} {\bibfnamefont {J.~M.}\ \bibnamefont
  {{Stone}}}, \bibinfo {author} {\bibfnamefont {P.~D.}\ \bibnamefont
  {{Mullen}}}, \bibinfo {author} {\bibfnamefont {D.}~\bibnamefont
  {{Fielding}}}, \bibinfo {author} {\bibfnamefont {P.}~\bibnamefont {{Grete}}},
  \bibinfo {author} {\bibfnamefont {M.}~\bibnamefont {{Guo}}}, \emph {et~al.},\
  }\href {https://doi.org/10.3847/1538-4365/ae3717} {\bibfield  {journal}
  {\bibinfo  {journal} {\apjs}\ }\textbf {\bibinfo {volume} {283}},\ \bibinfo
  {eid} {27} (\bibinfo {year} {2026})},\ \Eprint
  {https://arxiv.org/abs/2409.16053} {arXiv:2409.16053 [astro-ph.IM]}
  \BibitemShut {NoStop}%
\bibitem [{\citenamefont {{Lecoanet}}\ \emph {et~al.}(2016)\citenamefont
  {{Lecoanet}}, \citenamefont {{McCourt}}, \citenamefont {{Quataert}},
  \citenamefont {{Burns}}, \citenamefont {{Vasil}}, \citenamefont {{Oishi}},
  \citenamefont {{Brown}}, \citenamefont {{Stone}},\ and\ \citenamefont
  {{O'Leary}}}]{Lecoanet2016_KHI_benchmark}%
  \BibitemOpen
  \bibfield  {author} {\bibinfo {author} {\bibfnamefont {D.}~\bibnamefont
  {{Lecoanet}}}, \bibinfo {author} {\bibfnamefont {M.}~\bibnamefont
  {{McCourt}}}, \bibinfo {author} {\bibfnamefont {E.}~\bibnamefont
  {{Quataert}}}, \bibinfo {author} {\bibfnamefont {K.~J.}\ \bibnamefont
  {{Burns}}}, \bibinfo {author} {\bibfnamefont {G.~M.}\ \bibnamefont
  {{Vasil}}}, \emph {et~al.},\ }\href {https://doi.org/10.1093/mnras/stv2564}
  {\bibfield  {journal} {\bibinfo  {journal} {Monthly Notices of the Royal
  Astronomical Society}\ }\textbf {\bibinfo {volume} {455}},\ \bibinfo {pages}
  {4274} (\bibinfo {year} {2016})},\ \Eprint {https://arxiv.org/abs/1509.03630}
  {arXiv:1509.03630 [astro-ph.IM]} \BibitemShut {NoStop}%
\bibitem [{Sup()}]{SupplementalMaterial}%
  \BibitemOpen
  \href@noop {} {}\bibinfo {note} {See Supplemental Material at [URL will be
  inserted by publisher] for the run inventory, initial stretching model,
  energy and spectral diagnostics, and secondary-instability and vertical-field
  onset analysis.}\BibitemShut {Stop}%
\bibitem [{\citenamefont {Palenzuela}\ \emph {et~al.}(2022)\citenamefont
  {Palenzuela}, \citenamefont {Aguilera-Miret}, \citenamefont {Carrasco},
  \citenamefont {Ciolfi}, \citenamefont {Kalinani}, \citenamefont {Kastaun},
  \citenamefont {Mi{\~n}ano},\ and\ \citenamefont
  {Vigan{\`o}}}]{Palenzuela2022}%
  \BibitemOpen
  \bibfield  {author} {\bibinfo {author} {\bibfnamefont {C.}~\bibnamefont
  {Palenzuela}}, \bibinfo {author} {\bibfnamefont {R.}~\bibnamefont
  {Aguilera-Miret}}, \bibinfo {author} {\bibfnamefont {F.}~\bibnamefont
  {Carrasco}}, \bibinfo {author} {\bibfnamefont {R.}~\bibnamefont {Ciolfi}},
  \bibinfo {author} {\bibfnamefont {J.~V.}\ \bibnamefont {Kalinani}}, \emph
  {et~al.},\ }\href {https://doi.org/10.1103/PhysRevD.106.023013} {\bibfield
  {journal} {\bibinfo  {journal} {Physical Review D}\ }\textbf {\bibinfo
  {volume} {106}},\ \bibinfo {pages} {023013} (\bibinfo {year} {2022})},\
  \Eprint {https://arxiv.org/abs/2112.08413} {arXiv:2112.08413 [gr-qc]}
  \BibitemShut {NoStop}%
\bibitem [{\citenamefont {Zel'dovich}(1957)}]{Zeldovich1957}%
  \BibitemOpen
  \bibfield  {author} {\bibinfo {author} {\bibfnamefont {Y.~B.}\ \bibnamefont
  {Zel'dovich}},\ }\href {https://www.jetp.ras.ru/cgi-bin/dn/e_004_03_0460.pdf}
  {\bibfield  {journal} {\bibinfo  {journal} {Soviet Physics JETP}\ }\textbf
  {\bibinfo {volume} {4}},\ \bibinfo {pages} {460} (\bibinfo {year}
  {1957})}\BibitemShut {NoStop}%
\bibitem [{\citenamefont {Cowling}(1933)}]{Cowling1933}%
  \BibitemOpen
  \bibfield  {author} {\bibinfo {author} {\bibfnamefont {T.~G.}\ \bibnamefont
  {Cowling}},\ }\href {https://doi.org/10.1093/mnras/94.1.39} {\bibfield
  {journal} {\bibinfo  {journal} {Monthly Notices of the Royal Astronomical
  Society}\ }\textbf {\bibinfo {volume} {94}},\ \bibinfo {pages} {39} (\bibinfo
  {year} {1933})}\BibitemShut {NoStop}%
\bibitem [{\citenamefont {Matsumoto}\ and\ \citenamefont
  {Hoshino}(2004)}]{Matsumoto2004}%
  \BibitemOpen
  \bibfield  {author} {\bibinfo {author} {\bibfnamefont {Y.}~\bibnamefont
  {Matsumoto}}\ and\ \bibinfo {author} {\bibfnamefont {M.}~\bibnamefont
  {Hoshino}},\ }\href {https://doi.org/10.1029/2003GL018195} {\bibfield
  {journal} {\bibinfo  {journal} {Geophysical Research Letters}\ }\textbf
  {\bibinfo {volume} {31}},\ \bibinfo {pages} {L02807} (\bibinfo {year}
  {2004})}\BibitemShut {NoStop}%
\bibitem [{\citenamefont {Dixit}\ and\ \citenamefont
  {Govindarajan}(2010)}]{Dixit2010}%
  \BibitemOpen
  \bibfield  {author} {\bibinfo {author} {\bibfnamefont {H.~N.}\ \bibnamefont
  {Dixit}}\ and\ \bibinfo {author} {\bibfnamefont {R.}~\bibnamefont
  {Govindarajan}},\ }\href {https://doi.org/10.1017/S0022112009992515}
  {\bibfield  {journal} {\bibinfo  {journal} {Journal of Fluid Mechanics}\
  }\textbf {\bibinfo {volume} {646}},\ \bibinfo {pages} {415} (\bibinfo {year}
  {2010})}\BibitemShut {NoStop}%
\bibitem [{\citenamefont {Skoutnev}\ \emph
  {et~al.}(2021{\natexlab{a}})\citenamefont {Skoutnev}, \citenamefont {Most},
  \citenamefont {Bhattacharjee},\ and\ \citenamefont
  {Philippov}}]{Skoutnev2021_RTI_dynamo}%
  \BibitemOpen
  \bibfield  {author} {\bibinfo {author} {\bibfnamefont {V.}~\bibnamefont
  {Skoutnev}}, \bibinfo {author} {\bibfnamefont {E.~R.}\ \bibnamefont {Most}},
  \bibinfo {author} {\bibfnamefont {A.}~\bibnamefont {Bhattacharjee}},\ and\
  \bibinfo {author} {\bibfnamefont {A.~A.}\ \bibnamefont {Philippov}},\ }\href
  {https://doi.org/10.3847/1538-4357/ac1ba4} {\bibfield  {journal} {\bibinfo
  {journal} {The Astrophysical Journal}\ }\textbf {\bibinfo {volume} {921}},\
  \bibinfo {pages} {75} (\bibinfo {year} {2021}{\natexlab{a}})}\BibitemShut
  {NoStop}%
\bibitem [{Note1()}]{Note1}%
  \BibitemOpen
  \bibinfo {note} {Movies showing different slice orientations and
  three-dimensional fly-throughs are available \protect \href
  {https://astro-beattie.com/movies/}{here}.}\BibitemShut {Stop}%
\bibitem [{Note2()}]{Note2}%
  \BibitemOpen
  \bibinfo {note} {Initially $A= (\rho _h - \rho _l)/ (\rho _h + \rho _l) =
  5/11$, where $\rho _l$ and $\rho _h$ are the densities of the light and dense
  fluids.}\BibitemShut {Stop}%
\bibitem [{\citenamefont {Musolino}\ \emph {et~al.}(2025)\citenamefont
  {Musolino}, \citenamefont {Rezzolla},\ and\ \citenamefont
  {Most}}]{Musolino:2024sju}%
  \BibitemOpen
  \bibfield  {author} {\bibinfo {author} {\bibfnamefont {C.}~\bibnamefont
  {Musolino}}, \bibinfo {author} {\bibfnamefont {L.}~\bibnamefont {Rezzolla}},\
  and\ \bibinfo {author} {\bibfnamefont {E.~R.}\ \bibnamefont {Most}},\ }\href
  {https://doi.org/10.3847/2041-8213/adcd6d} {\bibfield  {journal} {\bibinfo
  {journal} {Astrophys. J. Lett.}\ }\textbf {\bibinfo {volume} {984}},\
  \bibinfo {pages} {L61} (\bibinfo {year} {2025})},\ \Eprint
  {https://arxiv.org/abs/2410.06253} {arXiv:2410.06253 [astro-ph.HE]}
  \BibitemShut {NoStop}%
\bibitem [{\citenamefont {Schober}\ \emph {et~al.}(2012)\citenamefont
  {Schober}, \citenamefont {Schleicher}, \citenamefont {Federrath},
  \citenamefont {Klessen},\ and\ \citenamefont
  {Banerjee}}]{Schober2012_highPm}%
  \BibitemOpen
  \bibfield  {author} {\bibinfo {author} {\bibfnamefont {J.}~\bibnamefont
  {Schober}}, \bibinfo {author} {\bibfnamefont {D.}~\bibnamefont {Schleicher}},
  \bibinfo {author} {\bibfnamefont {C.}~\bibnamefont {Federrath}}, \bibinfo
  {author} {\bibfnamefont {R.}~\bibnamefont {Klessen}},\ and\ \bibinfo {author}
  {\bibfnamefont {R.}~\bibnamefont {Banerjee}},\ }\href
  {https://doi.org/10.1103/PhysRevE.85.026303} {\bibfield  {journal} {\bibinfo
  {journal} {Physical Review E}\ }\textbf {\bibinfo {volume} {85}},\ \bibinfo
  {pages} {026303} (\bibinfo {year} {2012})},\ \Eprint
  {https://arxiv.org/abs/1109.4571} {arXiv:1109.4571} \BibitemShut {NoStop}%
\bibitem [{\citenamefont {{Grete}}\ \emph {et~al.}(2017)\citenamefont
  {{Grete}}, \citenamefont {{O'Shea}}, \citenamefont {{Beckwith}},
  \citenamefont {{Schmidt}},\ and\ \citenamefont
  {{Christlieb}}}]{Grete2017_transfer_funcs}%
  \BibitemOpen
  \bibfield  {author} {\bibinfo {author} {\bibfnamefont {P.}~\bibnamefont
  {{Grete}}}, \bibinfo {author} {\bibfnamefont {B.~W.}\ \bibnamefont
  {{O'Shea}}}, \bibinfo {author} {\bibfnamefont {K.}~\bibnamefont
  {{Beckwith}}}, \bibinfo {author} {\bibfnamefont {W.}~\bibnamefont
  {{Schmidt}}},\ and\ \bibinfo {author} {\bibfnamefont {A.}~\bibnamefont
  {{Christlieb}}},\ }\href {https://doi.org/10.1063/1.4990613} {\bibfield
  {journal} {\bibinfo  {journal} {Physics of Plasmas}\ }\textbf {\bibinfo
  {volume} {24}},\ \bibinfo {eid} {092311} (\bibinfo {year} {2017})},\ \Eprint
  {https://arxiv.org/abs/1706.06339} {arXiv:1706.06339 [physics.plasm-ph]}
  \BibitemShut {NoStop}%
\bibitem [{\citenamefont {{Beattie}}\ \emph
  {et~al.}(2025{\natexlab{a}})\citenamefont {{Beattie}}, \citenamefont
  {{Kolborg}}, \citenamefont {{Ramirez-Ruiz}},\ and\ \citenamefont
  {{Federrath}}}]{Beattie2025_slk}%
  \BibitemOpen
  \bibfield  {author} {\bibinfo {author} {\bibfnamefont {J.~R.}\ \bibnamefont
  {{Beattie}}}, \bibinfo {author} {\bibfnamefont {A.~N.}\ \bibnamefont
  {{Kolborg}}}, \bibinfo {author} {\bibfnamefont {E.}~\bibnamefont
  {{Ramirez-Ruiz}}},\ and\ \bibinfo {author} {\bibfnamefont {C.}~\bibnamefont
  {{Federrath}}},\ }\href {https://doi.org/10.3847/1538-4357/ae07cd} {\bibfield
   {journal} {\bibinfo  {journal} {\apj}\ }\textbf {\bibinfo {volume} {994}},\
  \bibinfo {eid} {193} (\bibinfo {year} {2025}{\natexlab{a}})},\ \Eprint
  {https://arxiv.org/abs/2501.09855} {arXiv:2501.09855 [astro-ph.GA]}
  \BibitemShut {NoStop}%
\bibitem [{\citenamefont {{Merlini}}\ \emph {et~al.}(2026)\citenamefont
  {{Merlini}}, \citenamefont {{Beattie}},\ and\ \citenamefont
  {{Valenzuela-Villaseca}}}]{Merlini2026_shock_interaction}%
  \BibitemOpen
  \bibfield  {author} {\bibinfo {author} {\bibfnamefont {S.}~\bibnamefont
  {{Merlini}}}, \bibinfo {author} {\bibfnamefont {J.~R.}\ \bibnamefont
  {{Beattie}}},\ and\ \bibinfo {author} {\bibfnamefont {V.}~\bibnamefont
  {{Valenzuela-Villaseca}}},\ }\href
  {https://doi.org/10.48550/arXiv.2605.21672} {\bibfield  {journal} {\bibinfo
  {journal} {arXiv e-prints}\ ,\ \bibinfo {eid} {arXiv:2605.21672}} (\bibinfo
  {year} {2026})},\ \Eprint {https://arxiv.org/abs/2605.21672}
  {arXiv:2605.21672 [physics.plasm-ph]} \BibitemShut {NoStop}%
\bibitem [{\citenamefont {{Beresnyak}}(2012)}]{Beresnyak2012_universal_ssd}%
  \BibitemOpen
  \bibfield  {author} {\bibinfo {author} {\bibfnamefont {A.}~\bibnamefont
  {{Beresnyak}}},\ }\href {https://doi.org/10.1103/PhysRevLett.108.035002}
  {\bibfield  {journal} {\bibinfo  {journal} {\prl}\ }\textbf {\bibinfo
  {volume} {108}},\ \bibinfo {eid} {035002} (\bibinfo {year} {2012})},\ \Eprint
  {https://arxiv.org/abs/1109.4644} {arXiv:1109.4644 [astro-ph.GA]}
  \BibitemShut {NoStop}%
\bibitem [{\citenamefont {Guilet}\ \emph {et~al.}(2017)\citenamefont {Guilet},
  \citenamefont {Bauswein}, \citenamefont {Just},\ and\ \citenamefont
  {Janka}}]{Guilet:2016sqd}%
  \BibitemOpen
  \bibfield  {author} {\bibinfo {author} {\bibfnamefont {J.}~\bibnamefont
  {Guilet}}, \bibinfo {author} {\bibfnamefont {A.}~\bibnamefont {Bauswein}},
  \bibinfo {author} {\bibfnamefont {O.}~\bibnamefont {Just}},\ and\ \bibinfo
  {author} {\bibfnamefont {H.-T.}\ \bibnamefont {Janka}},\ }\href
  {https://doi.org/10.1093/mnras/stx1739} {\bibfield  {journal} {\bibinfo
  {journal} {Mon. Not. Roy. Astron. Soc.}\ }\textbf {\bibinfo {volume} {471}},\
  \bibinfo {pages} {1879} (\bibinfo {year} {2017})},\ \Eprint
  {https://arxiv.org/abs/1610.08532} {arXiv:1610.08532 [astro-ph.HE]}
  \BibitemShut {NoStop}%
\bibitem [{\citenamefont
  {Gavassino}(2025)}]{Gavassino2025_shear_viscosity_radiation}%
  \BibitemOpen
  \bibfield  {author} {\bibinfo {author} {\bibfnamefont {L.}~\bibnamefont
  {Gavassino}},\ }\bibfield  {journal} {\bibinfo  {journal} {Physical Review
  D}\ }\textbf {\bibinfo {volume} {111}},\ \href
  {https://doi.org/10.1103/physrevd.111.l081306} {10.1103/physrevd.111.l081306}
  (\bibinfo {year} {2025})\BibitemShut {NoStop}%
\bibitem [{\citenamefont {{Reboul-Salze}}\ \emph {et~al.}(2025)\citenamefont
  {{Reboul-Salze}}, \citenamefont {{Barr{\`e}re}}, \citenamefont {{Kiuchi}},
  \citenamefont {{Guilet}}, \citenamefont {{Raynaud}}, \citenamefont
  {{Fujibayashi}},\ and\ \citenamefont
  {{Shibata}}}]{ReboulSalze2025_neutrino_viscosity}%
  \BibitemOpen
  \bibfield  {author} {\bibinfo {author} {\bibfnamefont {A.}~\bibnamefont
  {{Reboul-Salze}}}, \bibinfo {author} {\bibfnamefont {P.}~\bibnamefont
  {{Barr{\`e}re}}}, \bibinfo {author} {\bibfnamefont {K.}~\bibnamefont
  {{Kiuchi}}}, \bibinfo {author} {\bibfnamefont {J.}~\bibnamefont {{Guilet}}},
  \bibinfo {author} {\bibfnamefont {R.}~\bibnamefont {{Raynaud}}}, \emph
  {et~al.},\ }\href {https://doi.org/10.1051/0004-6361/202453126} {\bibfield
  {journal} {\bibinfo  {journal} {\aap}\ }\textbf {\bibinfo {volume} {699}},\
  \bibinfo {eid} {A4} (\bibinfo {year} {2025})},\ \Eprint
  {https://arxiv.org/abs/2411.19328} {arXiv:2411.19328 [astro-ph.HE]}
  \BibitemShut {NoStop}%
\bibitem [{\citenamefont {Ng}\ and\ \citenamefont
  {Most}(2026{\natexlab{b}})}]{NgMost2026}%
  \BibitemOpen
  \bibfield  {author} {\bibinfo {author} {\bibfnamefont {H.~H.-Y.}\
  \bibnamefont {Ng}}\ and\ \bibinfo {author} {\bibfnamefont {E.~R.}\
  \bibnamefont {Most}},\ }\bibfield  {journal} {\bibinfo  {journal} {arXiv
  e-prints}\ }\href {https://doi.org/10.48550/arXiv.2607.28556}
  {10.48550/arXiv.2607.28556} (\bibinfo {year} {2026}{\natexlab{b}}),\ \Eprint
  {https://arxiv.org/abs/2607.28556} {arXiv:2607.28556 [astro-ph.HE]}
  \BibitemShut {NoStop}%
\bibitem [{\citenamefont {Tripathi}\ \emph {et~al.}(2026)\citenamefont
  {Tripathi}, \citenamefont {Fraser}, \citenamefont {Terry}, \citenamefont
  {Zweibel}, \citenamefont {Pueschel},\ and\ \citenamefont
  {Fan}}]{Tripathi2026_jets}%
  \BibitemOpen
  \bibfield  {author} {\bibinfo {author} {\bibfnamefont {B.}~\bibnamefont
  {Tripathi}}, \bibinfo {author} {\bibfnamefont {A.~E.}\ \bibnamefont
  {Fraser}}, \bibinfo {author} {\bibfnamefont {P.~W.}\ \bibnamefont {Terry}},
  \bibinfo {author} {\bibfnamefont {E.~G.}\ \bibnamefont {Zweibel}}, \bibinfo
  {author} {\bibfnamefont {M.~J.}\ \bibnamefont {Pueschel}}, \emph {et~al.},\
  }\href {https://doi.org/10.1038/s41586-025-09912-0} {\bibfield  {journal}
  {\bibinfo  {journal} {Nature}\ }\textbf {\bibinfo {volume} {649}},\ \bibinfo
  {pages} {848} (\bibinfo {year} {2026})},\ \Eprint
  {https://arxiv.org/abs/2608.12530} {arXiv:2608.12530} \BibitemShut {NoStop}%
\bibitem [{\citenamefont {{Chan}}\ \emph {et~al.}(2026)\citenamefont {{Chan}},
  \citenamefont {{Gottlieb}}, \citenamefont {{Jacquemin-Ide}}, \citenamefont
  {{Cantiello}},\ and\ \citenamefont {{Renzo}}}]{Chan2026_jets_from_scratch}%
  \BibitemOpen
  \bibfield  {author} {\bibinfo {author} {\bibfnamefont {H.-S.}\ \bibnamefont
  {{Chan}}}, \bibinfo {author} {\bibfnamefont {O.}~\bibnamefont {{Gottlieb}}},
  \bibinfo {author} {\bibfnamefont {J.}~\bibnamefont {{Jacquemin-Ide}}},
  \bibinfo {author} {\bibfnamefont {M.}~\bibnamefont {{Cantiello}}},\ and\
  \bibinfo {author} {\bibfnamefont {M.}~\bibnamefont {{Renzo}}},\ }\href
  {https://doi.org/10.3847/2041-8213/ae9156} {\bibfield  {journal} {\bibinfo
  {journal} {\apjl}\ }\textbf {\bibinfo {volume} {1007}},\ \bibinfo {eid} {L41}
  (\bibinfo {year} {2026})},\ \Eprint {https://arxiv.org/abs/2605.12931}
  {arXiv:2605.12931 [astro-ph.HE]} \BibitemShut {NoStop}%
\bibitem [{Note3()}]{Note3}%
  \BibitemOpen
  \bibinfo {note} {The corresponding wavelength is \begin {align} \lambda
  _{\protect \mathrm {RTI}}^{\infty } &= 2\pi \left (\protect \frac {16\nu ^2
  R}{A u_\theta ^2}\right )^{1/3}. \end {align} Using $\Delta U=0.3c$,
  $u_\theta =\Delta U/2$, $\delta _0=50\protect \,\protect \mathrm {m}$,
  density ratio $\rho _h/\rho _l=8/3$, Atwood number $A=5/11\simeq 0.4545$, and
  kinematic viscosity $\nu _\nu =10^7\protect \,\protect \mathrm {cm^2\protect
  \,s^{-1}}$ \cite
  {Skoutnev2021_RTI,Gavassino2025_shear_viscosity_radiation,ReboulSalze2025_neutrino_viscosity,NgMost2026},
  for the BNS system we obtain \(\lambda _{\protect \rm RTI}^{\infty }\approx
  6\protect \,\protect \mathrm {cm}\), below the spatial resolution of current
  global and even local zoom-in simulations \cite
  {Gutierrez2026,Kiuchi2026}.}\BibitemShut {Stop}%
\bibitem [{\citenamefont {Guilet}\ \emph {et~al.}(2010)\citenamefont {Guilet},
  \citenamefont {Sato},\ and\ \citenamefont {Foglizzo}}]{Guilet2010_SASI}%
  \BibitemOpen
  \bibfield  {author} {\bibinfo {author} {\bibfnamefont {J.}~\bibnamefont
  {Guilet}}, \bibinfo {author} {\bibfnamefont {J.}~\bibnamefont {Sato}},\ and\
  \bibinfo {author} {\bibfnamefont {T.}~\bibnamefont {Foglizzo}},\ }\href
  {https://doi.org/10.1088/0004-637X/713/2/1350} {\bibfield  {journal}
  {\bibinfo  {journal} {The Astrophysical Journal}\ }\textbf {\bibinfo {volume}
  {713}},\ \bibinfo {pages} {1350} (\bibinfo {year} {2010})},\ \Eprint
  {https://arxiv.org/abs/0910.3953} {arXiv:0910.3953} \BibitemShut {NoStop}%
\bibitem [{\citenamefont {Roediger}\ \emph {et~al.}(2013)\citenamefont
  {Roediger}, \citenamefont {Kraft}, \citenamefont {Forman}, \citenamefont
  {Nulsen},\ and\ \citenamefont {Churazov}}]{Roediger2013_Virgo}%
  \BibitemOpen
  \bibfield  {author} {\bibinfo {author} {\bibfnamefont {E.}~\bibnamefont
  {Roediger}}, \bibinfo {author} {\bibfnamefont {R.~P.}\ \bibnamefont {Kraft}},
  \bibinfo {author} {\bibfnamefont {W.~R.}\ \bibnamefont {Forman}}, \bibinfo
  {author} {\bibfnamefont {P.~E.~J.}\ \bibnamefont {Nulsen}},\ and\ \bibinfo
  {author} {\bibfnamefont {E.}~\bibnamefont {Churazov}},\ }\href
  {https://doi.org/10.1088/0004-637X/764/1/60} {\bibfield  {journal} {\bibinfo
  {journal} {The Astrophysical Journal}\ }\textbf {\bibinfo {volume} {764}},\
  \bibinfo {pages} {60} (\bibinfo {year} {2013})},\ \Eprint
  {https://arxiv.org/abs/1211.4874} {arXiv:1211.4874} \BibitemShut {NoStop}%
\bibitem [{\citenamefont
  {{Ferri{\`e}re}}(2020)}]{Ferriere2020_plasma_turbulence}%
  \BibitemOpen
  \bibfield  {author} {\bibinfo {author} {\bibfnamefont {K.}~\bibnamefont
  {{Ferri{\`e}re}}},\ }\href {https://doi.org/10.1088/1361-6587/ab49eb}
  {\bibfield  {journal} {\bibinfo  {journal} {Plasma Physics and Controlled
  Fusion}\ }\textbf {\bibinfo {volume} {62}},\ \bibinfo {pages} {014014}
  (\bibinfo {year} {2020})},\ \Eprint {https://arxiv.org/abs/1912.08237}
  {arXiv:1912.08237 [astro-ph.GA]} \BibitemShut {NoStop}%
\bibitem [{\citenamefont {Fontane}\ and\ \citenamefont
  {Joly}(2008)}]{FontaneJoly2008}%
  \BibitemOpen
  \bibfield  {author} {\bibinfo {author} {\bibfnamefont {J.}~\bibnamefont
  {Fontane}}\ and\ \bibinfo {author} {\bibfnamefont {L.}~\bibnamefont {Joly}},\
  }\href {https://doi.org/10.1017/S0022112008002966} {\bibfield  {journal}
  {\bibinfo  {journal} {Journal of Fluid Mechanics}\ }\textbf {\bibinfo
  {volume} {612}},\ \bibinfo {pages} {237} (\bibinfo {year}
  {2008})}\BibitemShut {NoStop}%
\bibitem [{\citenamefont {Matsumoto}\ and\ \citenamefont
  {Seki}(2007)}]{MatsumotoSeki2007}%
  \BibitemOpen
  \bibfield  {author} {\bibinfo {author} {\bibfnamefont {Y.}~\bibnamefont
  {Matsumoto}}\ and\ \bibinfo {author} {\bibfnamefont {K.}~\bibnamefont
  {Seki}},\ }\href {https://doi.org/10.1029/2006JA012114} {\bibfield  {journal}
  {\bibinfo  {journal} {Journal of Geophysical Research: Space Physics}\
  }\textbf {\bibinfo {volume} {112}},\ \bibinfo {pages} {A06223} (\bibinfo
  {year} {2007})}\BibitemShut {NoStop}%
\bibitem [{\citenamefont {Skoutnev}\ \emph
  {et~al.}(2021{\natexlab{b}})\citenamefont {Skoutnev}, \citenamefont {Most},
  \citenamefont {Bhattacharjee},\ and\ \citenamefont
  {Philippov}}]{Skoutnev2021_RTI}%
  \BibitemOpen
  \bibfield  {author} {\bibinfo {author} {\bibfnamefont {V.}~\bibnamefont
  {Skoutnev}}, \bibinfo {author} {\bibfnamefont {E.~R.}\ \bibnamefont {Most}},
  \bibinfo {author} {\bibfnamefont {A.}~\bibnamefont {Bhattacharjee}},\ and\
  \bibinfo {author} {\bibfnamefont {A.~A.}\ \bibnamefont {Philippov}},\ }\href
  {https://doi.org/10.3847/1538-4357/ac1ba4} {\bibfield  {journal} {\bibinfo
  {journal} {The Astrophysical Journal}\ }\textbf {\bibinfo {volume} {921}},\
  \bibinfo {pages} {75} (\bibinfo {year} {2021}{\natexlab{b}})},\ \Eprint
  {https://arxiv.org/abs/2106.04787} {arXiv:2106.04787 [physics.flu-dyn]}
  \BibitemShut {NoStop}%
\bibitem [{\citenamefont {{Beattie}}\ \emph
  {et~al.}(2025{\natexlab{b}})\citenamefont {{Beattie}}, \citenamefont
  {{Federrath}}, \citenamefont {{Kriel}}, \citenamefont {{Hew}},\ and\
  \citenamefont {{Bhattacharjee}}}]{Beattie2025_bulk_viscosity}%
  \BibitemOpen
  \bibfield  {author} {\bibinfo {author} {\bibfnamefont {J.~R.}\ \bibnamefont
  {{Beattie}}}, \bibinfo {author} {\bibfnamefont {C.}~\bibnamefont
  {{Federrath}}}, \bibinfo {author} {\bibfnamefont {N.}~\bibnamefont
  {{Kriel}}}, \bibinfo {author} {\bibfnamefont {J.~K.~J.}\ \bibnamefont
  {{Hew}}},\ and\ \bibinfo {author} {\bibfnamefont {A.}~\bibnamefont
  {{Bhattacharjee}}},\ }\href {https://doi.org/10.1093/mnras/staf1318}
  {\bibfield  {journal} {\bibinfo  {journal} {\mnras}\ }\textbf {\bibinfo
  {volume} {542}},\ \bibinfo {pages} {2669} (\bibinfo {year}
  {2025}{\natexlab{b}})},\ \Eprint {https://arxiv.org/abs/2312.03984}
  {arXiv:2312.03984 [astro-ph.GA]} \BibitemShut {NoStop}%
\bibitem [{\citenamefont {{Berlok}}\ and\ \citenamefont
  {{Pfrommer}}(2019)}]{Berlok2019_KHI_linear_theory}%
  \BibitemOpen
  \bibfield  {author} {\bibinfo {author} {\bibfnamefont {T.}~\bibnamefont
  {{Berlok}}}\ and\ \bibinfo {author} {\bibfnamefont {C.}~\bibnamefont
  {{Pfrommer}}},\ }\href {https://doi.org/10.1093/mnras/stz379} {\bibfield
  {journal} {\bibinfo  {journal} {Monthly Notices of the Royal Astronomical
  Society}\ }\textbf {\bibinfo {volume} {485}},\ \bibinfo {pages} {908}
  (\bibinfo {year} {2019})},\ \Eprint {https://arxiv.org/abs/1902.01403}
  {arXiv:1902.01403 [astro-ph.GA]} \BibitemShut {NoStop}%
\bibitem [{\citenamefont {Beattie}\ and\ \citenamefont
  {Bhattacharjee}(2026)}]{Beattie2026_sda_prl}%
  \BibitemOpen
  \bibfield  {author} {\bibinfo {author} {\bibfnamefont {J.~R.}\ \bibnamefont
  {Beattie}}\ and\ \bibinfo {author} {\bibfnamefont {A.}~\bibnamefont
  {Bhattacharjee}},\ }\href {https://doi.org/10.1103/zp1g-rkjc} {\bibfield
  {journal} {\bibinfo  {journal} {Phys. Rev. Lett.}\ }\textbf {\bibinfo
  {volume} {137}},\ \bibinfo {pages} {165202} (\bibinfo {year}
  {2026})}\BibitemShut {NoStop}%
\bibitem [{\citenamefont {{Fielding}}\ \emph {et~al.}(2023)\citenamefont
  {{Fielding}}, \citenamefont {{Ripperda}},\ and\ \citenamefont
  {{Philippov}}}]{Fielding2023_plasmoid}%
  \BibitemOpen
  \bibfield  {author} {\bibinfo {author} {\bibfnamefont {D.~B.}\ \bibnamefont
  {{Fielding}}}, \bibinfo {author} {\bibfnamefont {B.}~\bibnamefont
  {{Ripperda}}},\ and\ \bibinfo {author} {\bibfnamefont {A.~A.}\ \bibnamefont
  {{Philippov}}},\ }\href {https://doi.org/10.3847/2041-8213/accf1f} {\bibfield
   {journal} {\bibinfo  {journal} {\apjl}\ }\textbf {\bibinfo {volume} {949}},\
  \bibinfo {eid} {L5} (\bibinfo {year} {2023})},\ \Eprint
  {https://arxiv.org/abs/2211.06434} {arXiv:2211.06434 [astro-ph.GA]}
  \BibitemShut {NoStop}%
\bibitem [{\citenamefont {{Beattie}}\ \emph
  {et~al.}(2025{\natexlab{c}})\citenamefont {{Beattie}}, \citenamefont
  {{Federrath}}, \citenamefont {{Klessen}}, \citenamefont {{Cielo}},\ and\
  \citenamefont {{Bhattacharjee}}}]{Beattie2025_nature_astronomy}%
  \BibitemOpen
  \bibfield  {author} {\bibinfo {author} {\bibfnamefont {J.~R.}\ \bibnamefont
  {{Beattie}}}, \bibinfo {author} {\bibfnamefont {C.}~\bibnamefont
  {{Federrath}}}, \bibinfo {author} {\bibfnamefont {R.~S.}\ \bibnamefont
  {{Klessen}}}, \bibinfo {author} {\bibfnamefont {S.}~\bibnamefont {{Cielo}}},\
  and\ \bibinfo {author} {\bibfnamefont {A.}~\bibnamefont {{Bhattacharjee}}},\
  }\href {https://doi.org/10.1038/s41550-025-02551-5} {\bibfield  {journal}
  {\bibinfo  {journal} {Nature Astronomy}\ }\textbf {\bibinfo {volume} {9}},\
  \bibinfo {pages} {1195} (\bibinfo {year} {2025}{\natexlab{c}})},\ \Eprint
  {https://arxiv.org/abs/2504.07136} {arXiv:2504.07136 [astro-ph.GA]}
  \BibitemShut {NoStop}%
\bibitem [{\citenamefont {{Stone}}\ and\ \citenamefont
  {{Gardiner}}(2007)}]{Stone2007_RTI}%
  \BibitemOpen
  \bibfield  {author} {\bibinfo {author} {\bibfnamefont {J.~M.}\ \bibnamefont
  {{Stone}}}\ and\ \bibinfo {author} {\bibfnamefont {T.}~\bibnamefont
  {{Gardiner}}},\ }\href {https://doi.org/10.1086/523099} {\bibfield  {journal}
  {\bibinfo  {journal} {\apj}\ }\textbf {\bibinfo {volume} {671}},\ \bibinfo
  {pages} {1726} (\bibinfo {year} {2007})},\ \Eprint
  {https://arxiv.org/abs/0709.0452} {arXiv:0709.0452 [astro-ph]} \BibitemShut
  {NoStop}%
\bibitem [{\citenamefont {Sullivan}\ and\ \citenamefont
  {Kaszynski}(2019)}]{Sullivan2019pyvista}%
  \BibitemOpen
  \bibfield  {author} {\bibinfo {author} {\bibfnamefont {B.}~\bibnamefont
  {Sullivan}}\ and\ \bibinfo {author} {\bibfnamefont {A.}~\bibnamefont
  {Kaszynski}},\ }\href {https://doi.org/10.21105/joss.01450} {\bibfield
  {journal} {\bibinfo  {journal} {Journal of Open Source Software}\ }\textbf
  {\bibinfo {volume} {4}},\ \bibinfo {pages} {1450} (\bibinfo {year}
  {2019})}\BibitemShut {NoStop}%
\end{thebibliography}
